\documentclass[aps,rmp,reprint,amsmath,amssymb,graphicx,longbibliography]{revtex4-2}
\usepackage{CJK}
\usepackage{amssymb}
\usepackage{amsmath}
\usepackage{amsfonts}
\usepackage{graphicx}
\usepackage{rotating}
\usepackage{bm}
\usepackage{color}
\usepackage{url}

\begin{document}

\title{Micius quantum experiments in space}
\author{Chao-Yang Lu}
\affiliation{Hefei National Laboratory for Physical Sciences at Microscale and Department
of Modern Physics, University of Science and Technology of China, Hefei, Anhui
230026, China}
\affiliation{CAS Center for Excellence and Synergetic Innovation Center in Quantum Information and Quantum Physics, University of Science and Technology of China, Shanghai, 201315, China}
\author{Yuan Cao}
\affiliation{Hefei National Laboratory for Physical Sciences at Microscale and Department
of Modern Physics, University of Science and Technology of China, Hefei, Anhui
230026, China}
\affiliation{CAS Center for Excellence and Synergetic Innovation Center in Quantum Information and Quantum Physics, University of Science and Technology of China, Shanghai, 201315, China}
\author{Cheng-Zhi Peng}
\affiliation{Hefei National Laboratory for Physical Sciences at Microscale and Department
of Modern Physics, University of Science and Technology of China, Hefei, Anhui
230026, China}
\affiliation{CAS Center for Excellence and Synergetic Innovation Center in Quantum Information and Quantum Physics, University of Science and Technology of China, Shanghai, 201315, China}
\author{Jian-Wei Pan}
\email{pan@ustc.edu.cn}
\affiliation{Hefei National Laboratory for Physical Sciences at Microscale and Department
of Modern Physics, University of Science and Technology of China, Hefei, Anhui
230026, China}
\affiliation{CAS Center for Excellence and Synergetic Innovation Center in Quantum Information and Quantum Physics, University of Science and Technology of China, Shanghai, 201315, China}


\begin{abstract}

Quantum theory has been successfully validated in numerous laboratory experiments. 
But would such a theory, which excellently describes the behavior of microscopic physical systems, and its predicted phenomena such as quantum entanglement, be still applicable on very large length scales? 
From a practical perspective, how can quantum key distribution---where the security of establishing secret keys between distant parties is ensured by the laws of quantum mechanics---be made technologically useful on a global scale?
Due to photon loss in optical fibers and terrestrial free space, the achievable distance using direct transmission of single photons has been limited to a few hundred kilometers. 
A promising route to testing quantum physics over long distances and in the relativistic regimes, and thus realizing flexible global-scale quantum networks is via the use of satellites and space-based technologies, where a significant advantage is that the photon loss and turbulence predominantly occurs in the lower \textasciitilde10 km of the atmosphere, and most of the photons' transmission path in the space is virtually in vacuum with almost zero absorption and decoherence. 
In this Article, we review the progress in free-space quantum experiments, with a focus on the fast-developing Micius satellite-based quantum communications. 
The perspective of space-ground integrated quantum networks and fundamental quantum optics experiments in space conceivable with satellites are discussed.

\end{abstract}
\maketitle
\tableofcontents

\section{Introduction}

Privacy and security are rooted in human beings since ancient times.
They underpin human dignity and are among the most important human rights.
With the exponential growth of the Internet and the use of \textit{e}-commerce, it is critically important to establish a secure global network.
Cryptography, the use of codes and ciphers to protect secrets, began thousands of years ago.
Traditional public-key cryptography usually relies on the perceived computational intractability of certain mathematical functions.
However, history shows that many advances in classical cryptography were subsequently defeated by advances in cracking.
Thus it has even been suggested that ``human ingenuity cannot concoct a cipher which human ingenuity cannot resolve'' \cite{Poe1841}.

It might come as a surprise that the very fundamental principle of quantum mechanics was exploited to solve this long-standing problem on information security that mathematicians struggled with for centuries.
The first idea along these lines was proposed in the 1970s by Stephen Wiesner who designed quantum banknotes using quantum two-state systems and conjugate encoding, which would be impossible to counterfeit.
The principal drawback of Wiesner's idea was that it would require quantum information in superposition states to be held captive and kept coherent for long periods of time, which appears to be beyond even the technological capability of today, 50 years later.
Inspired by Wiesner's idea, in the 1980s, Charles Bennett and Gilles Brassard put forward a feasible protocol of quantum key distribution (QKD) known as BB84 \cite{BB84}.
This protocol permits two distant communicating parties to produce a common, random string of secret bits, called a secret key.
This key can then be used alongside the one-time pad encryption which, as strictly proven by Shannon in 1949 \cite{Shannon1949Communication}, is an unconditionally secure method to encrypt and decrypt a message.

A remarkable surge of interest in the international scientific and industrial community has propelled quantum cryptography into mainstream computer science and physics in the past few years.
Particularly after the invention of Shor's quantum factoring algorithm in 1994, an important goal of the field is to transform the beautiful idea of QKD into a practically useful technology.
While significant progress has been made in small-scale demonstrations \cite{Bennett1989, BPMEWZ_97}, the real challenge is to increase the communication range to large distances, eventually to a global scale.
Due to photon loss in the channel, the range of secure QKD via direct transmission of single photons is limited to a few hundred kilometers (see Section II and III).
Unlike classical bits, the quantum signal of QKD cannot be noiselessly amplified due to the implications of the quantum non-cloning theorem \cite{nonclone}, where the security of the QKD is rooted \cite{GisinRevModPhys2002}.

One potential solution to this notorious scalability problem is via the use of quantum repeaters \cite{Briegel1998repeater}.
However, the current state of quantum memories \cite{Yang-memory} and quantum repeaters remains far from practical applications in realistic long-distance quantum communications (see Section IV).

A more promising solution for global-scale QKD is to utilize satellites, which can conveniently connect two distant locations on Earth.
An important advantage of satellite-based free-space quantum communications is that the photon loss induced by atmospheric absorption and scattering predominantly occur only in the lower $\sim$10 km of the atmosphere, with about a 3 dB loss on a clear day.
Most of the photon transmission is across a near-vacuum environment, with almost no absorption and decoherence.
The loss caused by beam diffraction is approximately proportional to the square of distance.
By contrast, the losses in fiber channels are predominantly due to the absorption and scattering of the fiber medium, which is proportional to the exponent of the distance.
Thus, for long communicating distances (typically hundreds to thousands of kilometers) the satellite-ground free-space channels will have advantages over fiber-based channels, in terms of channel losses (for more details, see section V).

In addition to QKD, the use of quantum communication in space would be beneficial for the testing of fundamental principles of quantum physics on a large scale.
The confirmation of the laws of physics is restricted by the boundaries of our experimental observations.
For instance, quantum mechanics predicts that it is possible to observe quantum entanglement over any distance; 
however, it is necessary to confirm such a prediction and verify whether unexpected effects (such as the influence of gravitational fields) place some bound on such distances \cite{Paoloreview2019}.
New satellite-based laboratories in space would create vast platforms for fundamental experiments in quantum optics at distances that are previously inaccessible on the ground, such as long-range Bell tests \cite{Bell1964} with human's free will \cite{Hallfreewill, Caofreewill, thebigbelltest}, and the probing of the interaction of quantum mechanics with general relativity \cite{rideout2012fundamental}.

Due to the low-gravity and ultra-stable conditions in space, many experiments related to quantum physics that cannot be undertaken on the ground can be achieved in space. 
High-precision measurements on quantum systems, especially those involves matter waves, in the presence of a gravitational field could challenge our understanding of general relativity and quantum mechanics.
Researchers from Germany, France, and the United States created Bose-Einstein condensate in space and conducted 110 experiments central to matter-wave interferometry, as part of the sounding-rocket mission MAIUS-1 in 2017 \cite{2018MAIUS}.

Exploiting the persistent free fall condition of a low Earth orbit, cold atom systems can take advantage of microgravity, which is high on the to-do-list in the next-generation experiments.
The Cold Atom Lab (CAL) is designed to provide the first ultracold quantum gas experiment aboard the ISS utilizing an apparatus developed, assembled, and qualified by NASA's Jet Propulsion Laboratory (JPL); this facility was successfully launched on 21th May, 2018 \cite{CAL_2018NASA} and has been installed on the ISS.
Recently, a proposal for the mission ``Space Atomic Gravity Explorer'' (SAGE) was presented to the European Space Agency.
It has the scientific objective to investigate gravitational waves, dark matter, and other fundamental aspects of gravity, alongside investigating the connection between gravitational physics and quantum physics using new quantum sensors, namely, optical atomic clocks and atom interferometers based on ultracold strontium atoms \cite{SAGE_2019}.
The successful operation of LISA Pathfinder \cite{LisaPathfinder2016} paves the way for the space optical Laser Interferometer Space Antenna (LISA), which will observe gravitational waves at low frequencies, with a peak sensitivity in the range 1-10 mHz. 
In the SAGE proposal, by using the atomic sensors, the observation of gravitational waves in the low frequency range (from $10^{-3}$ Hz to 10 Hz) is possible.
This proposal will complement the LISA as an alternative technology, and more importantly, it will provide a measurement method in the range between those of the LISA and terrestrial detectors \cite{Kolkowitz2016prd}.

\begin{figure}[bt]
	\centering
	\includegraphics[width=0.475\textwidth]{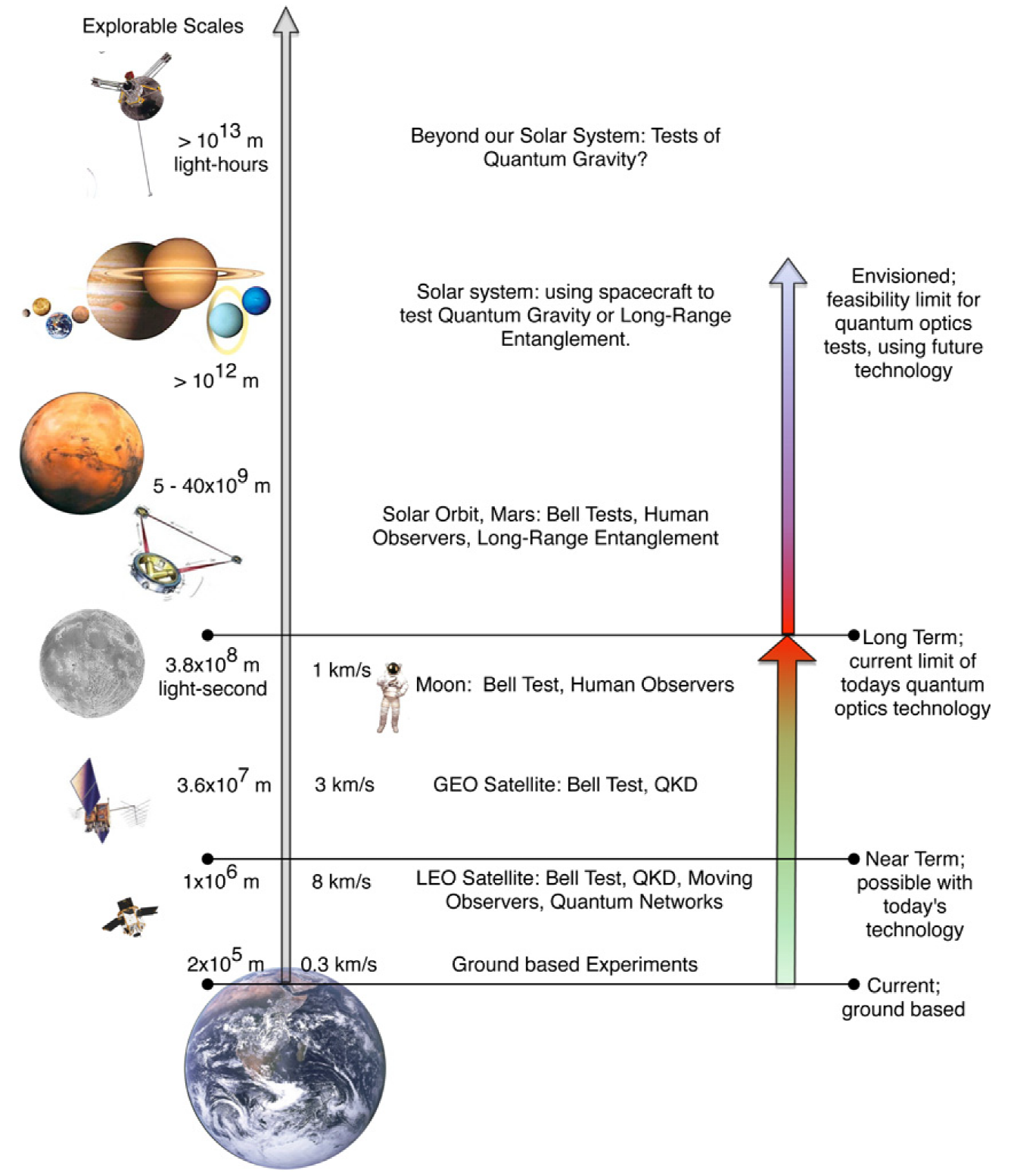}
	\caption{Overview of the distance scales and the corresponding conceived quantum experiments \cite{rideout2012fundamental}.}
	\label{spacescale}
\end{figure}

Using satellite-based free-space channels can be regarded as extending the space scale of quantum experiments from several meters to the magnitude of thousands of kilometers or more, as shown in Fig.\ref{spacescale}~\cite{rideout2012fundamental}, which allows us to explore the nature of the quantum world at increasingly large length scales.
It is of great interest to carry out quantum experiments on such a large space scale in terms of fundamental scientific interest as well as practical applications.

The focus of this article is to review the developments in satellite-based quantum communications and fundamental tests of quantum physics using the Micius satellite.
This review begins by introducing early ideas \cite{aRarity2002, Hwang2003} and preliminary tests \cite{Buttler1998a, aspelmeyer2003long, Kurtsiefer2002}.
To envision a practically working low-earth-orbit satellite under feasible budget, serious numerical analysis and engineering considerations were put forward and summarised in Section III and V. 
Step-by-step ground-based feasibility studies \cite{peng2005, ursin2007entanglement, Villoresi_2008, Jin2010tele, yin2012-100km, ma144kmtele, Yin2013oe, Cao2013oe, Wangdirect2013, Nauerth2013} and developments of the key technologies required are reviewed in Section VI.
For example, entanglement distribution over a terrestrial free-space channel with a length of 13 km was demonstrated in \cite{peng2005}, proving that the entanglement can extend over distances greater than the effective atmospheric thickness.
Bidirectional distribution of entangled photons \cite{yin2012-100km} was later performed over a distance of 102 km with an $\sim$80 dB effective channel loss, which was comparable to that of a two-downlink channel from a satellite to the ground.
Finally, the full verification, in conditions of rapid motion and random movement of satellites, attitude change, vibration, and a high-loss regime, addressing a wide range of all parameters relevant to low-Earth orbit satellites, has been carried out \cite{Wangdirect2013}.
During these feasibility studies, the necessary toolbox for satellite-based long-distance quantum communications has been gradually developed, including robust and compact quantum light sources, narrow beam divergence, time synchronization, and rapid acquiring, pointing, and tracking (APT) technologies, which are the key to the optimization of the link efficiency and overcoming the atmospheric turbulence. 
This body of work was sufficient groundwork to permit the funding of the Micius satellite project for quantum experiments in the framework of Chinese Academy of Sciences Strategic Priority Program.

The satellite, which is dedicated to quantum science experiments, named after the 4th century BCE Chinese philosopher \textit{Micius}, was launched from China in August 2016.
Within a year of the launch, three key milestones for a global-scale quantum communication network were achieved: (1) satellite-to-ground decoy-state QKD with KHz rate over a distance of up to 1200 km \cite{liao2017satellite} and satellite-replayed intercontinental key exchange \cite{Liao2018relay}, (2) satellite-based entanglement distribution to two locations on the Earth separated by 1205 km and subsequent Bell test \cite{yin2017satellite}, and (3) ground-to-satellite quantum teleportation \cite{ren2017satellite}.
These experiments established the possibility of effective link efficiencies through satellite of 12-20 orders of magnitudes greater than direct transmission through optical fibers over a distance of $\sim$1200 km.
A comprehensive survey and an analysis of these works are presented in Section VII, VIII, and IX.

Meanwhile, with the success of the Micius satellite, an international race related to quantum communication experiments in space is predicted to start.
Many satellite projects designed for quantum communications have been approved and funded, such as the Quantum Encryption and Science Satellite (QEYSSat) project in Canada \cite{Jennewein2014QEYSSAT}, and the CubeSat Quantum Communications Mission (CQuCoM) undertaking by a joint research team \cite{2017CubeSat}. 
Meanwhile, more extensive studies of quantum physics in space, including matter-wave in the presence of a gravitational field, are also being implemented or in the planning stages (more details can be found in Section X).

This review ends with an outlook on the future work that needs to be done to eventually build a global-scale practical quantum networks.
Outstanding challenges include enabling daytime operation of QKD \cite{Liao2017daylight}, increasing time and area coverage through the use of higher-orbit satellites, and constructing satellite constellations; these challenges are being considered and addressed by the ongoing research efforts in this emerging field.

\section{Small-scale quantum communications}

\subsection{First proof-of-concept demonstrations}

Quantum superposition is one of the fundamental principles of quantum mechanics; this property distinguishes a quantum state from a classical one.
In a classical two-value system, for example a coin, we find it in either one of its two possible states, that is, either heads or tails.
In its quantum counterpart, however, a two-state quantum system can be found in any superposition of the two possible basis states, e.g., $|\Psi\rangle=(1/\sqrt{2})(|0\rangle+|1\rangle)$, 
where the two orthogonal basis states are denoted by $|0\rangle$ and $|1\rangle$.
In the classical world, two coins can be found in one of the states heads/heads, heads/tails, tails/heads, or tails/tails, and we can identify these four possibilities with the four quantum states $|0\rangle_{1}|0\rangle_{2}$, $|0\rangle_{1}|1\rangle_{2}$, $|1\rangle_{1}|0\rangle_{2}$, and $|1\rangle_{1}|1\rangle_{2}$, which describes the two-state quantum systems.

The superposition principle applies also to more than one quantum system; the two quantum particles are no longer restricted to the four ``classical" basis states above, but it can be found in any superposition, for instance in the following four entangled states\\
\begin{equation}\label{formula1}
|\Psi^{\pm}\rangle = \frac{1}{\sqrt{2}}(|0\rangle_{1}|1\rangle_{2}\pm|1\rangle_{1}|0\rangle_{2}),
\end{equation}
\begin{equation}\label{formula1}
|\Phi^{\pm}\rangle = \frac{1}{\sqrt{2}}(|0\rangle_{1}|0\rangle_{2}\pm|1\rangle_{1}|1\rangle_{2}).
\end{equation}
These entangled states are referred to as ``Bell state" since they maximally violate a Bell inequality \cite{Bell1964, CHSH1969}, showing a stark contradiction between classical local hidden variable theory and quantum mechanics \cite{Bellnonlocality2014RMP}.
Quantum entanglement describes a physical phenomenon whereby the quantum states of a many-particle system cannot be factorized into a product of single-particle wave functions; this applies even when these particles are separated by large distances.
It was first recognized by \cite{EPR1935} and \cite{EPR_Schrodinger}, and experimentally generated by \cite{Wu_EPR} in the annihilation radiation; it was then applied to the Bell test \cite{Clauser1972,Aspect1982}.

\begin{figure*}
  \centering
  \includegraphics[width=0.77\textwidth]{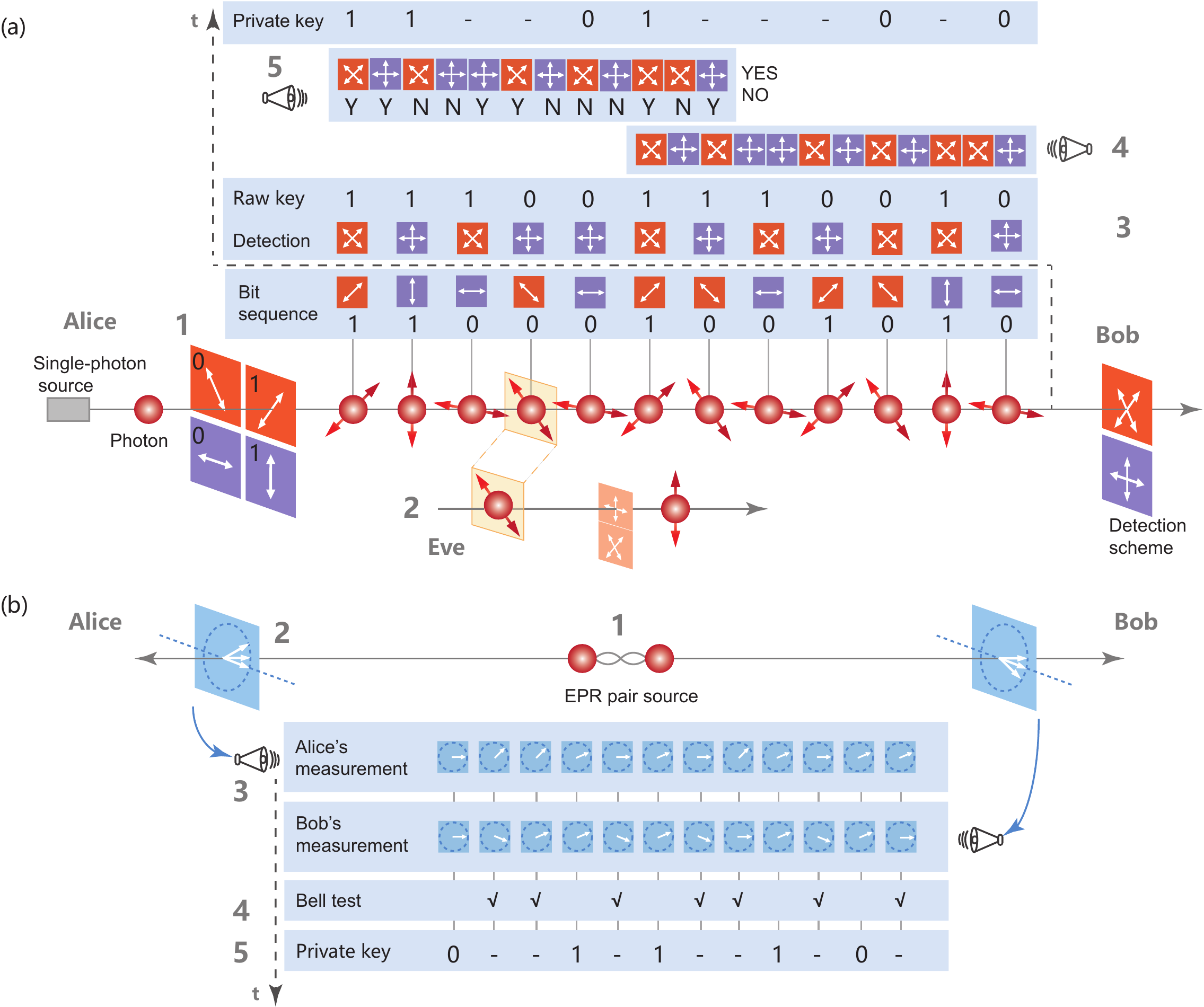}
  \caption{Quantum cryptographic protocols. (a)The BB84 Protocol. The aim of this protocol is for the sender (Alice) to send secret key to the receiver (Bob) by transferring single photons, encoding the information in the quantum states. This protocol exploits four polarization states of photons which span two bases (eg., the horizontal polarization $|H\rangle$ and vertical polarization $|V\rangle$, and the diagonal polarization $\left|45^{\circ}\right\rangle$ and anti-diagonal polarization $\left|-45^{0}\right\rangle$). In information encoding, they use $|H\rangle$ and $\left|-45^{\circ}\right\rangle$ represent bit $0,$ and $|V\rangle$ and $\left|45^{\circ}\right\rangle$ represent bit 1. The operation steps are as follows.
  (1) Alice chooses a group of bit sequence, and encodes these bits in the polarization of photons with random choice of encoding basis.
  (2) Alice sends the photons to Bob.
  (3) Bob randomly chooses the detection scheme to measure the state and obtains the raw key.
  (4) Bob broadcasts his choice of measurement bases for each photon through a classical information channel.
  (5) Alice responds yes or no for the same or different basis that they use for encoding and measurement for each photon.
  (6) They discard the events in which different bases were used and keep the remained data for private key (any eavesdropping in step (2) could be alerted in this final check).
  (b) The Ekert91 Protocol.  This protocol operated for sharing secret keys between Alice and Bob by distributing EPR pairs.
  (1) Alice and Bob firstly share an entangled photon pair in the singlet state $|\Psi^-\rangle$.
  (2) Alice and Bob receive the photon then randomly and independently choose their measurement bases, obtained by rotating the $\{|H\rangle, |V\rangle\}$ basis with angle from the set $\{0, \pi/4, \pi/8\}$ for Alice and $\{0, -\pi/8, \pi/8\}$ for Bob.
  (3) They measure and register a series of photon pairs, after that, broadcast the measurement bases they have used, while keeping the outcomes in secret.
  (4) They use the measurement outcomes with the same angles as raw keys and use the others for Bell inequality test.
  (5) If the Bell inequality is violated, the eavesdropping is excluded therefore the keys are safe (their keys are anti-aligned). Otherwise, they discard all the keys.}
  \label{Fig1}
\end{figure*}

Note that both single-particle and entangled-two-particle states can be applied to QKD.
Single-particle states represent the prepare-and-measure scheme, such as BB84 \cite{BB84}, in which Alice sends each quantum bit (qubit) in one of four states of two complementary bases; B92 \cite{B92} where Alice sends each qubit in one of two non-orthogonal states; six-state \cite{sixstate} in which Alice sends each qubit in one of six states of three complementary bases.
A detailed description of the BB84 protocol is given in Fig.~\ref{Fig1}(a). 
The security of QKD is ensured by the no-cloning theorem \cite{nonclone} which prohibits the precise copying of an unknown quantum state.
The no-cloning theorem prevents the eavesdropper from copying Alice's qubits; information gain is only possible at the expense of introducing disturbance to the signal in any attempt to distinguish between the two non-orthogonal quantum states.
Entanglement-based QKD, including schemes such as Ekert91 \cite{E91} in which entangled pairs of qubits are distributed to Alice and Bob, who then extract key bits by measuring their qubits, as shown in Fig.~\ref{Fig1}(b); BBM92 \cite{Bennett:BBM92:1992} where each party measures half of the EPR pair in one of two complementary bases. 
Note that in Ekert91, Alice and Bob estimate the Eve's information based on the Bell's inequality test; whereas in BBM92, as in the case of BB84, Alice and Bob make use of the privacy amplification to eliminate Eve's information regarding the final key \cite{LoUnconditional1999}.

\begin{figure}[bt]
	\centering
	\includegraphics[width=0.475\textwidth]{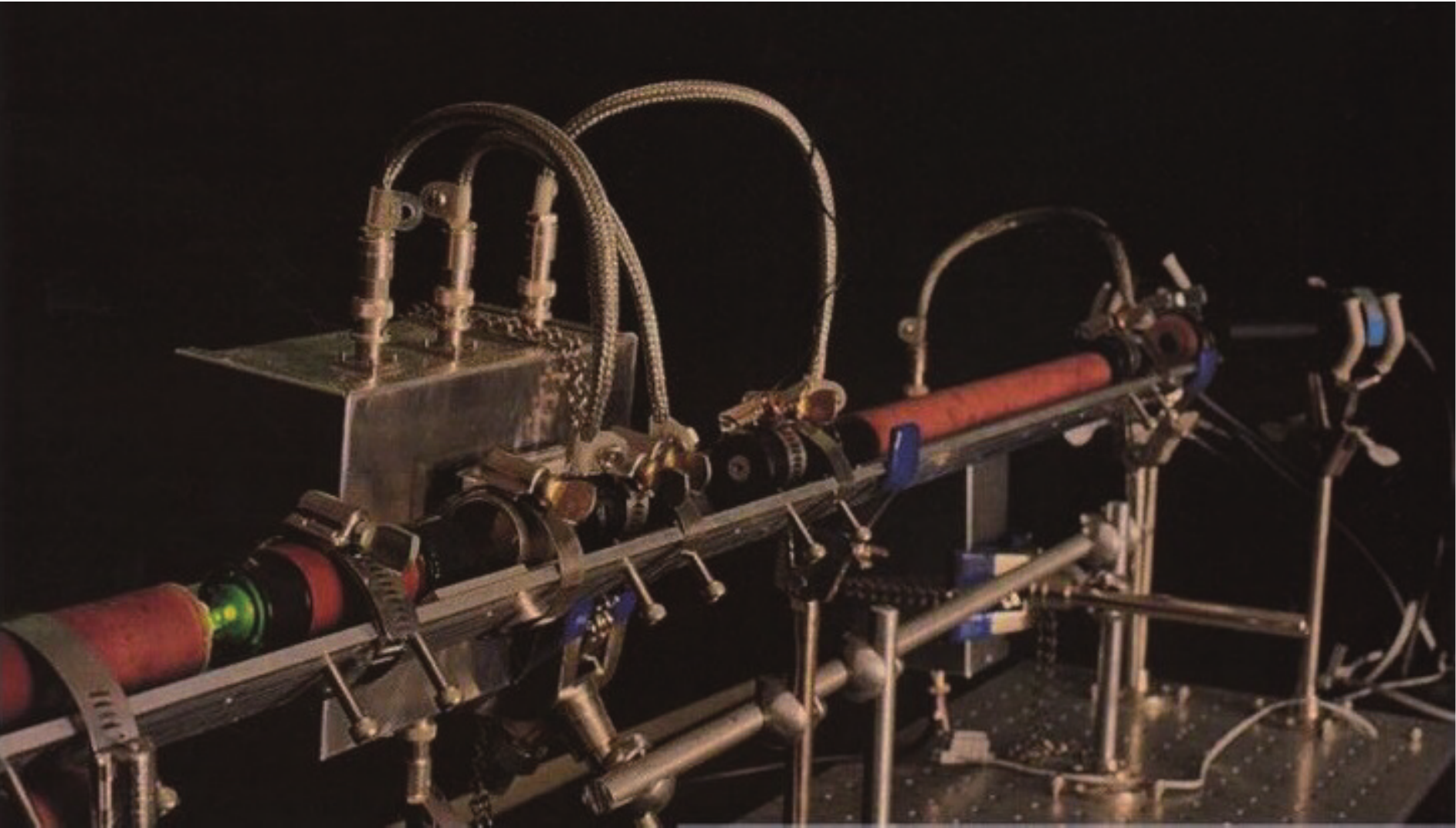}
	\caption{Photograph of the apparatus used in Bennett, \textit{et al}. (1989).}
	\label{fig2}
\end{figure}

After proposing the BB84 protocol, Charles Bennett and his colleagues performed the first proof-of-concept test in 1989 \cite{Bennett1989}.
A light emitting diode (LED) and photomultipliers fixed at two ends of the table represent the sender Alice and receiver Bob, i.e., the photon source and detectors, respectively (see Fig.~\ref{fig2}).
Pumped by current pulses, the LED emitted green light pulses, which were subsequently filtered by a 550 $\pm$ 20 nm bandpass filter and initialized in a horizontal polarization state.
For the encoding, the polarization of each light pulse was rotated to either a horizontal, vertical, left-circular or right-circular polarized state, randomly and independently, via the use of two Pockels cells to which modulated voltages were applied.
A circular basis was exploited instead of the diagonal basis, so that a lower voltage, quarter-wave voltage, was required to operate the Pockels cells.
After transmission through a 32-cm-long free-space quantum channel from Alice to Bob, each encoded pulse was then projected in either a rectilinear or circular basis randomly and independently, by another Pockels cell and a calcite Wollaston prism, depending on whether the applied quarter-wave voltage was turned on or off.
The calcite Wollaston prism was made of birefringent crystals, through which horizontally and vertically polarized light diverged onto two distinct paths, and could then be detected by one of two photomultipliers, one placed in each path.

In principle, a bit of key was successfully distributed when Alice encoded the bit and Bob measured it in the same basis.
Nevertheless, due to experimental imperfections, such as dark counts and differences in the basis alignment between Alice and Bob, errors still could exist after basis announcement.
Therefore, reconciliation, e.g., permutation and parity check, was performed on Alice's and Bob's bit sequence in order to discover and discard errors.
According to the error rate, in a conservative sense that all errors were induced by Eve, who might intercept/resending or beamsplit photons in the quantum channel, the number of bits of information leaked to Eve can be estimated.
By privacy amplification, shared consistent bits were compression encoded.
Although, fewer bits of shared secret key were left, the amount of information potentially leaked to Eve was decreased by several orders of magnitude.
As an example of data from their experiment, Alice sent Bob 715,000 pulses with the mean photon number of each pulse set to 0.12.
754 bits of secret key was finally shared between Alice and Bob.
The experiment was very preliminary as noted by one of the authors who recalled that they ``could literally hear the photons as they flew, and zeroes and ones made different noises'' \cite{Brassard2012Brief}.

\begin{figure}
\includegraphics[width=0.4\textwidth]
{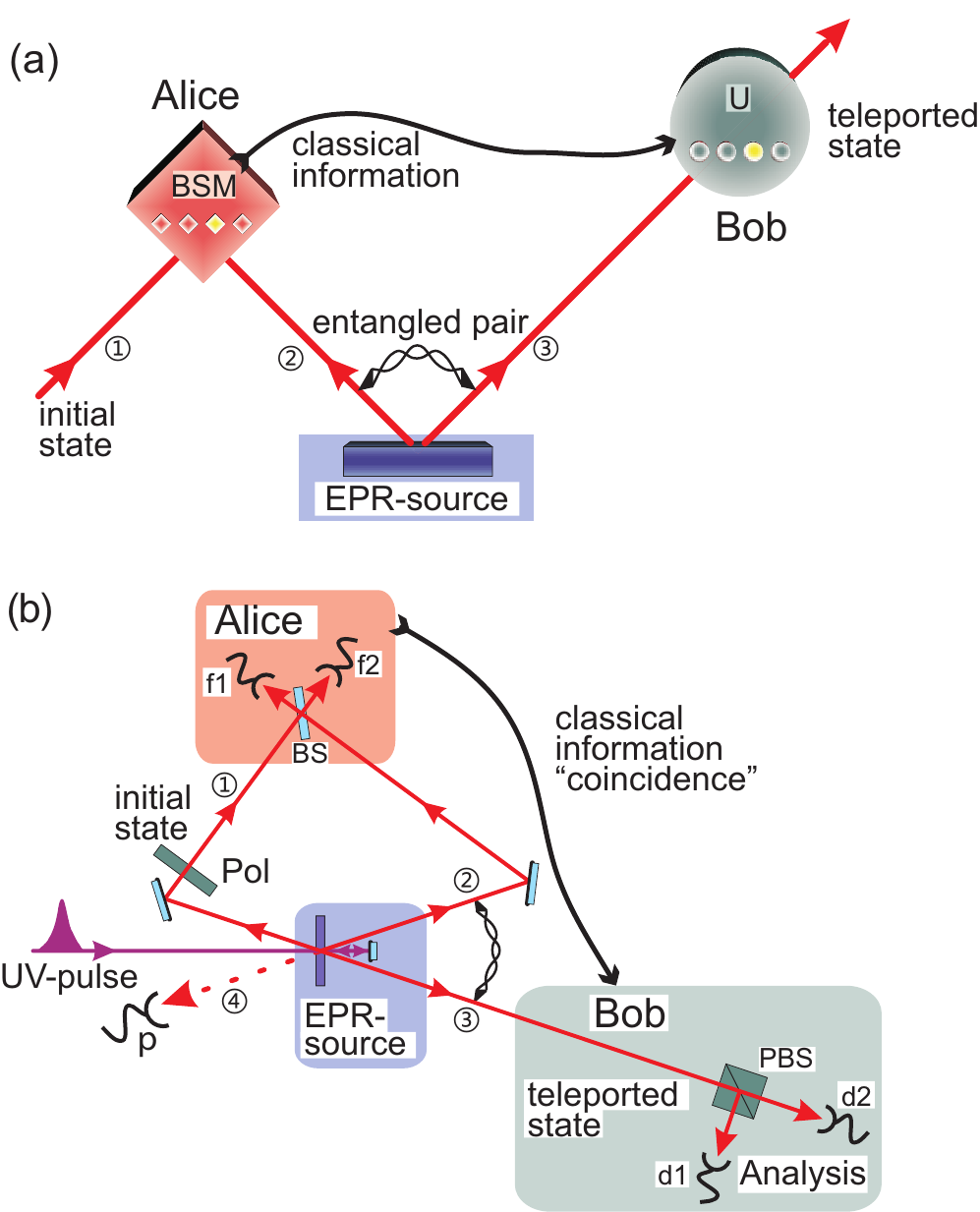}%
\caption{(a) Principle of quantum teleportation. (b) Setup of the Innsbruck
teleportation experiment (Bouwmeester \textit{et al}., 1997). A pulse of
ultraviolet light passing through a non-linear crystal creates an ancillary
pair of entangled photons 2 and 3 in a polarization state $|\psi^{-}\rangle_{12}$.
The pulse is reflected, and during its second
passage through the crystal another pair of
photons can be created.}%
\label{telidea}%
\end{figure}

While the QKD is for the secure transmission of a classical message, another important aspect of quantum communications is quantum teleportation \cite{Teleportation1993}, which is a way to faithfully transfer quantum states.
Quantum teleportation relies on both a classical channel and a quantum channel---entanglement---that are shared between the two communication parties (see Fig.~\ref{telidea}(a)).
The quantum state to be teleported, for example, can be the polarization of a single photon, which can be written as:  $|\chi\rangle_{1}=\alpha|H\rangle_{1}+\beta|V\rangle_{1}$, where $\alpha$  and $\beta$  are two unknown, complex numbers satisfying $|\alpha|^{2}+|\beta|^{2}=1$, and $|H\rangle$ and $|V\rangle$ denote the horizontal and vertical polarization states, respectively, which can be used to encode the basic logic $0$  and $1$  for a qubit.
The entangled state of a pair of photons can be written as  $\left|\phi^{-}\right\rangle_{23}=\left(|H\rangle_{2}|H\rangle_{3}-|V\rangle_{2}|V\rangle_{3}\right) / \sqrt{2}$, one of the four maximally entangled two-qubit Bell states.
Alice performs a joint measurement on the to-be-teleported photon 1 and the photon 2 from the entangled pair, projecting them into one of the four Bell states.
Then the joint three-photon system is in the product state,\\
\begin{equation}\label{formula1}
|\psi\rangle_{123} = |\chi\rangle_{1}\otimes|\phi^{-}\rangle_{23},
\end{equation}
which can be decomposed into\\
\begin{equation}\begin{aligned}\label{formula1}
 & |\psi\rangle_{123} = \frac{1}{2}[-|\psi^{-}\rangle_{12}(\alpha|H\rangle_{3}+\beta|V\rangle_{3})-|\psi^{+}\rangle_{12}(\alpha|H\rangle_{3}\\&~~~~~~~~~~~~~~~-\beta|V\rangle_{3})  +|\phi^{-}\rangle_{12}(\alpha|V\rangle_{3}+\beta|H\rangle_{3}) \\ & ~~~~~~~~~~~~~~~  + |\phi^{+}\rangle_{12}(\alpha|V\rangle_{3}-\beta|H\rangle_{3})].
\end{aligned}\end{equation}
Bob is then informed about the outcome of the Bell-state measurement (BSM) via classical communication, and accordingly applies a Pauli correction to the photon 3.
It should be noted that quantum teleportation does not violate the no-cloning theorem.
After successful teleportation, photon 1 is not available in its original state any longer, and therefore photon 3 is not a clone but instead  the result of teleportation.
Furthermore, the quantum state can only be recovered after Bob receives the classical information sent by Alice. 
According to relativity, the speed of the transmission of the classical information cannot exceed the speed of light. 
Thus, superluminal communication cannot occur as a result of quantum teleportation.


The first faithful demonstration of quantum teleportation was reported using the setup shown in Fig.~\ref{telidea}(b) \cite{BPMEWZ_97}.
The work developed a method for coherently controlling two independent pairs of down-converted entangled photons. 
Double passing the $\beta$-barium borate (BBO) crystal, the pulsed pump created two pairs of entangled photons.
One pair (2-3) was used as the entanglement channel, whereas the photon 1 is heralded prepared in the to-be-teleported state.
The BSM of photons 1 and 2 was achieved using a beam splitter, which, upon a coincidence detection of a single photon in each output of the beam splitter, unambiguously projected the two independent photons into the spatially asymmetric Bell state $|\psi^{-}\rangle_{12}=\left(|H\rangle_{1}|V\rangle_{2}-|V\rangle_{1}|H\rangle_{2}\right) / \sqrt{2}$.
An essential prerequisite to this method is to make the independently generated single photons 1 and 2 indistinguishable in their temporal, spatial, and spectral degrees of freedom.
To do so, the two pairs of entangled photons, pumped by a femtosecond laser, were time synchronized, spectrally narrowband filtered, and coupled into single-mode fibers to ensure a sufficient wave function overlap.
The verification of the success of teleportation was then done using a four-photon coincidence detection.
 
Due to its capacity to faithfully transfer of quantum states from one particle to another at a distance, quantum teleportation has been recognized as an important and elegant tool for long-distance quantum communication and distributed quantum networks.

\subsection{Early efforts toward longer distances}

After the first proof-of-principle demonstrations, naturally, early efforts were made to extend the implementations to longer distance, with the aim of transforming the schemes into practical applications.

\subsubsection{Quantum key distribution}

\begin{figure}[bt]
	\centering
	\includegraphics[width=0.44\textwidth]{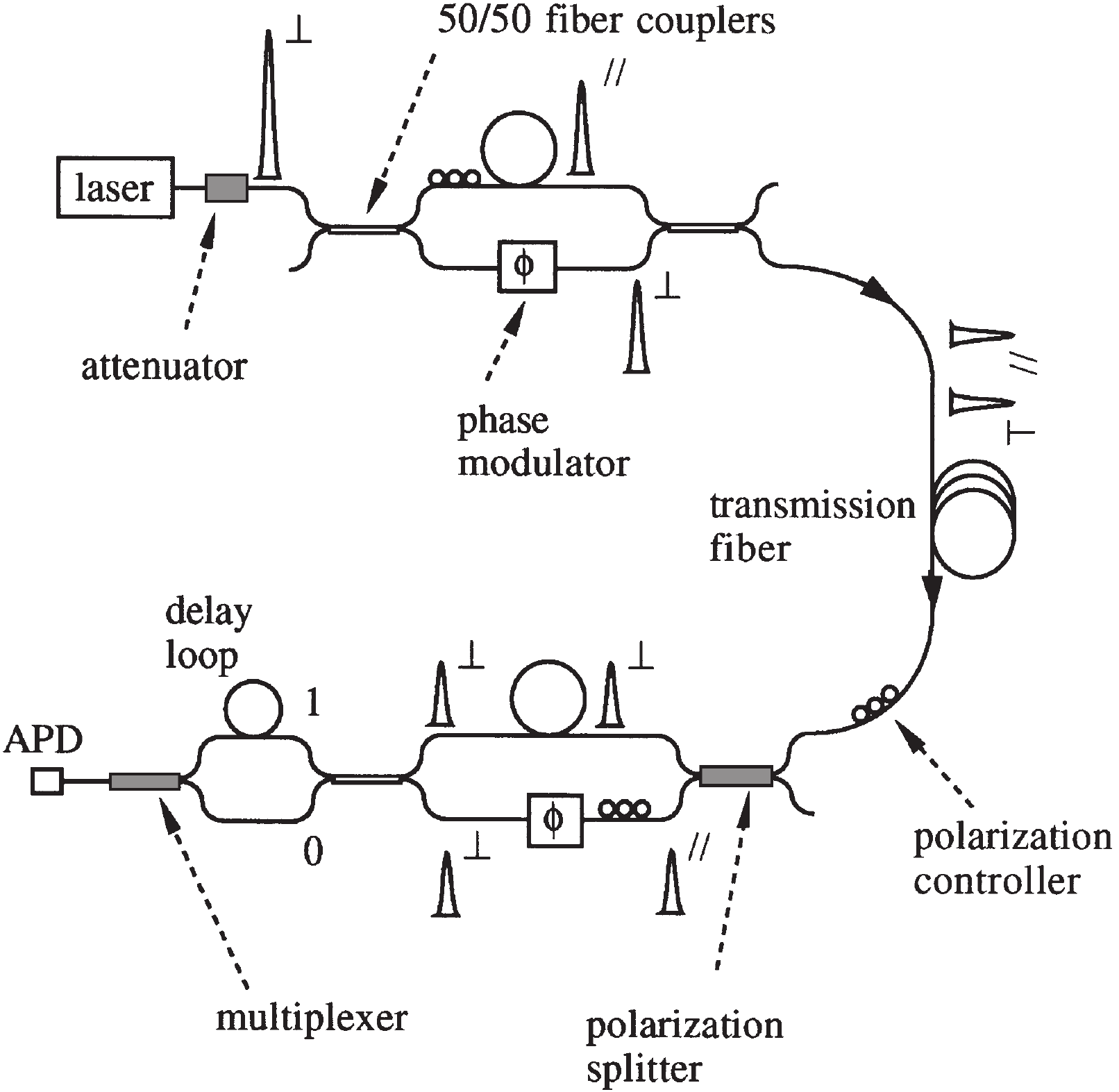}
	\caption{Phase-encoding quantum cryptography scheme over 30 km fiber \cite{townsend1995-30km}. The experimental system is based on a Mach-Zehnder interferometer. The source is a 1.3-$\mu$m-wavelength semiconductor laser with a pulse duration of 80 ps and repetition rate of 1 MHz. The laser output is strongly attenuated so that the average photon number of the pulse pairs entering the transmission fiber is $\sim$0.1. The received photons are detected by a liquid-nitrogen-cooled germanium avalanche photodiode.}
	\label{fig3}
\end{figure}

Optical fibers are convenient, commercially available, and widely used in telecommunications, which was a straightforward way to extend the experimental range beyond that achieved in the initial table-top free-space 32-cm-distance experiment.
Experimental groups from the British Telecom (BT) laboratories in the UK \cite{townsend1993single-a, townsend1993single-b, townsend1994} used phase encoding to demonstrate quantum cryptography over a 10-km-long optical fiber, which was later \cite{townsend1995-30km} extended to 30 km with raw key rate of 260 bits/s and an error rate of 4$\%$ (see Fig.~\ref{fig3}).
Later, a Swiss group \cite{gisin1993expQKD} exploited polarization encoding to test quantum cryptography over 1.1-km optical fiber.
To overcome the time-dependent polarization changes during the transmission of photons in optical fibers, methods were developed \cite{franson1994expQKD}.

The research efforts continued toward field tests using installed optical fibers in real-world environments.
The first step in this direction was taken by researchers at Los Alamos National Laboratory, who used the minimal Bennett protocol \cite{B92} to perform quantum key distribution over 14-km-long underground optical fiber \cite{Hughes1996} with an error rate of 1.2$\%$; the same group later increased this distances to 48 km \cite{Hughes2000} with an error rate of 9$\%$.
The Swiss group \cite{Muller1996} exploited a 23-km-long standard telecom optical fiber installed under a lake and demonstrated a quantum cryptography with an error rate of 3.4$\%$.
Note that there were serious practical security loopholes associated with these early QKD experiments, as discussed in Sec. III. 
For example, the photon-number-splitting (PNS) attack \cite{brassard2000limitations, Norbert2000} was proposed to target imperfect single-photon sources.
Thus, these early implementations of QKD couldn't be used for real-life applications with realistic devices.

\begin{figure}[bt]
	\centering
	\includegraphics[width=0.49\textwidth]{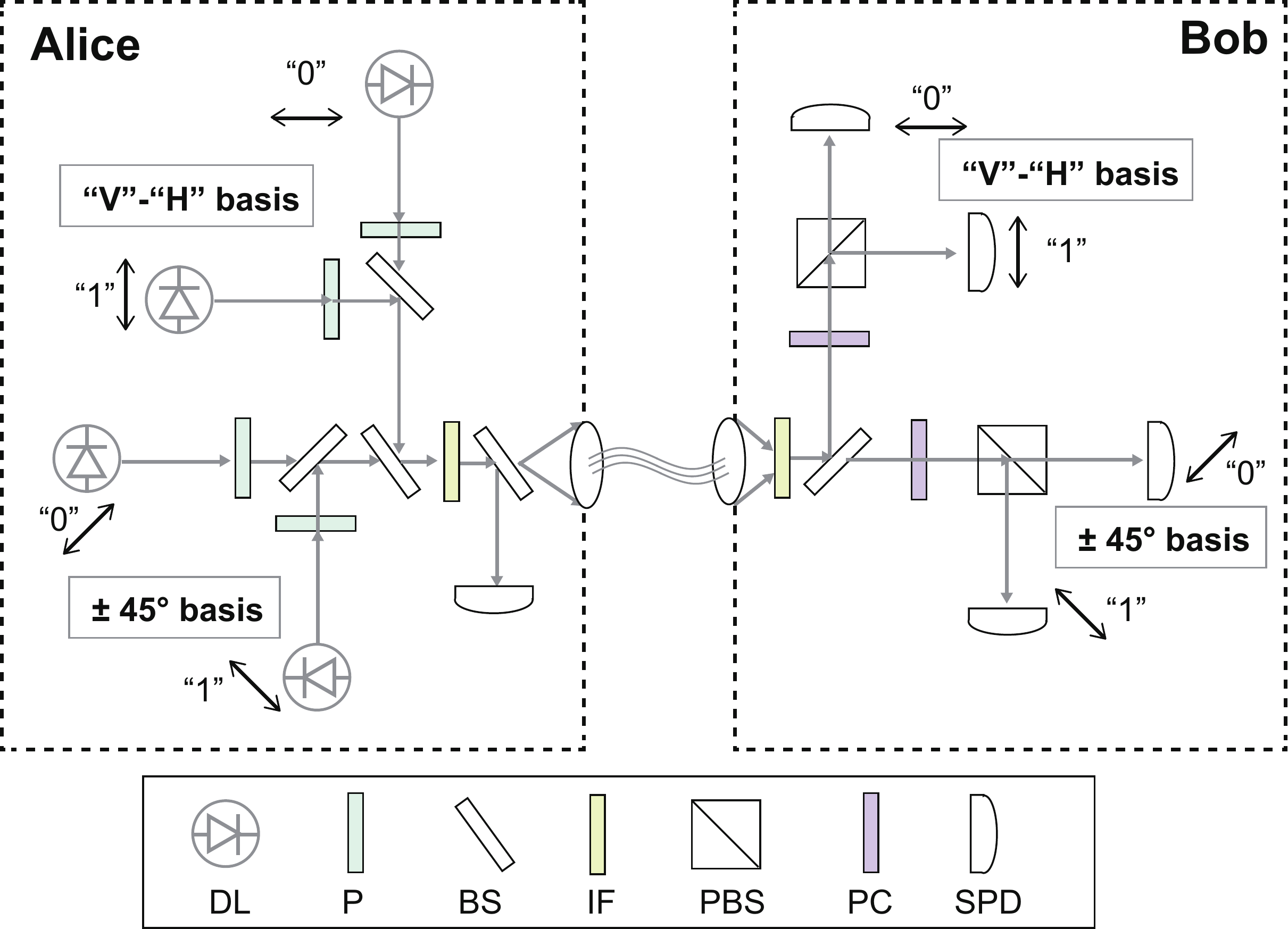}
	\caption{The polarization optics of the QKD transmitter and receiver used in \cite{Hughes2002freespace}. The lasers in the transmitter are attenuated to have an average photon number less than 1. The multi-detection events due to multi-photon component in the weak coherent pulse are recorded but not used for key generation.}
	\label{fig4}
\end{figure}

Jacob and Franson considered another propagation medium: free space, which has the advantage of negligible birefringence, thus permitting the faithful transmission of the photon polarization states.
The first attempt \cite{Jacobs1996freespace} was over a 150 m hallway illuminated with fluorescent light, and over 75 m outdoors in bright daylight conditions.
Also under fluorescent lighting conditions, Buttler \textit{et al}. performed the experiment over a 205-m indoor optical path, which was achieved by sending the emitted beam up and down a 17.1-m-long laboratory hallway 6 times via the use of 10 mirrors, and a corner cube \cite{Buttler1998freespace}.
These preliminary tests suggested that the signal-to-noise ratio in free-space transmissions could be improved using a combination of detection timing, narrowband filters, and spatial filtering.
Further experiments extended the work to outdoor environments with point-to-point free-space transmission of attenuated laser pulses over distances of 0.5 km \cite{Hughes2000freespace}, 1.6 km \cite{Buttler2000freespace},  10 km \cite{Hughes2002freespace}(see Fig.~\ref{fig4}), and 23.4 km \cite{Kurtsiefer2002}.

The straightforward method of directly sending signals (photons) through optical fibers or terrestrial free-space suffers from two fundamental problems.
First, channel losses cause a decrease in the transmitted photons that scales exponentially down with the length (see Sec. III B).
Second, there are additional problems associated with the security of these the practical implementations due to the imperfect single-photon source and detectors (see Sec. III A).

\subsubsection{Entanglement distribution}

The experiments described above used attenuated laser pulses to approximate the single photons. Such laser pulses have a non-vanishing probability of containing two or more photons per pulse, leaving the system susceptible to the PNS attack, see \cite{feihu2019rmp} for details.
Remarkably, Ekert (1991) proposed entanglement-based quantum cryptography using Bell's inequality to establish security, which results in an inherent source-independent security.
In addition to its use in quantum cryptography, distributed entanglement is an essential resource in fundamental studies, such as tests of Bell's inequality \cite{Bell1964} as well as in many quantum information tasks, such as quantum teleportation \cite{Teleportation1993}, distributed quantum networks for computing \cite{Gottesman:1999kw}, and metrology \cite{Togan2011}.

\begin{figure}[bt]
	\centering
	\includegraphics[width=0.486\textwidth]{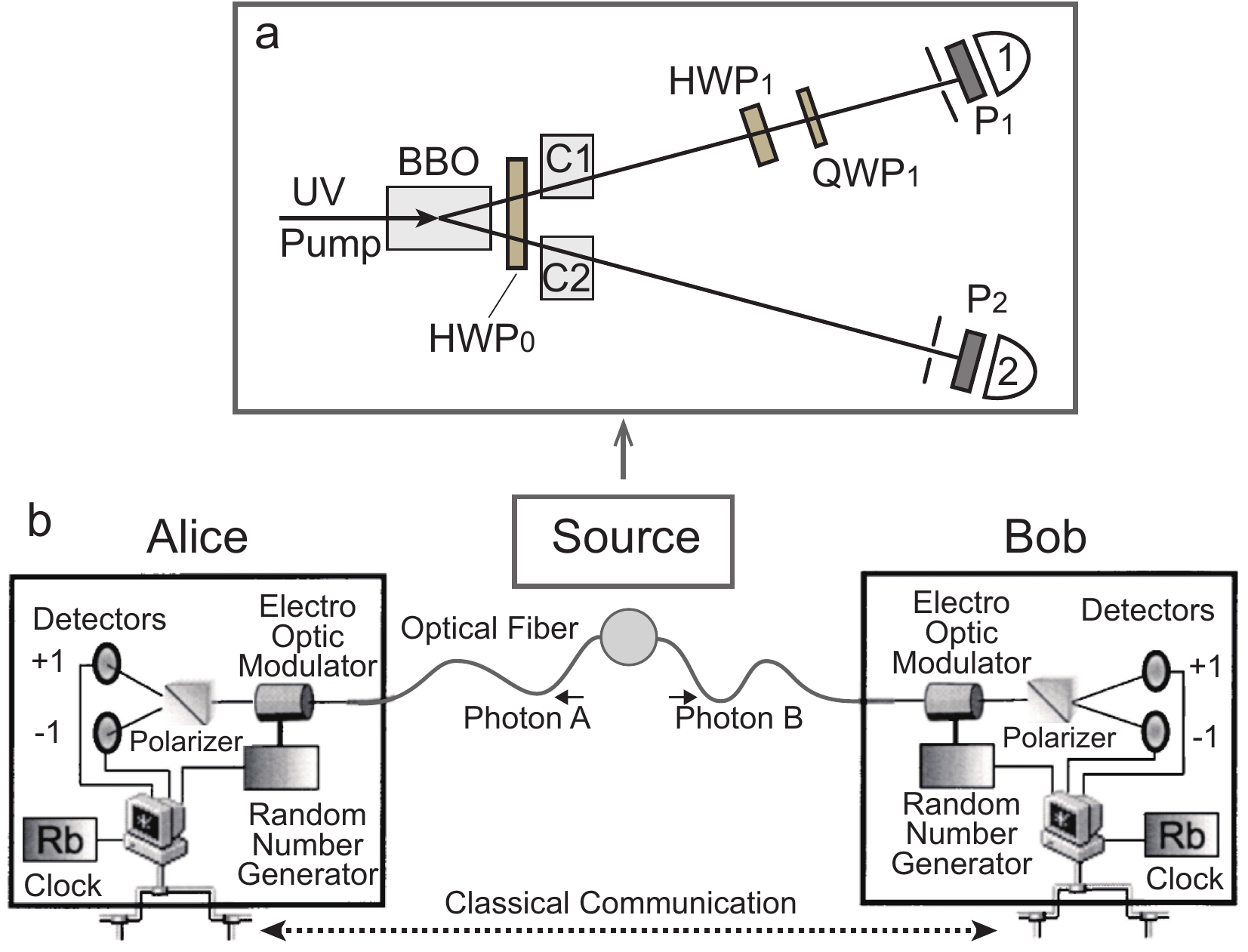}
	\caption{(a) The experimental configuration to produce entangled photon pairs using spontaneous parametric down-conversion \cite{kwiat1995new}. Type-II non-collinear phase matching in a $\beta$-barium borate (BBO) produces high-fidelity polarization entanglement. The extra birefringent crystals C1 and C2 along with the half-wave plate (HWP) are used to compensate the birefringent walk-off effect in the crystal. (b) The first full implementation of entanglement-based quantum cryptography over a distance of 360 m. The polarization-entangled photon pairs are transmitted via optical fibers which are then analyzed and detected. }
	\label{fig5}
\end{figure}

Producing (see Fig.~\ref{fig5}) and distributing entangled photons over long distances is of great importance both for fundamental understanding and for possible applications. 
In 2000, three independent groups simultaneously reported implementation of the Ekert-91 protocol, two using polarization encoding \cite{Jennewein2000entQKD, Naik2000entQKD} and one using time encoding \cite{Tittel_EnTime_2000}.

The experimental arrangement of the Vienna experiment \cite{Jennewein2000entQKD} is shown in Fig.~\ref{fig5}.
In this work, the entangled photons were produced by type-II spontaneous parametric down-conversion \cite{kwiat1995new}.
As shown in Fig.~\ref{fig5}(a), pumped by a 351 nm laser, a BBO crystal produced entangled photons at a wavelength of 702 nm with a two-photon coincidence rate of $\sim$1700 Hz.
The photons were coupled into 500 m long optical fiber and transmitted to Alice and Bob, respectively, who were physically separated by 360 m (Fig.~\ref{fig5}(b)).
Jennewein \textit{et al}. utilized Wigner's inequality to establish the security of the quantum channel, which has higher efficiency than the original Ekert's protocol.
The two communication parties varied their analyzers randomly between two settings, Alice: $-30^\circ$, $0^\circ$ and Bob: $0^\circ$, $30^\circ$.
The two-photon coincidence counts of the combinations of the settings of ($-30^\circ$, $30^\circ$), ($-30^\circ$, $0^\circ$), ($0^\circ$, $30^\circ$) were used to derive a violation of Wigner's inequality, thus ensuring the security of the key distribution.
The coincidence detection obtained at the parallel settings  ($-30^\circ$, $0^\circ$), which occurred in a quarter of all events, were used to generate keys.
Over a fiber distance of 500 m, the raw key rate was 420 bits/s and the quantum bit error rate was 3.4$\%$.

The entanglement-based quantum cryptography was later extended to a distance of 8.5 km using an optical fiber inside a laboratory \cite{Ribordy2000entQKD}.
This work utilized a time-bin encoding and at a wavelength of 1550 nm, the experiment generated a raw key rate of 134 Hz with an error rate of 8.6$\%$.
Poppe \textit{et al}. demonstrated a successful run of the entanglement-based quantum cryptography system where the produced key was directly handed over to an application that was used to send a quantum secured online wire transfer from the City Hall to the headquarters of Bank-Austria Creditanstalt over a physical distance of 650 m, showing a real-world application scenario outside ideal laboratory conditions \cite{Poppe:04}.

In 2003, entanglement distribution outside the laboratory over free space was preformed in Vienna city over a physical separation of 600 m \cite{aspelmeyer2003long}.
One of the entangled photons was distributed over 150 m and the other one over 500 m, to different sides of the Danube River.
The 600-m free-space link attenuation was estimated to be 12 dB.
The two-photon count rate was measured to be 10 kHz locally and 15 Hz after the free-space distribution.
Nevertheless, the survived two photons showed a state fidelity of $0.87\pm0.03$ and a violation of Bell's inequality.

\subsubsection{Quantum teleportation}

The first demonstration of quantum teleportation of quantum states \cite{BPMEWZ_97} inspired many subsequent experiments, such as extending to continuous-variable systems \cite{furusawa1998} and teleporting more complete photonic states \cite{Panfreely2003, TakedaDeterministic2013, Wang2015tele}, as also covered in a recent review \cite{Pirandola2015tele-review}. 
The restriction of post-selection, where teleported photons have to be detected (i.e., destroyed) to verify the success of the procedure in \cite{BPMEWZ_97}, was removed six years later by a new method \cite{Panfreely2003} reporting free propagating teleported single photons. 
The success probability ($1/2$) of the Bell-state measurement using beam splitters was improved to  near unity using a hybrid technique combining both discrete variable and continuous variable, resulting a deterministic quantum teleportation of photonic quantum bits \cite{TakedaDeterministic2013}. 
Quantum teleportation has also been extended to the composite state of two photons \cite{Zhang2006tele}, multiple degrees of freedom \cite{Wang2015tele}, and high dimensions of a single photon \cite{Luo2019telehd}.

In addition to photons, quantum teleportation has been demonstrated in other physical systems, such as nuclear magnetic resonance \cite{Nielsen1998NMRtele}, atomic ensembles \cite{Sherson:2006dt, Chen2008Memory}, trapped atoms \cite{Barrett2004atomtele, Riebe2004Deterministictele}, and various solid-state systems \cite{Steffen:2013ch, Pfaff:2014hy}. 
Typically, in pure matter-based systems, such as trapped atoms \cite{Barrett2004atomtele, Riebe2004Deterministictele}, teleportation is performed only over very short distances for atoms in proximity to each other. 
One method to extend the distance of matter qubit teleportation involves developing light-matter quantum interfaces. 
Using light-to-matter teleportation \cite{OlmschenkQtele2009}, the flying photonic qubits can be stored in stationary media, which is essential for the construction of a scalable quantum network based on the quantum repeater scheme (more details in Section IV). 
However, the achievable range of quantum teleportation is determined by the distance across which the entanglement can be distributed.
Thus, early experiments suffered similar limitations as the QKD experiments.
The early experiments were limited to, for example, a 2-km-long optical fiber linking two laboratories physically separated by 55 m \cite{Marcikic2003tele}, and over 600 m across the Danube river \cite{Ursin2004teleDanube}.

The capability of multi-photon manipulation \cite{pan2012rmp} opens the way to not only quantum teleportation \cite{BPMEWZ_97}, but also entanglement swapping \cite{Pan1998expswapping} and entanglement purification \cite{pan2003purification}, which, combined with quantum memory, are important tools for quantum repeaters (see Section IV).
Interestingly, quantum teleportation can be considered as a method for probabilistic quantum non-demolition measurement, which can be exploited as a quantum relay (a simpler version of quantum repeater without quantum memory) to extend the distance of quantum communications moderately \cite{Jacob2002quantumrelays, Waks2002}.
For example, using teleportation, one can increase the distance by approximately three times that of the direction transmission of a single photon at the same signal-to-noise ratio \cite{Riedmatten2004}.


\section{Challenges in practical and large-scale applications}

\subsection{Security loopholes}
There are practical security limitations associated with the early implementations using realistic QKD devices in Sec. II. 
Ideally, only when perfect single-photon sources and detectors are utilized, will quantum cryptography be secure. 
Unfortunately, ideal devices do not exist in practice.
In reality, the imperfections of realistic QKD implementations might introduce deviations (or side channels) from the idealized models used for security analysis.
Eve might exploit these imperfections and launch quantum attacks.
For this reason, an arms race has been ongoing for more than 20 years in quantum cryptography among quantum code-makers and quantum code-breakers.
The race participants aim to assess the deviations of the real system from the ideal system, thus establishing the \emph{practical security} for QKD with realistic devices \cite{feihu2019rmp}.

Right after the security proofs have been established in the early 2000s \cite{Mayers:2001ACM, LoUnconditional1999, shor2000prl}, a well-known quantum hacking strategy was proposed---photon number splitting (PNS) attack~\cite{brassard2000limitations,Norbert2000}, which targets on the practical laser source.
Since there would occasionally be two identical photon events in pulses from a quasi-single-photon source, the so-called PNS attack allows Eve to selectively suppress single-photon signals, and split two-photon signals by keeping one copy for herself without being noticed by Alice and Bob. Due to this loophole, the secure distance of QKD in optical fiber was limited to 10 km.

To resolve this problem, decoy-state QKD was theoretically proposed \cite{Hwang2003, Wang:Decoy:2005, Lo:Decoy:2005} and has been experimentally demonstrated \cite{Peng:ExpDecoy:2007, Zeilinger:Decoy:2007, Rosenberg:ExpDecoy:2007} to increase the practicality of QKD with standard weak coherent pulses (WCPs) generated by attenuated laser pulses.
Decoy-state QKD has become a standard technique in current QKD experiments.
After the decoy-state method, however, quantum code-breakers turned to attack other components, particularly the imperfect single-photon detectors \cite{Makarov:Bright:09}. 
For instance, one can use a specially tailored strong light to ``blind'' the single-photon detectors. 
Thus, the detector can only respond when the light intensity is greater than its threshold. 
If the eavesdropper sends a light pulse with an intensity higher than the threshold, the detector will only click when Bob's measurement basis is the same as that of the input light. 
Therefore, the eavesdropper can fully control the detectors. 
Fortunately, measure-device-independent QKD was proposed theoretically \cite{LoMDI2012} and experimentally \cite{Liu2013expMDI, Rubenok2013expMDI, Tang2014MDI200km, MDI-404km,CaofreespaceMDI2020} to increase the security of QKD against any detector imperfection.
Measure-device-independent QKD completely eliminates all security loopholes in detection systems and allows QKD networks to be secure with \emph{untrusted} relays.
See the reviews for further details \cite{entanglementRMP2009, sangouard2011rmp, pan2012rmp, Bellnonlocality2014RMP, Reiserer2015Cavity, feihu2019rmp}. 
Thus far, the security of the quantum cryptographic system, combing measurement-device-independent QKD and a self-calibrated homemade source, is believed to be sufficient for practical QKD at the level of the theoretical unconditional security.

\subsection{Long distances}

Having covered the typical early, small-scale experiments on quantum communications, here we raise the question, ``what limits the distance of quantum communications?''
In both the fiber optics and terrestrial free-space channels, there are inevitable photon losses, which scale exponentially up with the transmission length in optical fibers.
Gisin \textit{et al}. highlighted that at 1000 km, even with a perfect single-photon source of 10 GHz, ideal photon detectors, and 0.2 dB/km fiber losses, one would detect only 0.3 photons on average per century \cite{Gisin2010Quantum}.
In classical communications, it is possible to amplify the signal 0 and 1.
In contrast, an unknown quantum superposition state cannot be noiselessly amplified.
This is known as the quantum no-cloning theorem, a fundamental no-go theorem in quantum mechanics.
While it underpins the security of QKD, it excludes the possibility of simply amplifying quantum signals over long-distance quantum communications.

\section{Quantum repeater and its progress}

\begin{figure*}[t]
\centering
\includegraphics[width=0.9\textwidth]{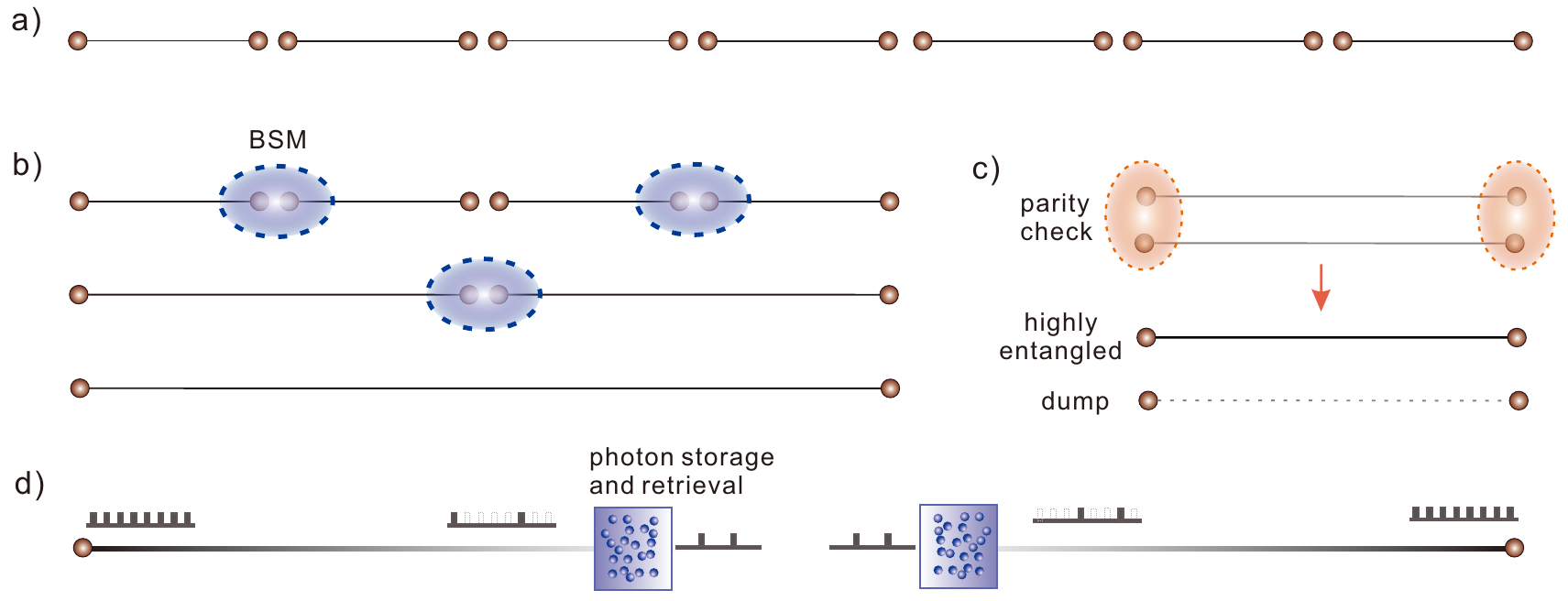}
\caption{
Principle of quantum repeaters \cite{Briegel1998repeater, DLCZ2001}. 
(a) Dividing the communication channel into \textit{N} segments. 
(b) Using entanglement swapping to establish entanglement between remote nodes. 
(c) Using entanglement purification to distill highly entangled pairs between remote nodes. 
(d) Using quantum memory to store single photons for a sufficiently long time and retrieve the single photon for time synchronized two-photon interference.
}
\label{fig:repeater}
\end{figure*}

One strategy for extending the distance of quantum communications is using the divide-and-conquer strategy. 
Unlike classical repeaters that work only for classical bits, quantum repeaters (see Fig.~\ref{fig:repeater}) combine entanglement swapping, entanglement purification, and light storage, and in principle, can enable quantum communication at arbitrarily large scales.
This section will briefly cover the principle and progress of quantum repeaters.

The key mechanism behind the extension of quantum communication distance by quantum repeaters is to change the notorious exponential scaling of the photon loss to a polynomial relation as a function of the channel length.
To do so, the whole channel must be divided into $N$ segments (see Fig.~\ref{fig:repeater}(a)), such that within each segment, direct transmission can yield a reasonably good signal-to-noise ratio.
The problem now sits on how to efficiently connect these segments in a quantum compatible way. 
It turns out that entanglement swapping \cite{entswapping1993}---a variant of quantum teleportation, where the particle to be teleported is itself part of an entangled pair---provides an ideal method for entangling the remote particle without any direct interaction (see Fig.~\ref{fig:repeater}(b)).
The key step in entanglement swapping involves performing a Bell-state measurement for two single photons, each from an entangled pair. 
The two photons should arrive simultaneously through a beam spitter and must be quantum-mechanically indistinguishable.

\begin{figure}[t]
\centering
\includegraphics[width=0.49\textwidth]{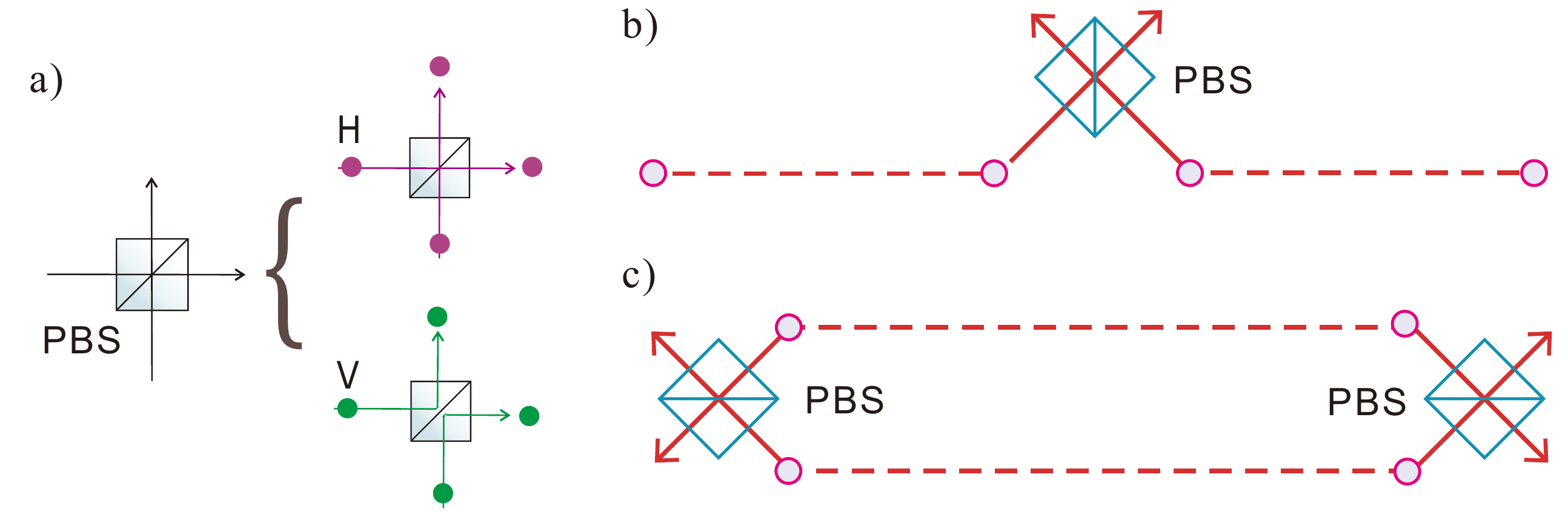}
\caption{
Experimental scheme for entanglement swapping and purification with high-precision polarization beam splitters (PBSs). (a) The PBS transmits horizontally (H) polarized photons and reflects vertically (V) polarized photons. Suppose two single photons are superposed on a PBS. If we postselect the output case where there is one photon in each side, there are two quantum mechanically indistinguishable possibilities: both photons are transmitted or both photons are reflected. (b) Entanglement swapping by interfering two photons on a PBS. The two distant photons become entangled upon the Bell-state measurement on the PBS. (c) Entanglement purification by locally two-photon interferences on the two PBSs. See \cite{pan2001} for details.
 }
\label{fig:pbs}
\end{figure}

The first experimental demonstration of entanglement swapping was reported by Pan \textit{et al}., who showed the possibility of entangling two independent photons without direction interaction \cite{Pan1998expswapping}, as shown in Fig.~\ref{fig:pbs}.
The precision of entanglement swapping was improved to be 0.98 in later experiments \cite{pan2003purification}, meeting the stringent requirement of fault-tolerant quantum repeaters \cite{Briegel1998repeater}.
Entanglement swapping has been used as an ubiquitous tool to entangle distant qubits such as in trapped ions \cite{monroe2007ions}, cold atomic ensembles \cite{yuan2008}, and nitrogen-vacancy centers in diamond \cite{Hensen2015Loophole}, by the joint projection of two flying photons.
Furthermore, it has been extended to field tests over tens of kilometers optical fibers by the fine synchronization of two arriving photons \cite{sun2017, xiaohui2019}.

The quantum repeater protocol also considers the presence of noise and decoherence as the entangled photons propagate through the channel (e.g., optical fibers that can induce undesired polarization rotation), which can induce the degradation of the entanglement fidelity compared to those of locally prepared ones.
To overcome the decoherence, entanglement purification \cite{Bennett1996Purification} was proposed to distill highly entangled pairs remotely separated from an ensemble of less entangled pairs.
The protocol of entanglement purification uses only local operation and classical communication.

In the protocol proposed, the local operations involved controlled-NOT (CNOT) logic gates between the qubits.
Due to the weak interaction between independent photons, however, the CNOT gates were difficult to implement.
Although, eventually, there were some experimental demonstrations \cite{brien2003, Gasparoni2004, zhao2005}, the photonic CNOT gates required ancillary photons and exhibited only moderate gate fidelity, as well as very low efficiency.

Using only linear optics, Pan \textit{et al}. put forward a feasible scheme for entanglement purification with high precision and efficiency. 
It was observed that the CNOT gates were not necessary \cite{pan2001}.
In fact, the polarizing beam splitter (PBS), an off-the-shell high-precision linear optical element that transmits horizontal polarization and reflects vertical polarization, was exploited by Pan \textit{et al}. to function as a parity checker that was sufficient to perform entanglement purification.
Interesting, the parity checking function of the PBS was found to apply to resource-efficient linear optical quantum computation \cite{browne2005resource}.
Two years afterwards, Pan \textit{et al}. experimentally demonstrated entanglement purification, achieving an enhanced entanglement fidelity after the purification \cite{pan2003purification}.
Walther \textit{et al}. further achieved the violation of Bell's inequality after the purification of two copies of less entangled pairs that were insufficient to violate Bell's inequality \cite{walther2005}.
Late, Wineland's group used trapped ions to demonstrate entanglement purification \cite{Reichle2006Experimental}.

Notably, the development of multiphoton interferometry and entanglement \cite{BPMEWZ_97, Pan2000BellGHZ} has also laid the technological foundation for entanglement swapping \cite{Pan1998expswapping}, entanglement purification \cite{pan2003purification}, and measurement-device-independent QKD \cite{Liu2013expMDI, Rubenok2013expMDI}, which are essential for long-distance quantum communications.
In fact, quantum teleportation and entanglement swapping can be considered methods for probabilistic quantum nondemolition measurement, which can be exploited to extend the distance of quantum communications by up to four times. 
This can be viewed as a simplified version of a quantum repeater without quantum memory, which has also been referred to as quantum relay \cite{Riedmatten2004}.

The last element of the quantum repeater \cite{Briegel1998repeater} is the quantum memory \cite{Lvovsky2009Qmemory, Afzelius2015Qmemory}, which can convert the quantum information carried in fast-flying photons into stationary matter qubits, store it for a certain time, and convert it back into photons on demand.
Several physical systems \cite{Simon2010review} that have been considered as candidates for quantum memory, such as cold atomic ensembles \cite{sangouard2011rmp}, rare earth elements \cite{Tittel2010Photon}, single trapped atoms \cite{Reiserer2015Cavity} and ions \cite{Duan2010review}, and color centers in diamond \cite{Gao2015NP, Atature2018}.
These systems can be categorized into single-particle approaches and ensemble approaches.
With single particles, it is advantageous to scale up to multiple qubits and perform gate operations.
Nevertheless, challenges still exist in improving the coupling with single-photons by either employing a high finesse cavity or a high numerical-aperture lens.
Conversely, with atomic ensembles, it is much easier to achieve efficient coupling with single photons because of the collectively enhanced interaction from large number of atoms.
Nevertheless, one must tackle inhomogeneous decoherence. 
After many years of extensive experimental investigations, the performance of each system was improved significantly.
To be employed in quantum repeaters, a quantum memory must possess many crucial properties, such as a long lifetime and high storage efficiency.
A system must exhibit high values of these parameters simultaneously to be suitable for use in quantum repeaters.
As a milestone, the approach of utilizing cold atomic ensembles afforded an efficient quantum memory with a sub-second lifetime in 2016, with the two parameters fulfilling the requirement of long-distance quantum repeaters for the first time \cite{Yang-memory, Bao2012, Zhao2009}.
Further development needs to incorporate efficient entanglement creation, telecom interface, multiqubit storage, gate operation, etc. 

Based on numerical analysis \cite{rozpedek2019, wu2020b, singh2021}, achieving a 1000 km quantum repeater requires demanding parameters that are still significantly beyond the capabilities of the state-of-the-art, such as trapped ions and cold atomic ensembles. 
Therefore, it is unlikely that quantum repeaters will become practical within 10 years, so alternative and more efficient routes for global-scale quantum communications are necessary.
Note that new quantum repeater concepts were developed recently, such as all-optical repeaters \cite{Azuma2015}, and repeaters that do not require two-way classical communications \cite{PhysRevLett.112.250501}.

\section{Satellite-based free-space channels}
\begin{figure}[t]\centering
\resizebox{8cm}{!}{\includegraphics{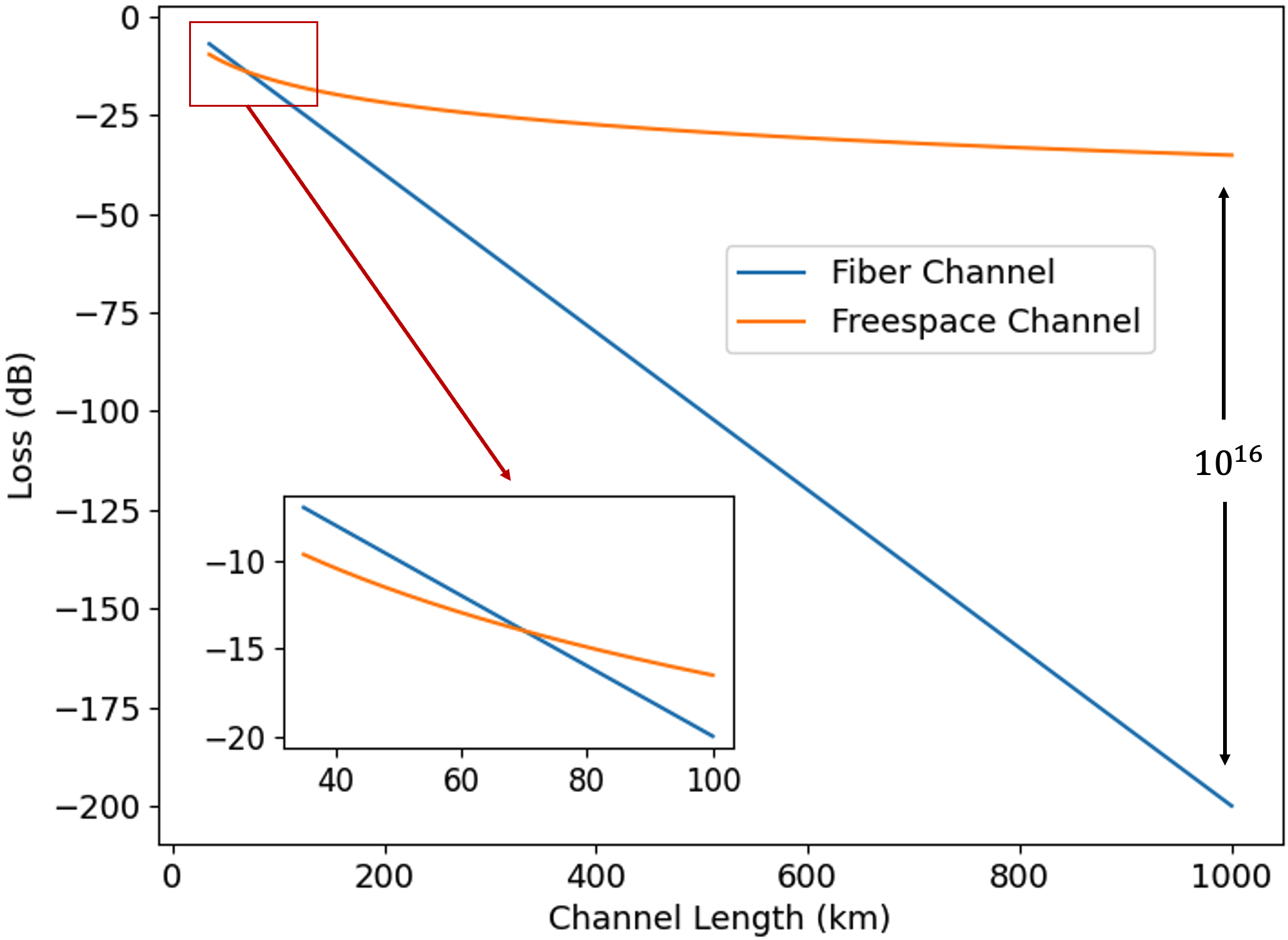}}
\caption{Typical losses in fiber and free-space channels. The attenuation parameter of fiber is $\sim 0.2~dB/km$. The parameters of free-space channel are based on the design of Micius satellite. The free-space channel shows advantage for a distance over $\sim$ 70 km}
\label{Fig:fiberfreespacechannelloss}
\end{figure}

Generally, the attenuation in free space is lower than that in fiber for optical signals. 
For instance, values of 0.07 dB/km can be achieved at 2400 m above sea-level \cite{Zeilinger:Decoy:2007} with high attenuations at relatively low altitudes and negligible attenuations in the vacuum above the Earth's atmosphere. 
Furthermore, the almost nonbirefringent character of the atmosphere guarantees the preservation of polarization state to a high degree. 
However, terrestrial free-space channels alone are not enough. 
They suffer from obstruction by objects in the line of sight, from possible strong attenuations due to weather conditions and aerosols, as well as effects from the Earth's curvature \cite{satellite2003}. 
Therefore, to fully exploit the advantages of free-space links, it is necessary to exploit space and satellite technologies.
Furthermore, the effective thickness of the atmosphere is approximately 5-10 km, and most of the photon's propagation path is in empty space with negligible absorption and turbulence, which is crucial for transmitting single photons that cannot be amplified.
Therefore, in the application of QKD networking with global-scale coverage, satellite-based free-space channels hold promise as a potential route and a new platform for quantum optics experiments at an astronomical scale.
The attenuation of the fiber and free-space channel with distance is simulated according to the parameters of the commercial fiber and the Micius satellite, as shown in Fig.~\ref{Fig:fiberfreespacechannelloss}.
We can see that for long-distance communication, the satellite-based free-space channel is superior to the fiber-based channel in terms of total losses.

To build the QKD network, trusted-node topologies are being built for fiber networks. 
However, ground-based nodes are at fixed locations, which lack flexibility and are vulnerable to constant surveillance and probes. 
By placing a satellite above the Earth's atmosphere, direct links can be established between ground stations and the satellite, thus enabling communication between any distant points on the planet. 
Furthermore, many more network topologies can be easily implemented with satellites by employing the downlink, uplink, and intersatellite channels. 
Moreover, more ambitious projects aimed at extending the network-scale to deeper space, such as to the Moon or planets in the solar system, can be also realized with satellites.

Global-scale classical communications have already employed satellites to conveniently cover the whole world.
In principle, quantum communications have a bright future, considering that the advantages of using satellites and free-space channels are even more physically significant.
There are enormous challenges to actually move the quantum optics setup inside well-shielded laboratories to space. 
To realize satellite-based quantum experiment platforms in space, intensive engineering efforts should be devoted to its design and strict step-by-step verifications. 
Considering a feasible low-Earth-orbit satellite within a realistic budget, one should carefully analyze the link efficiency and transmission fidelity of the free-space channels, and every relevant parameter such as the designed quantum light sources, which will be reviewed in this Section. 
The quantum science satellite, \textit{Micius}, is within the framework of Chinese Academy of Science (CAS) Strategic Priority Research Program on space science. 
The major scientific goals include: (1) satellite-to-ground QKD, (2) quantum entanglement distribution from satellite to two ground stations, and (3) ground-to-satellite quantum teleportation. 
Strictly following such engineering requirements for each of the three goals, ground-based experimental tests of the feasibility of satellite-based quantum communications were then carried out, as reviewed in Sec. VI.

\subsection{Analysis of space-ground links}

Several factors influence the attenuation of quantum communication channels when photons propagate between a ground station and a satellite. 
The fixed attenuations are affected mainly by the efficiencies of optical transmitting system $\eta _t$ and receiving system $\eta _r$, including optical and detection efficiencies.
During transmission, the optical beam will be broadened or deflected by diffraction, air turbulence and mispointing, which will induce losses $\eta _d$, $\eta _{at}$, and $\eta _p$, respectively.
Atmospheric absorption brings the attenuation $\eta _{as}$ relay to photon wavelength and air composition.
Considering of all these factors, one-way channel attenuation between a satellite and a ground station can be simulated as follows:

\begin{equation}\label{formula1}
\eta  = {\eta _t}{\eta _r}{\eta _d}{\eta _{at}}{\eta _p}{\eta _{as}}.
\end{equation}

\textbf{Beam diffraction}.
In free-space quantum communication, the diffraction of an optical beam depends mainly on its spatial mode, wavelength, and the telescope aperture.
We generally assume that the beam from transmitting antenna is Gaussian with waist radius of $\omega _0$. At a distance of $z$, the spot radius $\omega _d (z)$, will be

\begin{equation}\label{formula2}
\omega _d( z ) = {\omega _0}\sqrt {1 + {{(z/{z_R})}^2}} ,
\end{equation}
where Rayleigh range $z_R = \pi \omega _0^2/\lambda$  relays to the wavelength $\lambda$. For a telescope with aperture radius $r$, the receiving efficiency $\eta _d$ can be as high as:

\begin{equation}\label{formula3}
\eta _d ( r ) = 1 - \exp ( - \frac{{2{r^2}}}{{\omega _d^2}}).
\end{equation}
Note that the spot radius $\omega _d( z ) $ increases linearly with $z$ when the distance $z \gg {z_R}$. 
The divergence half-angle of the beam far from the waist is given by $\theta \approx \lambda {\rm{/}}( \pi \omega _{\rm{0}} )$.
Therefore, the diffraction loss in long-distance quantum communications can be mitigated by choosing relatively shorter photon wavelengths or a large waist radius.
But the transmitting antenna truncates the beam sent to ground, causing significant losses if the beam waist is larger than the telescope radius.
For a downlink, it is recommended to set the full width at half maximum (FWHM) beam waist to be half of the transmitting telescope diameter \cite{Stutzman2012}.

\textbf{Air turbulence}
Air turbulence is one of the main factors limiting the channel efficiency in free-space quantum communication.
It induces the atmospheric refractive index inhomogeneity, which changes the direction of the propagating beam.
In terms of the beam size, larger-scale turbulent causes beam deflection, while small-scale turbulence induces beam broadening \cite{Vasylyev2016}.
For the receiver, by accumulating the randomly moving spots, the average long-term spot theoretically tends to follow a Gaussian intensity distribution \cite{Dios2004}.
The equivalent radius of this spot is given by:
\begin{equation}\label{formula4}
\omega _{at}(z){\rm{ = }}\omega _d(z) \sqrt{1+ \omega _a}{\rm{ = }}\omega _d(z)\sqrt{1+1.33 \sigma_R^{2}\Lambda^{5/6}},
\end{equation}
where $\omega _a$ represents the effect of air turbulence on the optical beam, $\sigma_R^{2}$ represents the Rytov variance for a plane wave, and $\Lambda$ represents the Fresnel ratio of the beam at the receiver. More details are provided in ~\cite{laserbeam2005}.

\textbf{Pointing error}
To establish the link between the ground and the high-speed moving satellite, a high-precision and high-bandwidth acquiring, pointing, and tracking (APT) system should be developed, which generally consists of coarse and fine tracking systems.
The closed-loop coarse tracking usually works at a frequency of several Hz, and the field of view (FOV) is relatively large, which induces a large pointing error.
A fine tracking system with a frequency of up to kHz can point precisely, and the FOV is small.
The combination of coarse and fine tracking can provide a large FOV, high closed-loop bandwidth, and pointing precision.
The pointing error will induce spot jitter, where the instantaneous spot can be described using the Rice intensity distribution.
We can define the $\eta _p$ as the expected value of the mispointing loss and given by,
\begin{equation}\label{formula5}
\eta _p = \frac{{\omega _{at}^2}}{{\omega _{at}^2 + 4\sigma _p^2}}.
\end{equation}
where we assume the pointing probability density follows a Gaussian distribution with variance $\sigma _p$ \cite{Toyoshima1998}.
\textbf{Atmosphere transmittance}.
Atmospheric transmittance is reduced by the air absorption and the scattering of the propagating beam.
The atmospheric components mainly include gas molecules and tiny particles (such as water droplets, dust and aerosols).
The gas molecules have certain specific absorption lines, and the scattering can be described as Rayleigh scattering $I\left( \lambda  \right) \propto {I_0}\left( \lambda  \right)/{\lambda ^4}$ for the incident light intensity ${I_0} ( \lambda )$.
The particles in the atmosphere are mainly distributed on the ground surface, with a decreasing concentration as the altitude increases.
The transmittance is related to the visibility. 
When the particle size is equivalent to or larger than the wavelength, Mie scattering theory can be employed to describe the scattering phenomenon.
The scattering intensity is proportional to ${\lambda ^2}/r_p^2$, where $r_p$ is the particle radius.

\begin{figure*}[t]
\centering
\includegraphics[width=0.84\textwidth]{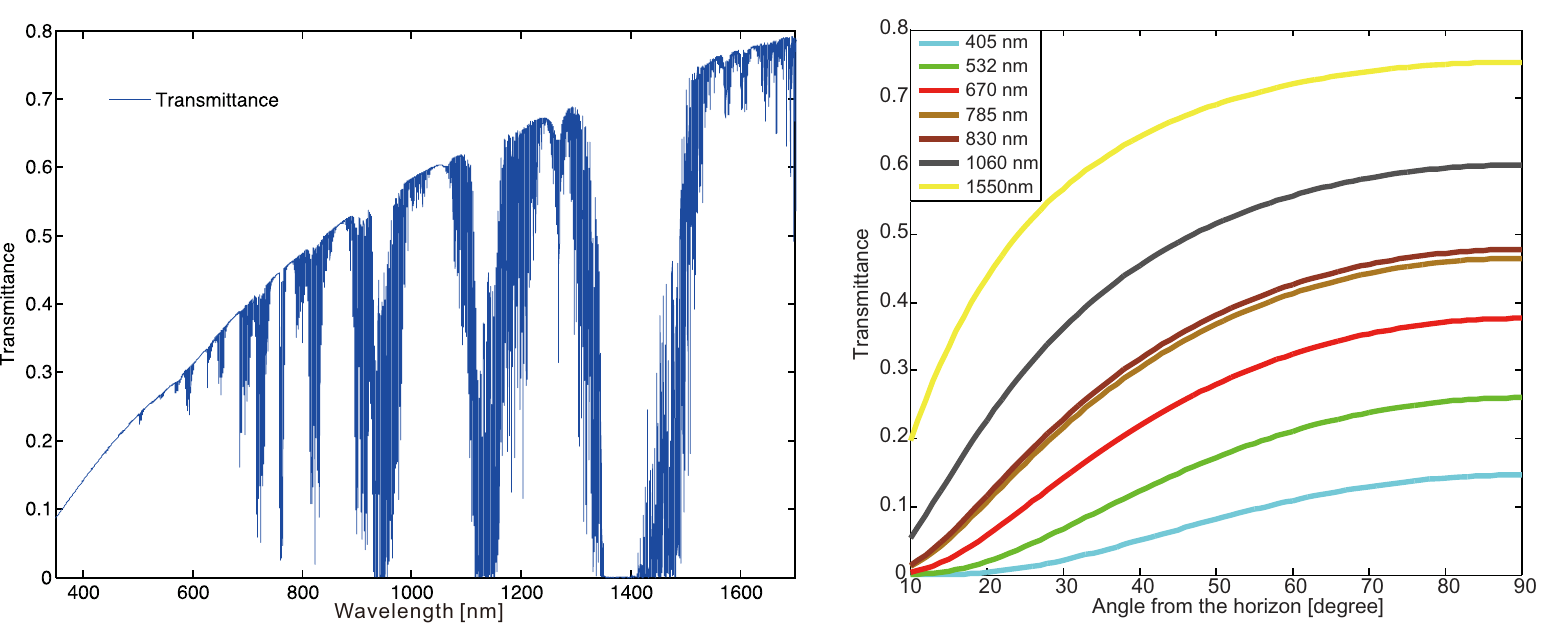}
\caption{
The simulated atmospheric transmittance for zenith with different wavelengths (left) and different altitude angles (right).
}
\label{fig:transmittance}
\end{figure*}

Software MODTRAN can be used to simulate and analyze the atmospheric transmittance, which was developed by the Spectral Sciences Inc. and the Air Force Research Laboratory.
Figure. \ref{fig:transmittance} shows the simulated atmospheric transmittance in a rural sea-level location with 5-km visibility \cite{Bourgoin2013}.
The transmittance is good at a high altitude angle because of the relatively short propagation time in the atmosphere.

\textbf{Downlink and uplink channels}.
Satellite-based quantum science experiments can be performed using two different channels, namely downlink (from the satellite to the ground) and uplink (from the ground to the satellite).
For downlink transmissions, the beam reaches the air turbulence with a large size and is received immediately after crossing the atmosphere, thus the impact are marginal for the beam broadening and deflection induced by turbulence.
On the contrary, for the uplink, photons encounter air turbulence at the beginning of propagation and subsequently transmit to the satellite.
Therefore, turbulence induced distortion will significantly increase the beam divergence angle and result in a larger channel attenuation than that in the case of the downlink transmission.
Although the downlink channel has more advantages in terms of efficiency, the uplink channel is still competitive in some cases, such as when one wished simplify the payloads.
Note that the effects of turbulence can be partly compensated in theory, by using an adaptive optics system with feedback devices.

\subsection{Feasible channel parameters for the low-Earth orbit satellite}
Typically, due to the low channel attenuation of downlink, the satellite-based QKD and entanglement distribution are suitable for one-downlink and two-downlink channels respectively.
Conversely, considering the limitation of satellite resources and the flexibility of the ground system, some experiments may be more suitable for uplink transmission, such as quantum teleportation based on a multiphoton entanglement source.
Therefore, one-downlink, two-downlink, and one-uplink channels would be the basic elements for constructing a global-scale quantum network.

\begin{table*}\center
\begin{tabular}{|c|c|c|c|c|}
\hline & {One-downlink} & {Two-downlink} & {One-uplink}\\
&{for QKD}&{~for entanglement distribution~}&{~for quantum teleportation~}\\
\hline ~~Geometry attenuation $\eta_G$~~ & ~~~~~~0.0128~~~~~~ & ~~~~~0.000164~~~~ & ~~~~0.00045~~~~\\
\hline Atmosphere attenuation: $\eta_A$ & 0.5 & 0.25 & 0.5\\
\hline Transmitter efficiency: $\eta_T$ & 1 & 0.25 & 0.5\\
\hline Receiver efficiency:  $\eta_R$& 0.4 & 0.16 & 0.4\\
\hline Coupling efficiency: $\eta_C$& 0.5 & 0.25 & 0.5\\
\hline Detector efficiency: $\eta_D$& 0.5 & 0.25 & 0.5 \\
\hline Mispointing: $\eta_M$& 0.5 & 0.25 & 0.5 \\
\hline Total loss ($\eta_G \eta_A  \eta_T \eta_R \eta_C \eta_D \eta_M$) & \textbf{-35 dB} & \textbf{-75 dB} & \textbf{-53 dB}\\
\hline
\end{tabular}
\caption{
The estimated total loss for one and two downlink, and one-uplink channels, respectively.
In this estimation, the distance is set to 1000 km for three types channels.
The divergence angles of transmitter for down-link channels are set to about 15 $\mu rad$, and for up-link channels will be extended to 20 $\mu rad$ due to the stronger effect of turbulence.
The diameter of the receiving telescope is set to 1.2 m and 0.3 m for ground-based and satellite-based respectively.
The geometry attenuation is evaluated by the distance, the divergence angle, and the diameter of the receiving telescope.
}
\label{tab:channalloss}
\end{table*}

According to Eqs.~\eqref{formula1}-\eqref{formula5} and using typical parameters, the expected total loss of above three types of channels can be evaluated, as shown in Tab.~\ref{tab:channalloss}.
When estimating the satellite-based channel loss, the relevant constrains, including technology, resources, and cost, must be considered.
Considering the technology's maturity and affordability, the 1-m-diameter ground-based telescope and the satellite-borne transmitter with a divergence angle of 15 $\mu rad$ and central wavelength of 850 nm is reasonable.
For the QKD based on one-downlink channel, the distance (Z) is 500 km to 1200 km, the optical diameter of the receiving telescope (D) on the ground is 1.2 m, the efficiency of optical transmitting system can be set to 100\% only for QKD based on the weak coherent state source (approximately 25\% for the two-downlink entanglement distribution); the efficiency of the optical receiving system (including optical, filter and detection efficiencies) is approximately 0.2, the effective divergence angle of the transmitter ($\theta$) after considering the effect of diffraction and atmospheric turbulence is $\sim$ 15 $\mu rad$ (approximately 20-30 $\mu rad$ for uplink channels), the atmospheric transmittance is typically 0.5, and the mispointing loss is 0.5.
The geometry attenuation can be evaluated to about 19 dB when Z = 1000 km, through the approximation formula, $2({D}/{(\theta \cdot Z}))^{2}$.
Thus, the total channel loss is estimated to be -35 dB.
Similarly, for the entanglement distribution based on two-downlink channels and the quantum teleportation based the on one-uplink channel, the total channel losses are expected to be approximately -75 dB and -53 dB, respectively, as shown in Tab.~\ref{tab:channalloss}.

Using the estimated total channel loss and the requirements of the satellite mission as the input condition, one can output the specifications of key technologies.
For example, when considering the mission of a satellite-based entanglement distribution with the two-downlink channels, the total experimental time will be limited by the conditions of satellite altitude angle and the weather of the two ground stations.
If the total experimental time is set to 10,000 s, according to the total loss of -75 dB, the output brightness of satellite-borne entangled photon source must be over $2 \times 10^6$ pair/s to obtain more than 1000 coincidence events.
Furthermore, to achieve the raw key rate of 1 kbps, the repetition frequency of the satellite-borne decoy state source must be at least 100 MHz when the channel loss is approximately -35 dB.
The mispointing value of 0.5 indicates the tracking accuracy of the APT system should be achieved to 4 $\mu rad$.

\begin{figure}[t]\centering
\resizebox{8cm}{!}{\includegraphics{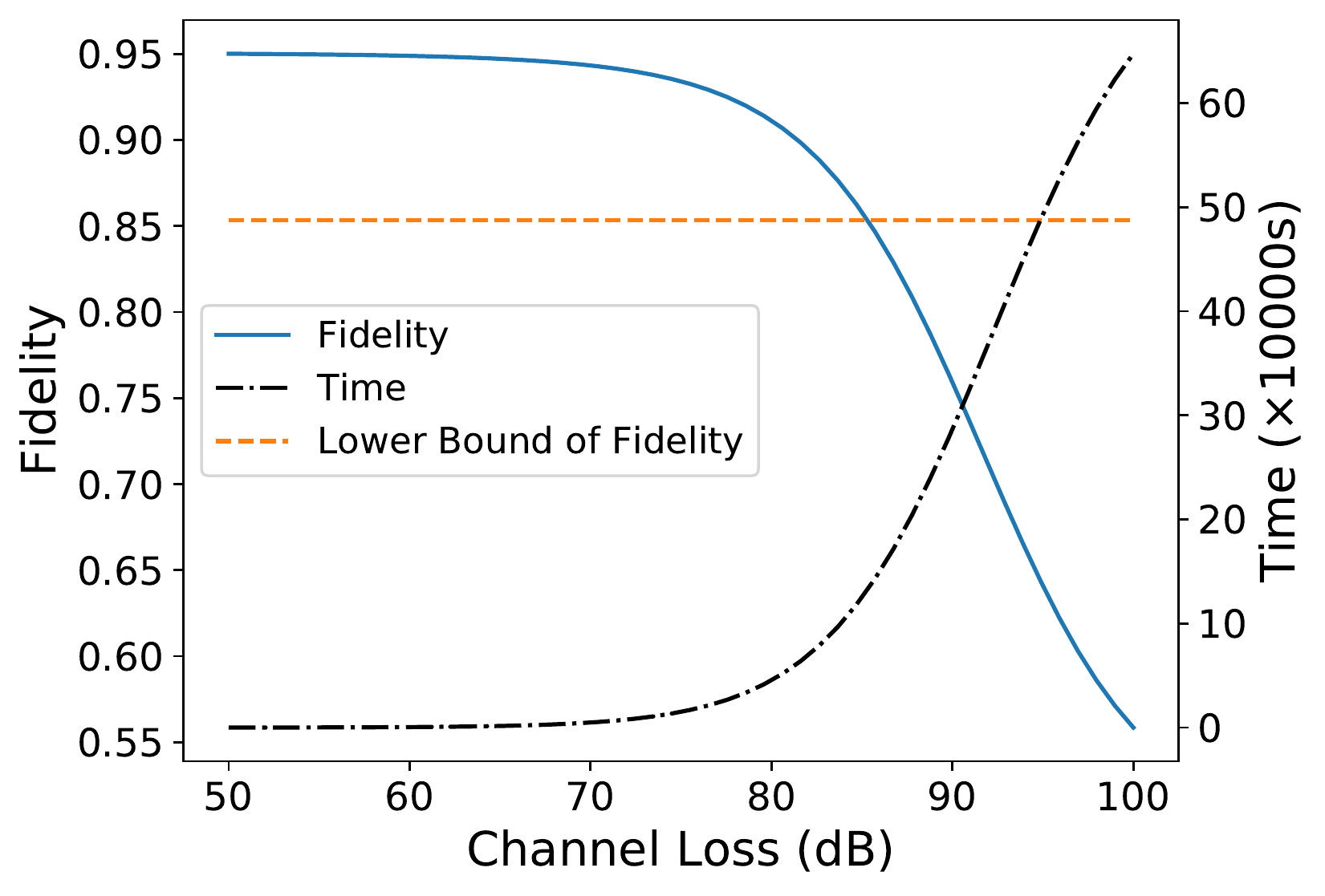}}
\caption{The effective fidelity and total experimental time with different two-downlink channel loss. The original state fidelity of the entangled photon source is 95\%, which will degrade with the increase of channel loss due to the inevitable noise. The lower bound of fidelity is defined as the minimum value for a violation of Bell inequality. The time coordinate represents the total required experimental time of accumulating 1000 coincident events on the ground.}
\label{Fig:twodownlinkloss}
\end{figure}

Furthermore, Fig.~\ref{Fig:twodownlinkloss} shows the complete evaluation of the total required experimental time and effective fidelity of the received entangled photon pairs on the ground with different total channel losses, when we consider the specific parameters of the background noise, the target detected coincident events on the ground, the fidelity of the satellite-borne entangled photon source, and the time synchronization accuracy.
According to this estimation, we can conclude further that the fidelity of the satellite-borne entangled photon source should be larger than 95\%, and the total channel loss should be less than 80 dB (naturally it will be 40 dB for the one-downlink channel) when the total experimental time is limited to $<$ 10,000 s.

\begin{figure}[t]\centering
\resizebox{8cm}{!}{\includegraphics{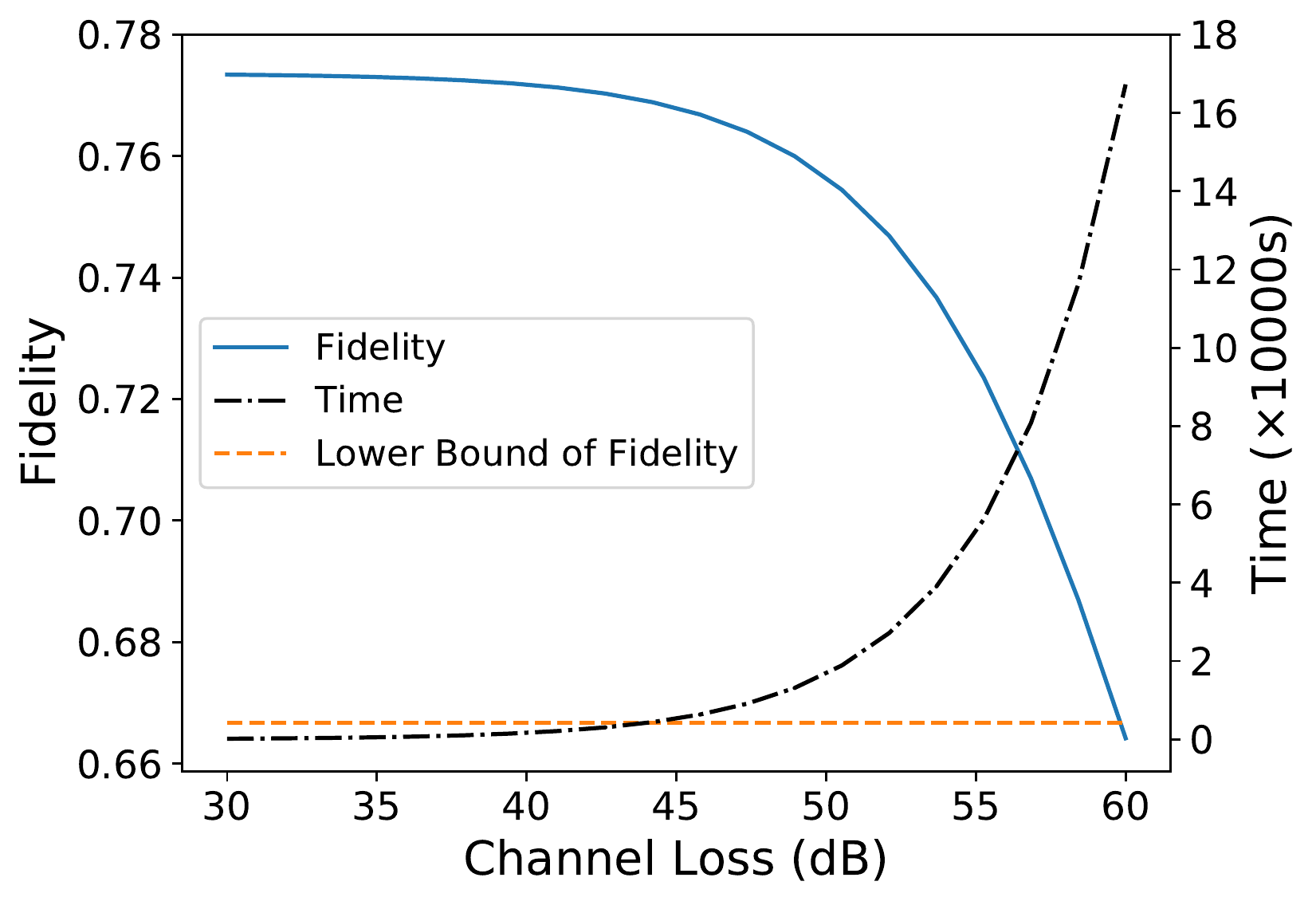}}
\caption{The effective fidelity and total experimental time with different one-uplink channel loss. The lower bound of fidelity is the classical limit of fidelity on a single copy of qubit. The time coordinate represents the total required experimental time of accumulating 400 coincident events.}
\label{Fig:oneuplinkloss}
\end{figure}

Regarding ground-to-satellite quantum teleportation, a similar simulation study is shown in Fig.~\ref{Fig:oneuplinkloss}. To realize the mission target of 400 coincident counts, the total channel loss should be less than 55 dB and the fidelity should be more than 75\%, when the total experimental time is limited to $<$ 40,000 s.

Finally, based on the above analysis and simulation of the feasible channel parameters, we can summarize the main requirements of satellite-based quantum science experiments, as shown in Tab.~\ref{tab:mainrequirement}.
These specific technical indicators can be directly used as verification targets for the ground-based feasibility studies (see Section VI), and the design input can be employed to develop payloads (see Section VII).
Furthermore, we would like to emphasize that only when we have performed this comprehensive analysis can we conduct the ground-based feasibility study and the develop key technologies following a systematic approach.

\begin{table}\center
\begin{tabular}{|l|l|l|}
\hline \multicolumn{2}{|l|}{\textbf{List of feasible design baseline:}} \\
\hline {Satellite operating lifetime} & {$\geqslant$ 2 years} \\
\hline Time synchronization accuracy~~ & $\leqslant$ 1 ns (1 $\sigma$)~\\
\hline \multicolumn{2}{|c|}{\textbf{Satellite-to-ground QKD}}\\
\hline Raw key rate & $\geqslant$ 1 kbps\\
\hline QBER & $\leqslant$ 3.5\% \\
\hline Total experimental time & $\geqslant$ 20000 s \\
\hline Total Channel loss & $\leqslant$ 40 dB \\
\hline \multicolumn{2}{|c|}{\textbf{Satellite-based entanglement distribution}}\\
\hline Received coincident count & $\geqslant$ 1000 \\
\hline Effective fidelity & $\geqslant$ 85\% \\
\hline Total experimental time & $\geqslant$ 10000 s \\
\hline Total Channel loss & $\leqslant$ 80 dB \\
\hline \multicolumn{2}{|c|}{\textbf{~Ground-to-satellite quantum teleportation~}}\\
\hline Received coincident count & $\geqslant$ 400  \\
\hline Effective fidelity & $\geqslant$ 75\% \\
\hline Total experimental time & $\geqslant$ 40000 s \\
\hline Total Channel loss & $\leqslant$ 55 dB \\
\hline
\end{tabular}
\caption{
Main practical requirements of satellite-based quantum science experiments.
}
\label{tab:mainrequirement}
\end{table}

\section{Ground-based feasibility studies and key technologies}

This section reviews the systematic ground-based feasibility studies and key technologies based on the requirements for the three Micius satellite missions as specified in Section V.

Before constructing and launching a costly satellite, thorough preliminary studies and simulations on the ground must be systematically performed to verify the scientific possibilities, evaluate the risks, and develop the technologies. 
The key questions of interest include, ``Can the single and entangled photons pass through the effective thickness (which is about 10 km) of the atmosphere?'',
``Can the quantum optics experiments be performed (with a sufficient count rate and signal-to-noise ratio) under the conditions of large attenuations and various turbulence on moving platforms?'' 
The ground-based feasibility demonstration and the development of key technologies cover the following five aspects: overcoming the effective atmospheric thickness, testing the feasibility of satellite-ground channels, testing moving objects, time synchronization, and polarization maintenance and compensation. 
Emphatically, only after developing all these technologies and combining them together compatibly, the Micius program was officially approved and the construct of the satellite started.

\subsection{Overcoming the effective atmospheric thickness}

\begin{figure}
\includegraphics[width=0.44\textwidth]
{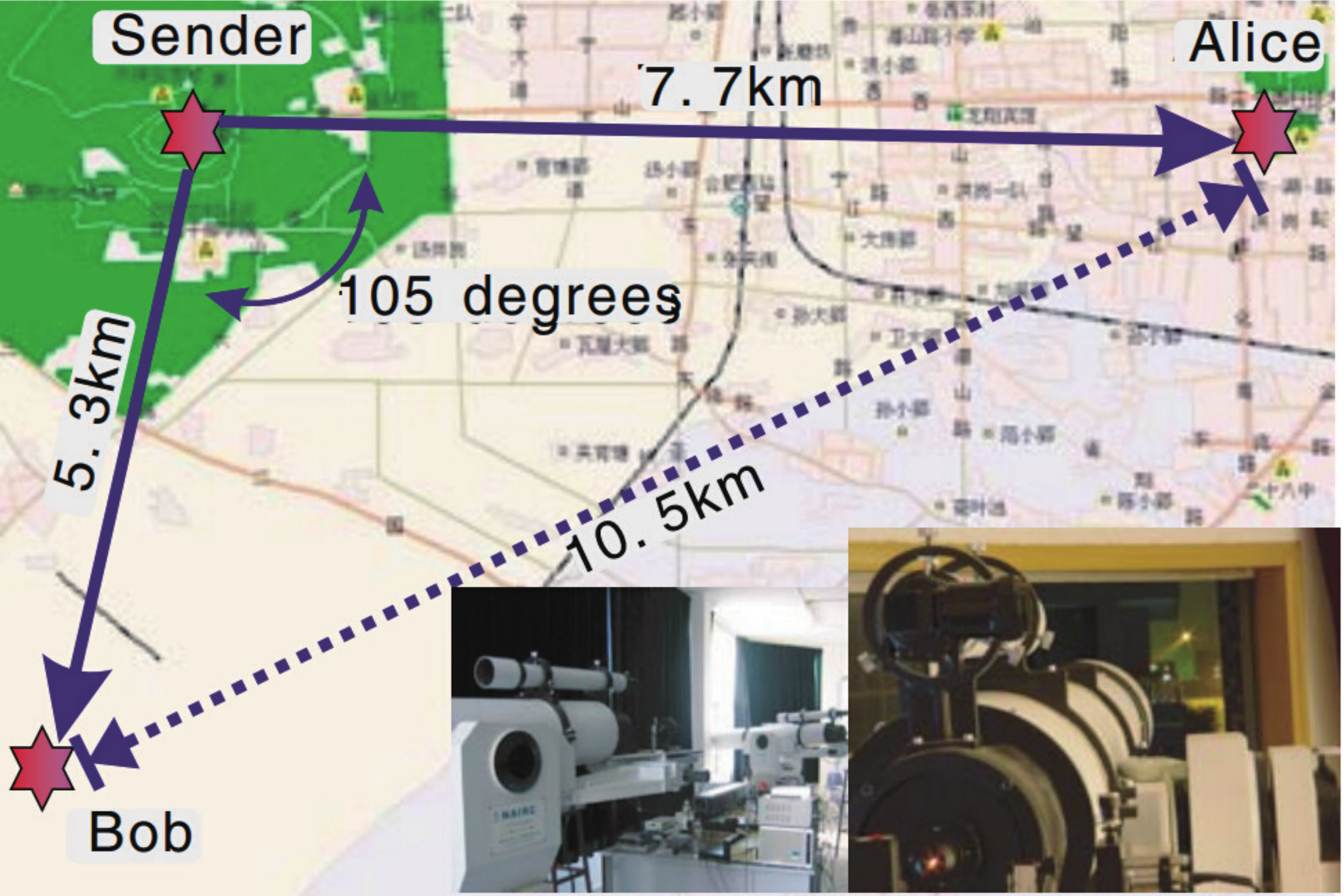}%
\caption{Overview of the Hefei 13 km entanglement distribution experiment. \cite{peng2005}}%
\label{fig:peng2005}%
\end{figure}

The first step involves verifying if the effective atmospheric thickness is favorable for the passage of single and entangled photons.
In 2005, entangled photon pairs were bidirectionally distributed over Hefei city, one arm across 5.3 km and the other across 7.7 km, conclusively exceeding the effective atmospheric thickness \cite{peng2005}, as shown in Fig.~\ref{fig:peng2005}.
Narrowband (2.8 nm) filtering and time synchronization (with a precision of 20 ns) were employed to reduce the background counts from the noisy city environment. 
The two-photon count rates were 10 kHz and 150-300 Hz at the sender and the receivers, respectively.
This corresponded to channel attenuation values in the range of 15.2-18.2 dB, depending on the weather conditions.
The physical separation between the two receivers was 10.5 km, which enabled the performance of a space-like Bell test with a measured $S$ value of $2.45\pm0.09$.
Three years later, single photons were transmitted over the Great Wall of China with an optical free-space distance of 16 km \cite{Jin2010tele}. 
These studies demonstrated that entanglement can still survive after both entangled photons have passed through the noisy ground atmosphere with a distance beyond the effective thickness of the aerosphere.
This is a step towards low-Earth-orbit satellite-to-ground downlink quantum experiments.

\subsection{Feasibility of satellite-to-ground one and two-downlink channels, and ground-to-satellite uplink channel}

To experimentally verify the feasibility of satellite-based quantum key distribution (QKD) through one-downlink channel, Wang \textit{et al}. conducted a full verification study of the decoy-state QKD over the 97-km free-space link and demonstrated the possibility of achieving a high signal-to-noise ratio and overcoming the obstacle of a high-loss environment~\cite{Wangdirect2013}. 
The total loss achieved in the experiment was over 50 dB, which is more than the expected value of $\sim$ 35 dB with the one-downlink channel mentioned in Section V.

\begin{figure*}[t]
  \centering
\includegraphics[width=0.85\textwidth] {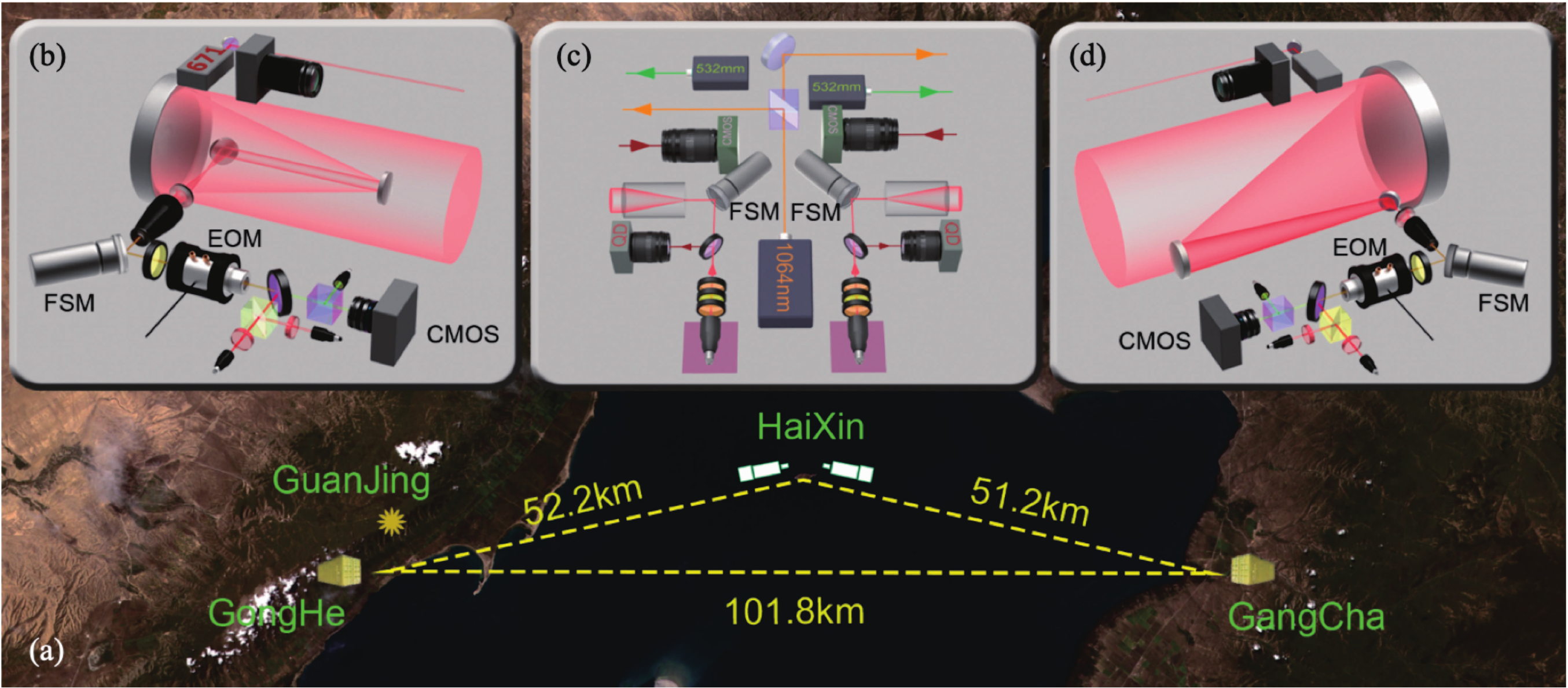}%
\caption{Bidirectional two-link entanglement distribution over Qinghai Lake with a distance of 101.8 km. \cite{yin2012-100km}}%
\label{fig:yin2012}%
\end{figure*}

Two-downlink channels are required for the satellite-to-ground entanglement distribution, both of which require two independent channels between three different locations.
From 2008 to 2010, a ground-based feasibility study was performed in Qinghai Lake, China \cite{yin2012-100km}.
In the study, entangled photon pairs were distributed over a two-link free-space channel with distances of 51.2 km and 52.2 km to two receivers separated by 101.8 km (Fig.~\ref{fig:yin2012}).
For the study, a crucial enabling technology was developed, i.e., closed-loop tracking, which was operated with a bandwidth and precision of 150 Hz and 3.5 $\mu$rad, respectively.
This tracking bandwidth was sufficient to overcome most of the atmospheric turbulence.
The authors obtained a two-photon rate of 6.5 MHz and measured two-photon correlation functions, violating Bell's inequality by 2.4 standard deviations.
The average two-link overall attenuation was measured to be 79.5 dB, which is higher than the estimated two-downlink loss (approximately 75 dB, as mentioned in Sec. V) based on the LEO satellite.

Going beyond the above mentioned up to two-photon experiments, multi-photon platform can be employed to test more sophisticated experiment on the quantum teleportation of independent single photons.
In terms of satellite-based teleportation over the one-uplink channel, more tolerance for the channel attenuation is required.
Yin \textit{et al}. investigated quantum teleportation through a free-space channel over 97 km in Qinghai Lake, China \cite{yin2012-100km}.
Employing a high-brightness multi-photon interferometry~\cite{Yao2012} with a four-photon count rate of 2 kHz, after transmitting the teleported photon over 97 km with a channel loss of 35-53 dB, a final count rate of 0.08 Hz was obtained.

For the three missions, the total channel attenuation in these works was higher than the estimated values in Sec. V, which provided the sufficient verification for three types of quantum channels.

\begin{figure*}[t]\centering
\resizebox{16cm}{!}{\includegraphics{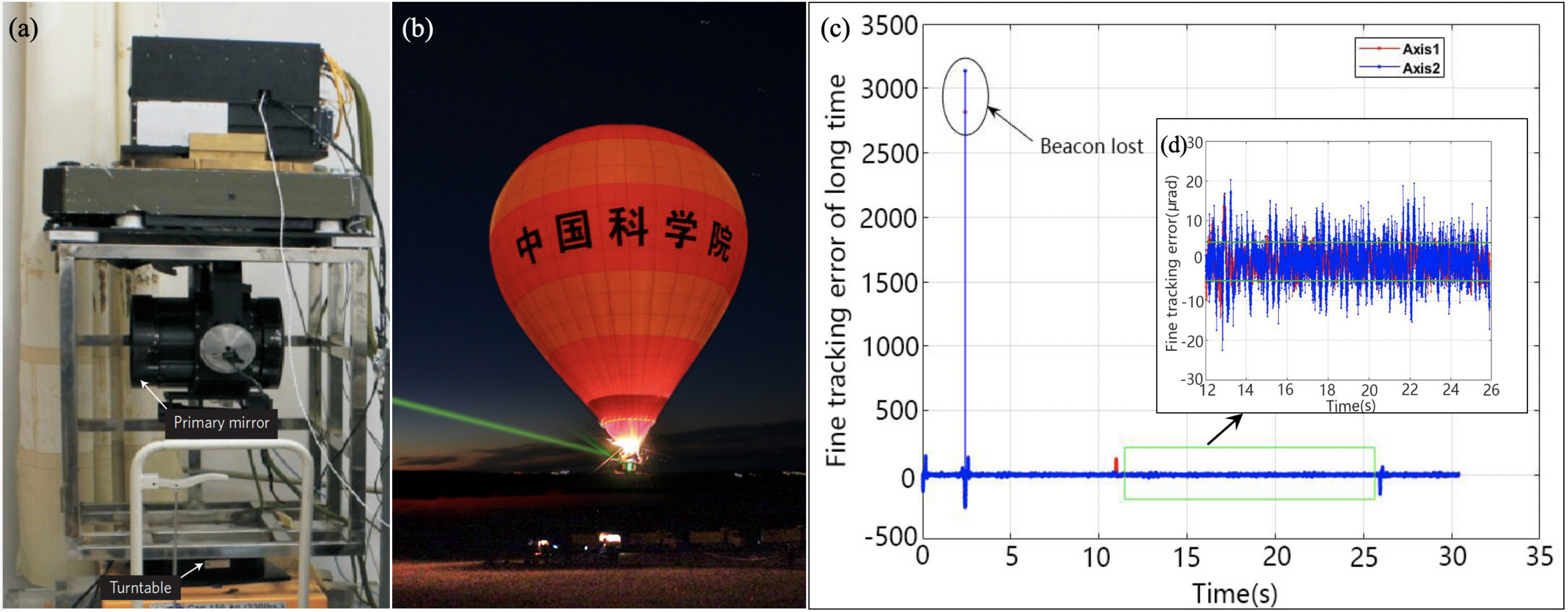}}
\caption{(a) The QKD transmitter is mounted on a turntable which has approximate dimensions of 500 mm$\times$450 mm$\times$600 mm (but 500 mm$\times$500 mm$\times$1,000 mm including the supporting metal frame). This is used for simulation of satellite orbiting. When the turntable undergoes a complex motion, the transmitter terminal will move accordingly. The simulation was done with an angular velocity and angular acceleration larger than that in a typical LEO. (b) Photo of the rising and erupting hot-air balloon in the floating platform experiment. (c) Tracking error in the hot-air balloon simulation. Long-time fine tracking error with the beacon light sometimes out of the field of view. Inset: Fine tracking error in the stabilized time area. When the beacon light is in the tracking field, tracking accuracy is about 5 $\mu$rad. \cite{Wangdirect2013}}
\label{Fig:100km-tracking}
\end{figure*}

\subsection{Testing with moving and vibrating objects}

Real implementations of satellite-based quantum communications require high-bandwidth and high-accuracy acquiring, pointing, and tracking (APT).
All the ground-based feasibility studies mentioned thus far are based on stationary systems.
For eventual satellite-based quantum communications, one should consider that the satellite performs rapid, relative angular motions with respect to the ground stations, which may include unwanted random motion.
For a typical LEO satellite at an altitude of 400-800 km, the angular velocity can reach 20 mrad/s and the angular acceleration can reach 0.23 mrad/s$^2$.
A verification environment that incorporates all possible motion modes and simulations of extreme events, including vibration, random motion, and attitude change, is highly desirable.

To this end, Wang \textit{et al}. carried out two other experiments, in addition to the 97-km high loss one mentioned in Subsection VI-B, for the direct and full-scale experimental verifications towards satellite-ground QKD \cite{Wangdirect2013}.
Simulation experiments with a turntable and a hot-air balloon were implemented to simulate the platform in a rapidly moving orbit as well as the vibration, random motion and attitude change related to the LEO satellite.
The turntable (see Fig.~\ref{Fig:100km-tracking}(a)) provides motion with a maximum angular velocity of 21 mrad/s and a maximum angular acceleration of 8.7 mrad/s$^2$.
With a distance of 40 km between the transmitter and receiver, such a motion regime completely covers the possible range of motion parameters for a 400-800 km LEO satellite.
The floating hot air balloon (see Fig.~\ref{Fig:100km-tracking}(b)), with a distance of 20 km from the ground receiver, was employed as a randomly vibrating and floating platform, which afforded an average angular velocity of 10.5 mrad/s and an average angular acceleration of 1.7 mrad/s$^2$ owing to its random motion.
The balloon could perform random and dramatic motions, which positioned it out of the view of the field.
This motivated the researchers to recapture the target rapidly, typically within 3-5 s (see Fig.~\ref{Fig:100km-tracking}(c)).
The performance of the acquiring, tracking and positioning system, including both coarse control and fine control, is summarized in Tab.~\ref{tab:APT}.
These verification environments incorporate more extreme events, including vibration, random movement, and attitude change, compared to what would suffer from an actual LEO satellite.

On the other hand, by utilizing the cube-corner retrorefector on satellites to simulate the quasi-single-photon source, one can verify the feasibility of establishing a quantum signal link between the satellite and the ground station \cite{Villoresi_2008}.
Yin \textit{et al}. performed an experimental simulation of a quasi-single-photon transmitter on the satellite with an average photon number of 0.85 per pulse and a full divergence angle of 38 $\mu rad$ sending to the ground \cite{Yin2013oe}.

\begin{table}
\resizebox{0.5\textwidth}{!}{
\begin{tabular}{@{}llll@{}}
\textbf{Components} & & \textbf{Transmitter terminal} & \textbf{Receiver terminal} \\
\hline
Telescope diameter & & 200 mm & 300 mm\\
Coarse pointing&Type& Two-axis gimbal&Two-axis\\
mechanism & & mount&gimbal mirror\\
&Tracking range & Azimuth:$\pm 45^{\circ}$ & Both:$\pm 5^{\circ}$ \\
& &Elevation:$\pm 70^{\circ}$& \\
Coarse camera & Field of view & $2^{\circ}$&$1^{\circ}$\\
&Size& $1000 \times1000$& $640\times480$\\ & &pixels&pixels \\
Fine pointing&Type&Fast steering&Fast steering\\
machanism& &mirror&mirror \\
& Tracking range & $\pm 0.7$ mrad &$\pm 0.7$ mrad \\
Fine camera & Field of view & $512~\mu rad$ & $512~\mu rad$\\ &size\&frames& $128 \times 128$ \&& $128 \times 128$ \&\\ &per second& 2300 Hz&2300 Hz\\
Tracking errors & Coarse & $\pm 200~\mu rad$ & \\
& tracking error& & \\
& Fine tracking& $\pm 5~\mu rad$ & \\
&error& &
\end{tabular}}
\caption{The parameters of APT system. \cite{Wangdirect2013}}
\label{tab:APT}
\end{table}


\subsection{Time synchronization}

\begin{figure}
\centering
\includegraphics[width=8.5cm]{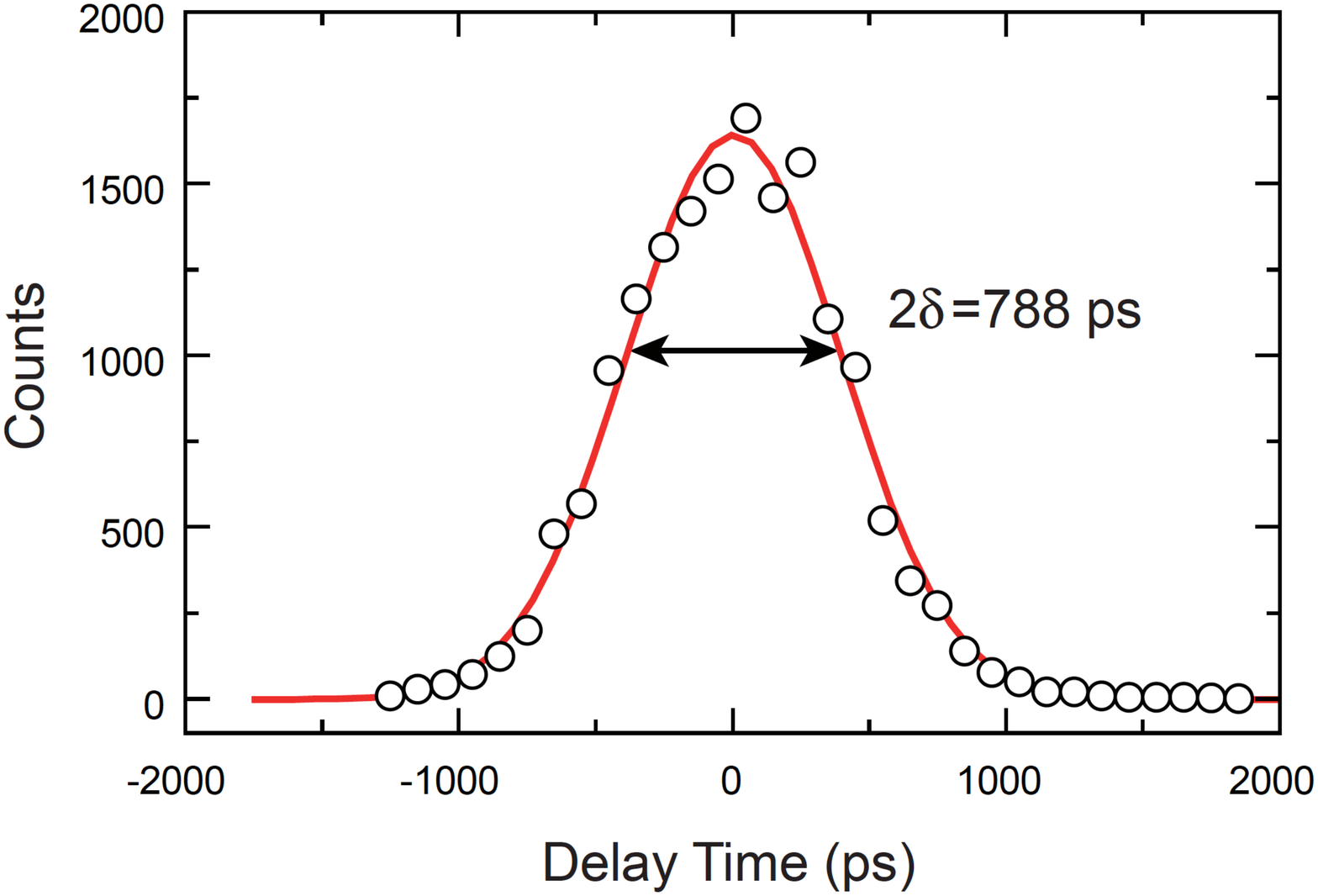}
\caption{
A typical recorded arrival time of the 1064 nm laser signals at both stations.
Synchronization accuracy of 788 ps was observed \cite{yin2012-100km}.
}
\label{fig:synchronization}
\end{figure}

Since the transmitter and receiver are separated by a large distance and have independent reference clocks, time synchronization is conducted to determine the absolute photon number and distinguish the signal photons from the noise.
As the distance between the transmitter and receiver changes all the time when the satellite passes over the ground station, both the GPS pulse-per-second (PPS) signal and an assistant pulse laser are employed in the typical synchronization scheme.
From the works of \cite{peng2005} to \cite{yin2012-100km}, this synchronization method has been continuously developed and improved.

In the experiment of \cite{yin2012-100km}, a pulsed laser (1064 nm, 10 kHz, 50 mW, 200 $\mu$rad) is employed for time synchronization.
A pulse length of this laser was 2.65 ns (full width at half maximum), with a rising edge of 2 ns.
The GPS PPS is added to synchronize the starting time.
They addressed the forefront of the detected signals with the constant fraction discriminator technique and utilized a high-accuracy time-to-digital converter (TDC) with a time resolution of 100 ps to record the arrival time of the synchronization signals at both Alice's and Bob's stations.
Through the above methods, in most cased, the jitter caused by the laser pulse itself and energy shaking is reduced.
Finally, the time synchronization of better than 1 ns for the quantum channel is achieved, as shown in Fig.~\ref{fig:synchronization}.
The technique of time synchronization developed in these works can be applied directly in satellite-based quantum science experiments.

\subsection{Polarization maintenance and compensation}
Polarization encoding is employed in the long-distance free-space quantum communication experiments. 
The relative motion of the satellite and the ground station can induce a time-dependent rotation of the photon polarization observed by the receiver. 
Thus, the polarization maintenance and compensation are necessary.
Wang \textit{et al}. developed the following methods to improve a system's polarization visibility: (1) all the reflection mirrors are coated with tailored films to maintain high polarization ($\geq$ 1,000:1), (2) using two mirrors for phase matching in polarization deflection, and (3) other optical elements (PBS and so on) are custom-made with a high extinction ratio \cite{Wangdirect2013}.
When all the above features were considered when setting up the optical system, a polarization extinction ratio of 200:1 was obtained for both transmitter and receiver systems.
Further, when the transmitter and receiver were connected during an experiment, the total polarization extinction ratio was as high as 100:1, which meets the requirements of the satellite-based QKD.

Another crucial challenge associated with satellite-ground quantum communication, attributed to the kinematical reference system, is the polarization compensation in real-time.
While the satellite reference frame keeps changing during the satellite motion, the satellite and ground are relatively static when the two sides face each other at any specific time in this frame.
In most cases, the polarization state changes over time because of the two-dimensional rotation of the telescope.
Following the telescope azimuth or elevation rotation, a pair of mirrors changes its position, which leads to changes in polarization.
This problem can be solved by inputing the real-time data of the orbit prediction into an auto-rotatable half-wave plate (HWP) for polarization tracking, which has been verified by \cite{Yin2013oe} and later employed by ground stations for the Micius satellite.
The extinction rate of the polarization under tracking was tested to be more than 100:1 with this method.

\subsection{Other parallel ground-based free-space experiments}

In parallel to the above-reviewed, comprehensive and systematic ground-based verification works dedicated to the Micius satellite, many other experiments were implemented around the world, and the scientific conclusions have been consolidated. 
In the experiment \cite{Resch2005freespace} performed in Vienna, one of the entangled photons was detected locally, while the other one was sent through free space across 7.8 km. 
A new testbed for free-space quantum communications---the link between the Canary islands of La Palma and Tenerife---was employed by \cite{Zeilinger:Decoy:2007} for testing decoy-state QKD over a distance of 144 km.  
Later, Ursin \textit{et al}. adopted the same experimental configuration as in \cite{Resch2005freespace} to send triggered single photons across a distance of 144 km (one link) using a free-space link between the Canary Islands La Palma and Tenerife  \cite{ursin2007entanglement}. 
Due to the various atmospheric influences at such a long distance scale, the apparent bearing of the receiver station varied in tens of seconds to minutes.
To maximize and stabilize the link efficiency, an active stabilization of the optical link was implemented via a closed-loop tracking system to correct the beam drifts induced by atmospheric changes, reducing the beam drift from 70 $\mu$rad (10 m) to 7 $\mu$rad (1 m).
Using this one-link free-space channel, quantum teleportation has also been demonstrated ~\cite{ma144kmtele} with a channel attenuation of approximately 30 dB.
A further experiment from the same group upgraded the BBO crystal into a more efficient down-conversion crystal, i.e., periodically poled KTiOPO$_4$, which generated entangled photons at $\sim$1 MHz and sent the two photons through the free-space channel across 144 km \cite{Fedrizzi:Entangle:2007, Fedrizzi2009High}.

To test the APT with moving objects, Nauerth \textit{et al} conducted a QKD experiment from an airplane to the ground \cite{Nauerth2013}.
The airplane was moving at a speed of 290 km/h, at a distance of 20 km, which corresponds to an angular velocity of 4 mrad/s.
The transmitting beam was narrowed by a divergence of 180 $\mu$rad.
To establish a stable link with this divergence, fine-pointing assemblies were implemented and optimized on both sides, with a precision upper bound of 150 $\mu$rad.
With these efforts, the experiment yielded an asymptotically secure key at a rate of 7.9 bit/s.

In addition to employing moving platforms to demonstrate the feasibility of downlink channels, uplink channels were verified by the research team from Canada with a truck and airplane.  
In 2015, they reported the first demonstration of QKD from a stationary transmitter to a receiver platform located on a moving truck \cite{Bourgoin2015}. 
In this experiment, QKD was implemented with a moving receiver at an angular speed similar to that of a satellite at an altitude of 600 km. 
Furthermore, they equipped a receiver prototype on an airplane to demonstrate QKD via an uplink channel \cite{Pugh_2017}. 
They specifically designed the receiver prototype to consist of many components that were compatible with the environment and resource constraints of a satellite. 
Their above ground-based feasibility experiments on uplink channels with moving platforms provided solid technical support for the follow-up satellite project---QEYSSat.

In 2017, \cite{Takenaka2017Satellite} used a classical laser source on the LEO satellite SOCRATES to test the feasibility of the satellite-to-ground quantum-limited link.
Additionally, \cite{Gunthner:17} completed a similar experiment using the classical laser source from a geostationary satellite.

\section{Developing and testing of the satellite payloads}

The Micius program was officially approved in 2011.
Construction of the first prototype satellite started in 2012, and was completed in 2014.
Thereafter, the project turned to build the flight model of the satellite, which was completed in November, 2015.
After a series of environmental tests, including thermal vacuum, thermal cycling, shock, vibration, and electromagnetic compatibility tests, etc., the Micius satellite, weighing 635 kg, was well-prepared and ready to be launched.
On August 16, 2016, the Micius satellite was successfully launched by the Long March-2D rocket, from the Jiuquan Satellite Launch Centre, China.
The orbit was circular and sun-synchronous with an altitude of 500 km.

\begin{figure*}[t]\centering
\resizebox{15cm}{!}{\includegraphics{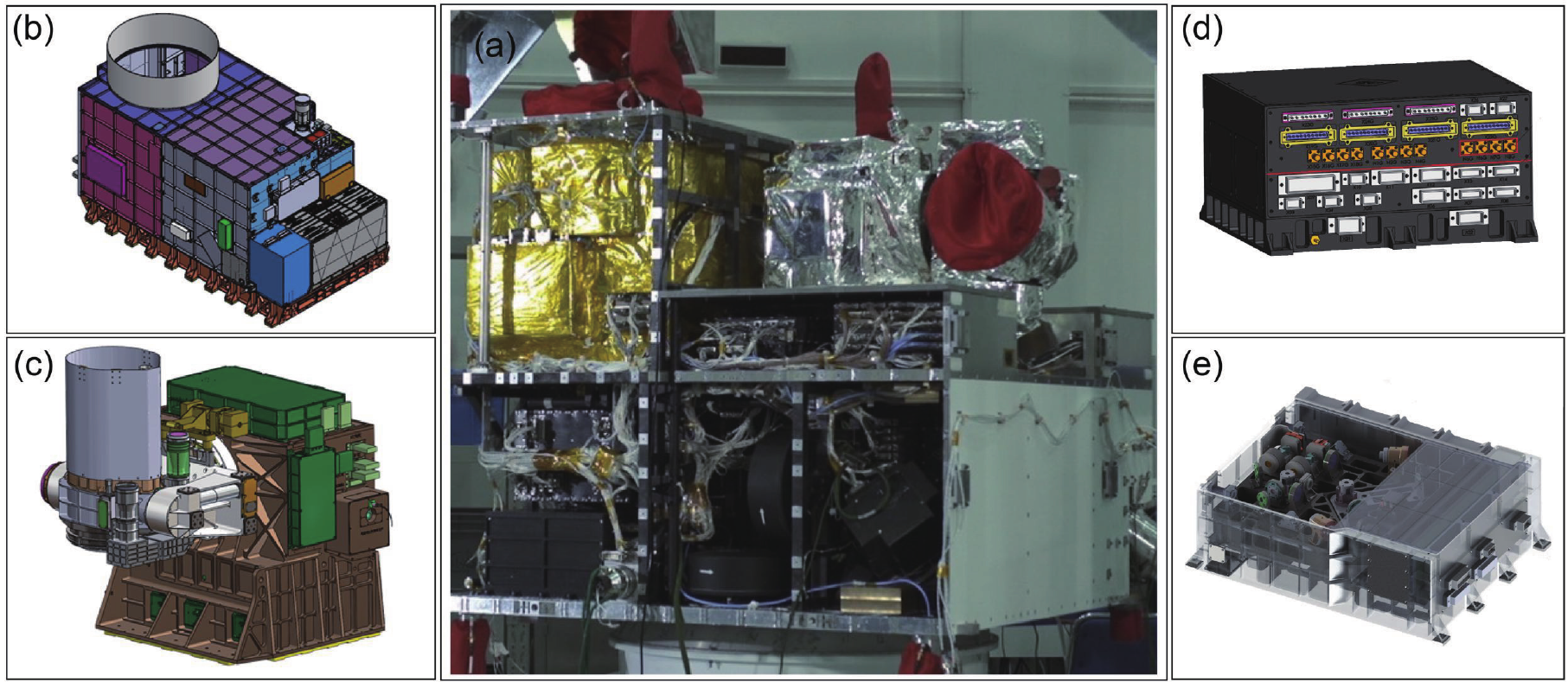}}
\caption{
The full view of the Micius satellite and main payloads.
(a) The photograph of the Micius satellite before launching. (b) The transmitter 1 for QKD, entanglement distribution and teleportation. (c) The transmitter 2 specially for entanglement distribution. (d) The experimental control box. (e) The entangled-photon source.}
\label{Fig:satellite}
\end{figure*}

\subsection{Payload design}

The Micius satellite has a double-decker design (see Fig.~\ref{Fig:satellite}).
The payloads for the science experiments are composed of two optical transmitters (transmitter 1 \& 2), a space-borne entangled photon source (the upper layer of the satellite), an experimental control processor, and two APT control boxes (the lower layer of the satellite), as shown in Fig.~\ref{Fig:satellite}(a).

The dimension of the experimental control box, as shown in Fig.~\ref{Fig:satellite}(d), is $280~\textrm{mm} \times 264~\textrm{mm} \times 150~\textrm{mm}$, weighing 7.5 kg.
The experimental control box has six main functions: experimental process management, random-number generation and storage, modulation of the decoy-state photon source, synchronization-pulse recording, QKD post-processing (including raw-key sifting, error correction and privacy amplification to obtain the secure final keys), and encryption management.

\begin{figure}[!t]\centering
\resizebox{7.52cm}{!}{\includegraphics{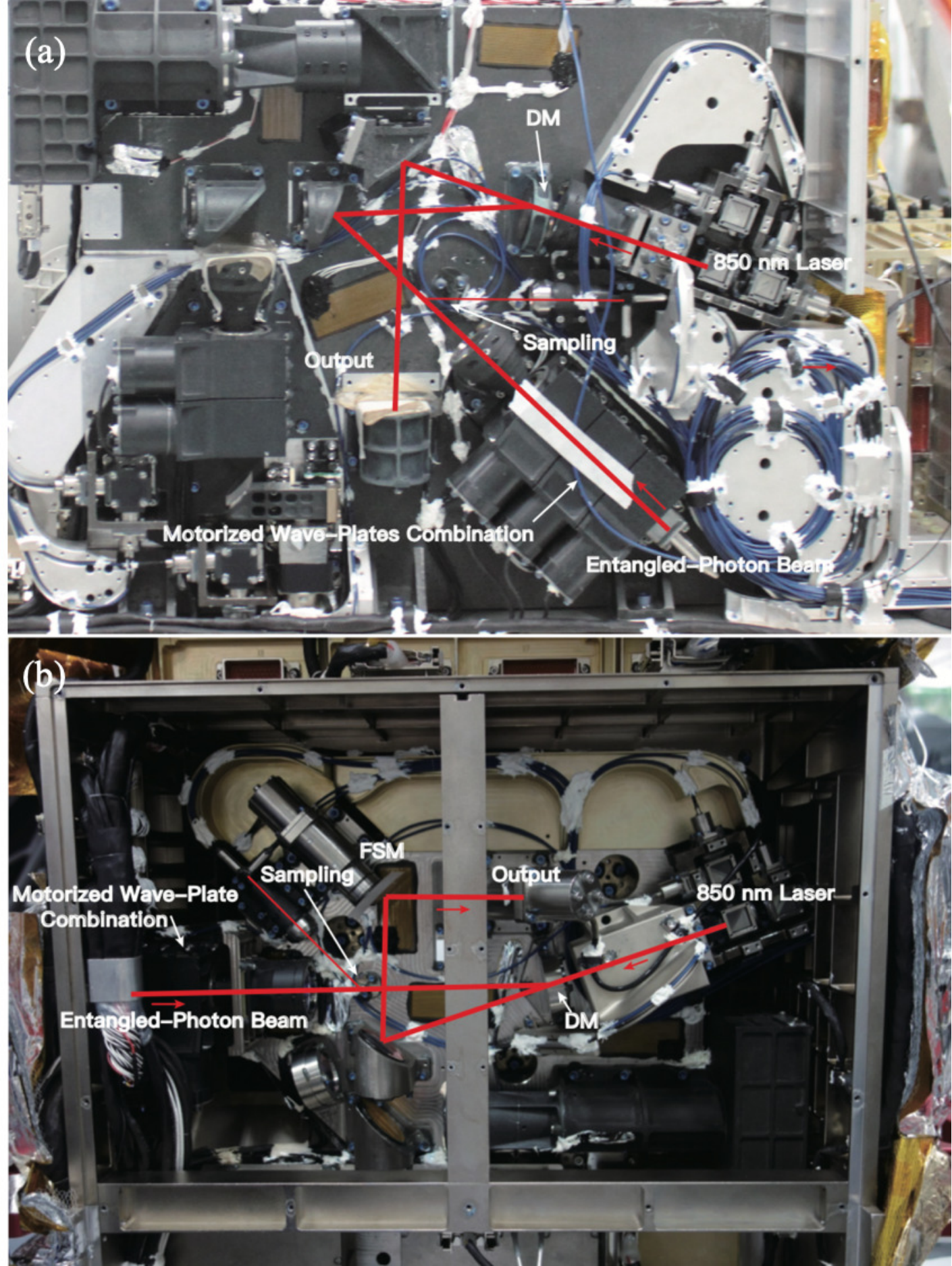}}
\caption{
The top view of transmitter's optics head. \cite{liao2017satellite}
The collimated beam from the entangled-photon source, passes through a motorized wave-plates combination, a beam expander, and then is combined with the 850 nm synchronization laser by a dichroic mirror.
(a) The transmitter 1's optics head.
(b) The transmitter 2's optics head.
}
\label{Fig:MTJF}
\end{figure}

Transmitter 1, with a diameter of 300 mm and a total weight of 115 kg, as shown in Fig.~\ref{Fig:satellite}(b), is involved in all the three main scientific goals.
It comprises eight laser diodes with drivers, a BB84 polarization encoding module, a telescope, a receiving module (including a QWP, HWP, PBS, and two single-photon detectors), and the APT system (including a beacon laser, a coarse camera, two-axis mirror, a fine camera, a fast steering mirror (FSM), etc.), as shown in Fig.~\ref{Fig:MTJF}(a).
Transmitter 2, with a diameter of 180 mm and a total weight of 83 kg, as shown in Fig.~\ref{Fig:satellite}(c), is specially designed for the quantum entanglement distribution from the satellite to two separate ground stations.
Further, it can serve as transmitter 1's backup for the satellite-based QKD.
Both transmitters contain a telescope and an optical box.
To reduce the emission loss, an off-axis telescope design is employed in transmitter 2.
The optical box mainly consists of a fine tracking system, an integrated receiving module for sampling measurement, and a motorized wave-plate combination for polarization correction, as shown in Fig.~\ref{Fig:MTJF}(b).

Additionally, both transmitters contain a multistage APT system.
The first stage involves satellite attitude control, where the system keeps the photons pointing to the ground station with an error less than 0.5 degree.
The second stage is the coarse control loop involving the two-axis gimbal mirror for transmitter 1 and the two-dimensional rotatable telescope for transmitter 2.
The third stage is the fine control loop, involving a FSM driven by piezoceramics and a camera.

\begin{figure}[!t]\center
\resizebox{7.5cm}{!}{\includegraphics{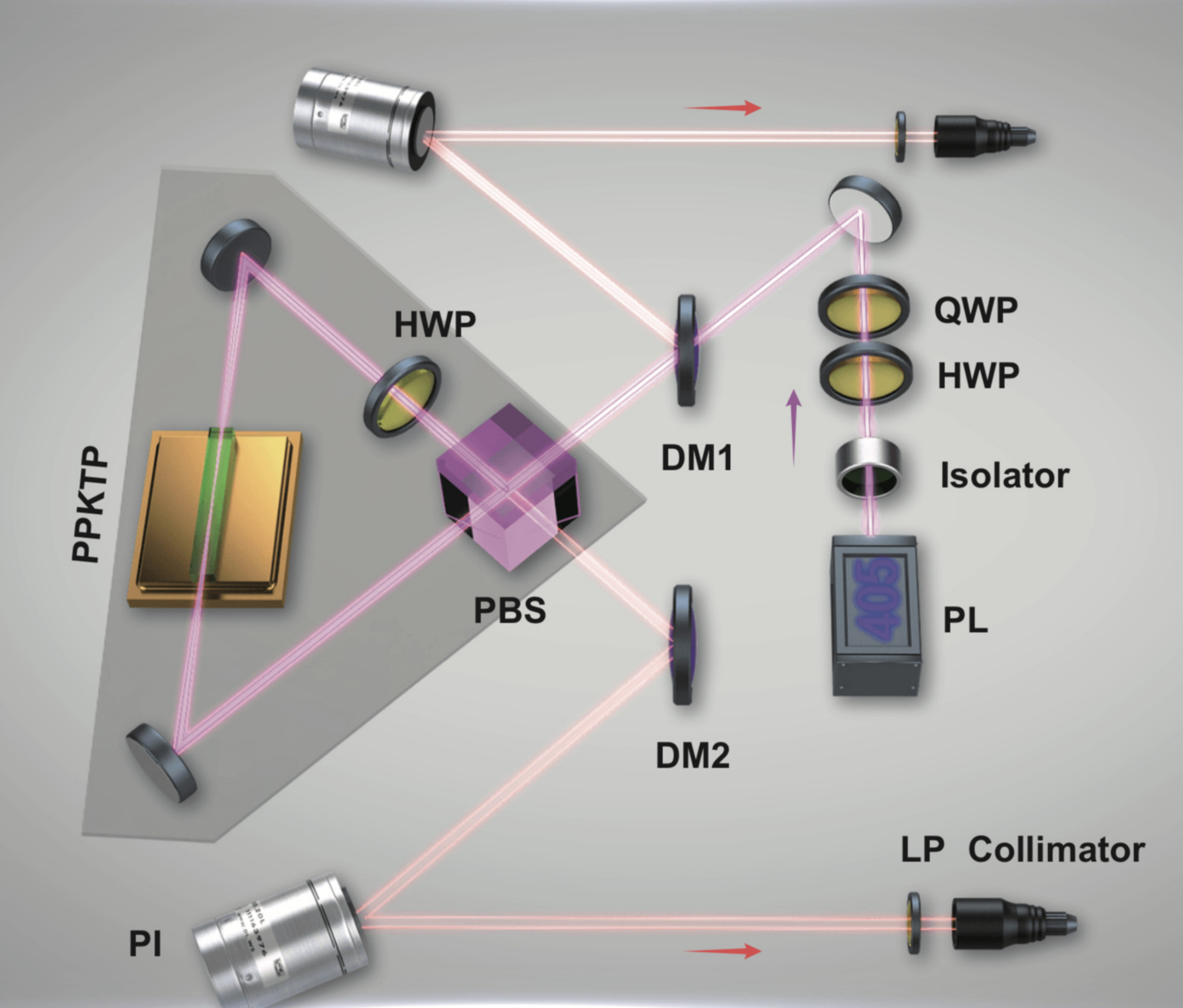}}
\caption{
The schematic of the satellite-borne entangled-photon source. \cite{yin2017satellite}
The entangled-photon source in Micius satellite is based on type-II PPKTP (periodically poled $KTiOPO_{4}$) crystal and the Sagnac interferometer.
PL, pump laser; DM, dichromatic mirror; PI, piezo steering mirror; QWP, quarter-wave plate; HWP, half-wave plate; PBS, polarizing beam splitter.
}
\label{Fig:entanglesource}
\end{figure}

The space-borne entangled-photon source (SEPS) is an optomechatronics integration payload with a dimension of $430~\textrm{mm} \times 355~\textrm{mm} \times 150~\textrm{mm}$ and a total weight of 23.8 kg, as shown in Fig.~\ref{Fig:satellite}(e).
The schematic of SEPS is shown in Fig.~\ref{Fig:entanglesource}.
A continuous-wave laser diode with a central wavelength of 405 nm and a linewidth of 160 MHz was employed to pump a periodically poled KTiOPO$_4$ (PPKTP) crystal inside a Sagnac interferometer.
The pump laser, split by a PBS, passes through the nonlinear crystal in clockwise and anticlockwise directions simultaneously, producing down-converted photon pairs at a wavelength of 810 nm, as polarization-entangled states close to the form ($|H\rangle_1|V\rangle_2+|V\rangle_1|H\rangle_2)/\sqrt{2}$, where $H$  and $V$  denote the horizontal and vertical polarization states, respectively, and the subscripts 1 and 2 denote the two output spatial modes.
This source is robust against various vibrations, temperatures, and electromagnetic conditions.
With a pump power of 30 mW, the source emits 5.9 million entangled-photon pairs per second, with a state fidelity of 0.907~$\pm$~0.007.

The optical elements are mounted and glued on the both sides of a 40 mm-thick optical bench.
The optical bench is composed of titanium alloy, which has a good balance of rigidity, thermal expansion, and density.
The most sensitive structure of SEPS, the Sagnac interferometer, is integrated into a palm-sized invar material plate with a thickness of 15 mm and thermo-insulated embedded into the titanium plate to achieve an optimal stability when the satellite is launched and in orbit.
The input and output beams of the Sagnac interferometer are collimated because of the confocal design, which relaxes the requirement of a high mounting accuracy for the couplers.
In addition, two piezo steering mirrors (PI) are employed to correct the pointing offset of the beam.
The Sagnac interferometer module in the SEPS has passed a series of space environment adaptability tests, such as the thermal-vacuum and vibration tests, which help release thermodynamic stress in advance and enhance system stability.
The SEPS is guided to two optical transmitters through two single-mode fibers with lengths of 280 mm and 410 mm, respectively.
After utilizing a 5$\times$ beam expander, a mirror is placed at the edge of the beam to sample the entangled photons by $1\%$.
An integrated BB84 receiving module, consisting of two Wollaston prisms and one beam splitter, realized a random measurement of four polarizations (0, 90, 45 and 135 degree).
Through the $1\%$ sampling in both entangled-photon transmitters, the source brightness can be estimated in orbit.

The APT control box dimension is $326~\textrm{mm} \times 244~\textrm{mm} \times 242~\textrm{mm}$ with a weight of 10 kg, and it mainly contains the control electronics for the coarse tracking loop and the fine tracking loop.
Its functions specifically as a motor driver, FSM driver, coarse feedback loop controller and fine feedback loop controller.

\subsection{Testing the payload under various conditions}
\begin{figure*}[!t]\centering
\resizebox{16cm}{!}{\includegraphics{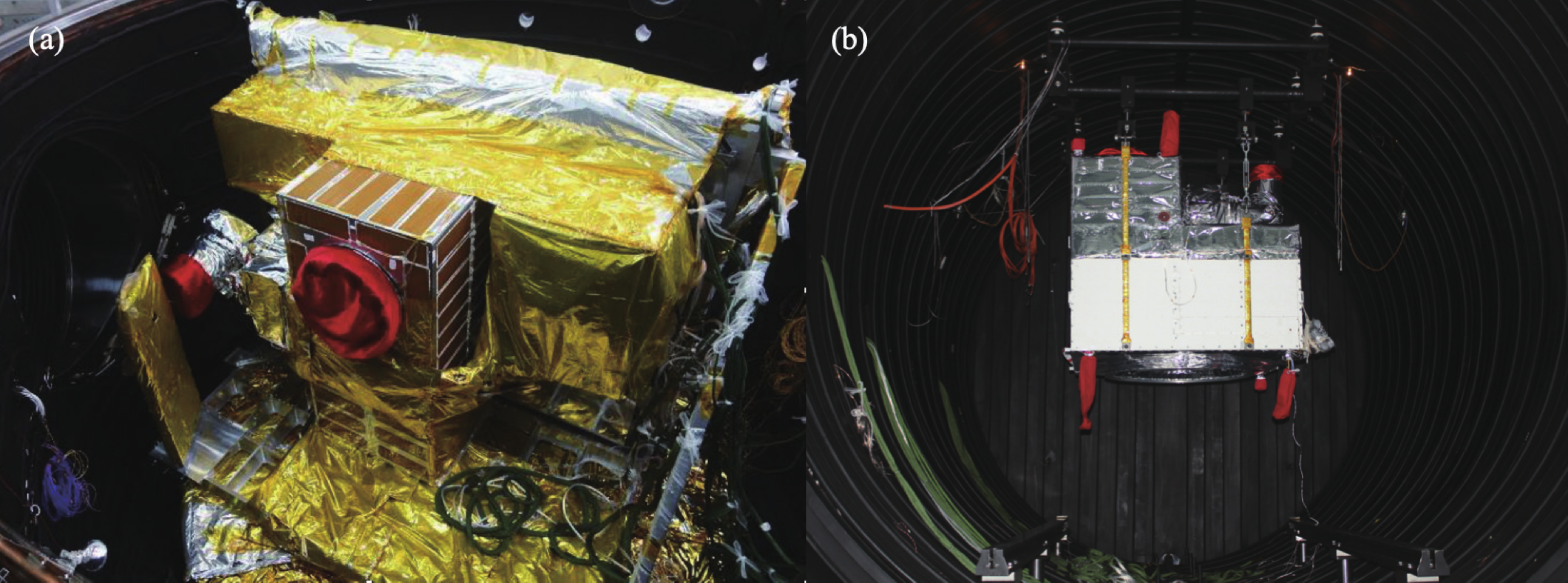}}
\caption{The thermal-vacuum-optical measurement (a) and the thermal balancing testing (b) for payload combination.
}
\label{Fig:thermalphoto}
\end{figure*}

For the design of the payloads, besides the functional requirements, it is also very important to ensure adaptability in space environment.
In general, there are three main types of space environment tests: vibration, thermal, and vacuum tests.
The vibration test can be subdivided into sinusoidal, random vibration, and impact test.
The thermal and vacuum tests include thermal vacuum, thermal cycling, and thermo-vacuum-optical measurements.
Considering all the payloads described in subsection VII.A, each payload must be subjected to these three space environmental simulation tests.
After every payload has passed the test separately, all the payloads are combined for the final test in the thermal-vacuum environment.
During the development of the Micius satellite, two integral space environment tests for the payload combination, thermal-vacuum-optical measurement and the thermal balance testing for the whole satellite, are implemented (see the Fig.~\ref{Fig:thermalphoto}).

\begin{figure}[!t]\centering
\resizebox{8cm}{!}{\includegraphics{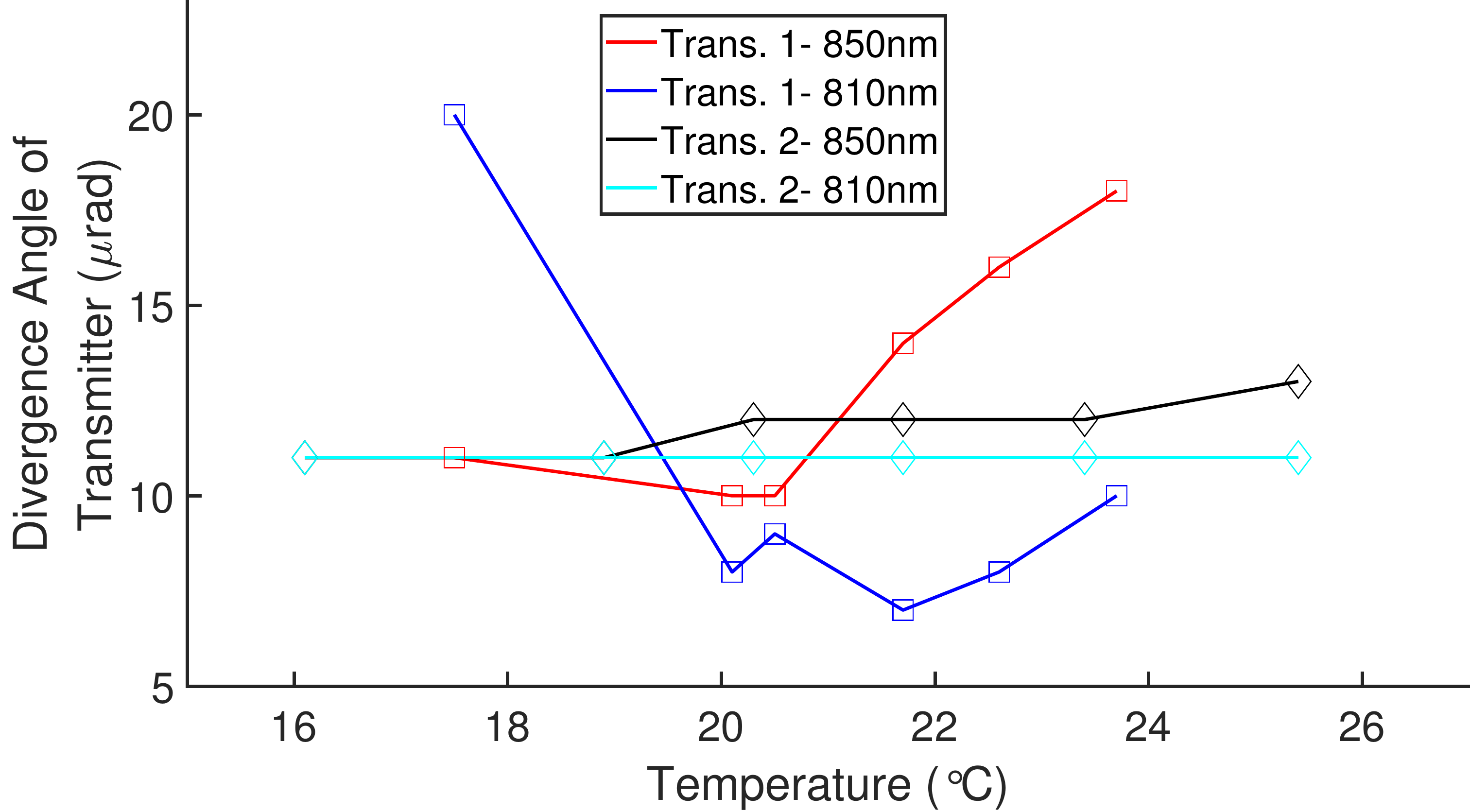}}
\caption{Test results of divergence angles in thermal vacuum conditions
}
\label{Fig:thermalresult}
\end{figure}

For the thermal-vacuum-optical measurement, as shown in Fig.~\ref{Fig:thermalphoto}(a), the integration of all the scientific payloads are integrated into a thermal-vacuum chamber.
The special design of this type thermal vacuum chamber is the optical window connected with large aperture collimator, which can provide measurement on the optical parameters in real time in various thermal vacuum environments.
The divergence angle is one of the key typical optical parameters for the two satellite-based transmitters, and it directly affects the channel loss.
The test results of the divergence angles are shown in Fig.~\ref{Fig:thermalresult}.
The figure shows the optimum operating temperature of the two transmitters at 810 nm and 850 nm, respectively.
Furthermore, for transmitter 1 and 2, the divergence angle meets the requirement of $<$ 15 $\mu rad$ when the temperature ranges from 18$^{\circ}$C to 22$^{\circ}$C.

Figure~\ref{Fig:thermalphoto}(b) shows the photo of the whole system of the Micius satellite when carrying out the thermal-balance testing, which is typically a part of the whole system thermal vacuum testing.
It mainly has two objectives: obtaining thermal data for analytic thermal model correlation and verifying the function of the thermal control of the whole system or the subsystems.
The test for the whole system of the Micius satellite lasted for 18 days, and six thermal balance operating conditions and four thermal cyclings were completed in total.

In addition to the above three types of space environmental tests, the radiation test is also very important.
The Micius satellite contains the silicon avalanche photodiodes (Si APD) in both transmitters for implementing the experiment of the ground-to-satellite quantum teleportation, and for sampling the entangled photons to test the in-orbit performance of the satellite-borne entangled photon source.
The Si APD is the radiation-sensitive device; thus, radiation degrades their performance by increasing dark currents, as well as decreasing responsivity and gain.
The radiation test was performed to the space-borne Si APDs of the Micius satellite, using 50 MeV proton source, which effectively penetrates the APD package front glass window.
According to the results of the radiation test, the in-orbit radiation-induced dark count rate (DCR) increment of the silicon APD increased by $\sim$219 cps/day.
To mitigate the APD radiation-induced DCR increment rate and guarantee the detector reliability, several solutions were developed, including multistage cooling technologies and specialized driver electronics for the space-borne, low-noise detectors.
Through these solutions, the expected detector in-orbit DCR increment rate can be reduced to less than 1 cps/day, satisfying the requirement of satellite-based quantum science experiments \cite{yang2019radiation, ren2017satellite}.

\section{Construction of cooperative ground stations}

\begin{figure*}[!t]\centering
\resizebox{15.2cm}{!}{\includegraphics{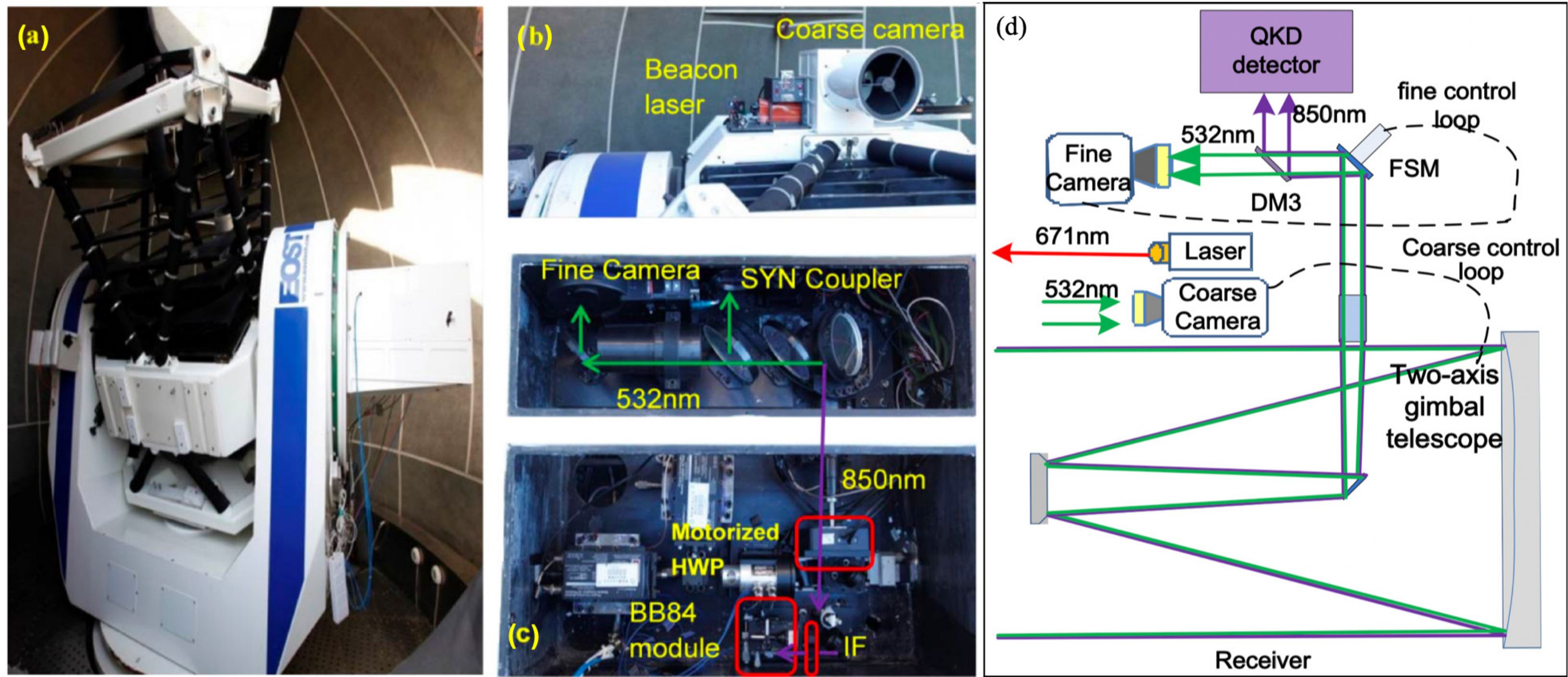}}
\caption{
The typical receiving ground station for the Micius satellite. \cite{liao2017satellite}
(a) The two-axis gimbal telescope.
(b) Beacon laser and coarse camera.
(c) One of the two layers of the optical receiver box.
(d) The typical optical design of the receiver, including the receiving telescope, the ATP system, and the QKD-detection module.}
\label{Fig:groundstation1}
\end{figure*}

To coordinate with the quantum science satellite in implementing the experiments, new ground stations need to be built or existing ones upgraded.
There are totally five ground stations in China: four of them are for receiving via downlink channels, while one single station for transmitting via uplink channels.
The receiving ground stations, each with a large-diameter telescope, i.e., the 1-meter-diameter telescope at the Xinglong Station of National Astronomical Observatories, CAS/NAOC (upgraded and rebuilt) for the satellite-to-ground QKD, the 1.8-meter-diameter telescope at Lijiang Station of Yunnan Astronomical Observatory/YAO (upgraded and rebuilt) for the entanglement distribution experiment, the 1.2-meter-diameter telescope at the Nanshan Observatory of Xinjiang Astronomical Observatory/XAO (newly built) for the entanglement distribution and QKD experiments, the 1.2-meter-diameter telescope at the Qinghai Station of Purple Mountain Observatory/PAO (newly built) for the entanglement distribution and QKD experiments (see Fig.~\ref{Fig:groundstation1}).

The transmitting station with three small transmitter telescopes located in Ngari (Ali) was specially constructed for the quantum teleportation experiment from the ground to the satellite, as shown in Fig.~\ref{Fig:groundstation2}.

The ground station in Xinglong was upgraded and rebuilt in 2014 for satellite-to-ground QKD experiments.
As shown in Fig.~\ref{Fig:groundstation1}(b), it consists of a Ritchey-Chretien telescope, a red beacon laser (671 nm, 2.7 W, 0.9 mrad), a coarse camera (field of view (FOV): $0.33^{\circ} \times 0.33^{\circ}$, pixels of $512 \times 512$, frames-per-second (FPS) of 56 Hz), and an optical receiver box which is installed on the arm of the gimbal.
A two-axis gimbal in a control loop with a coarse camera is employed to realize the coarse tracking function.
The 532 nm beacon laser coming from the satellite is detected by the coarse camera.
Guided by the 532 nm beacon laser, the 671 nm beacon laser transmitted from the ground telescope can point to the satellite precisely.
The fine tracking system and the BB84 measurement module are mounted in the receiver box.
The fine tracking system mainly includes a FSM based on voice-coil and a fine camera (FOV: 1.3 $\times$ 1.3~mrad, pixels: 128 $\times$ 128, FPS: 212 Hz).
A dichroic mirror was used to separate the 850 nm photons from the 532 nm beam.
A beam splitter (BS) was used to divide the 532 nm beam into two parts: one was detected by the fine camera for tracking, while the other was coupled into the fiber linked to a single-photon detector for the time synchronization.
The 850 nm photons are measured by a customized BB84 polarization analysis module, after passing through a beam expander, a motorized HWP and an interference filter.
The detectors' electric output pulses and the GPS PPS signal are fed into a time-to-digital convertor (TDC), which records the detection time and the channel number of the detectors.

\begin{figure*}[!t]\centering
\resizebox{11.5cm}{!}{\includegraphics{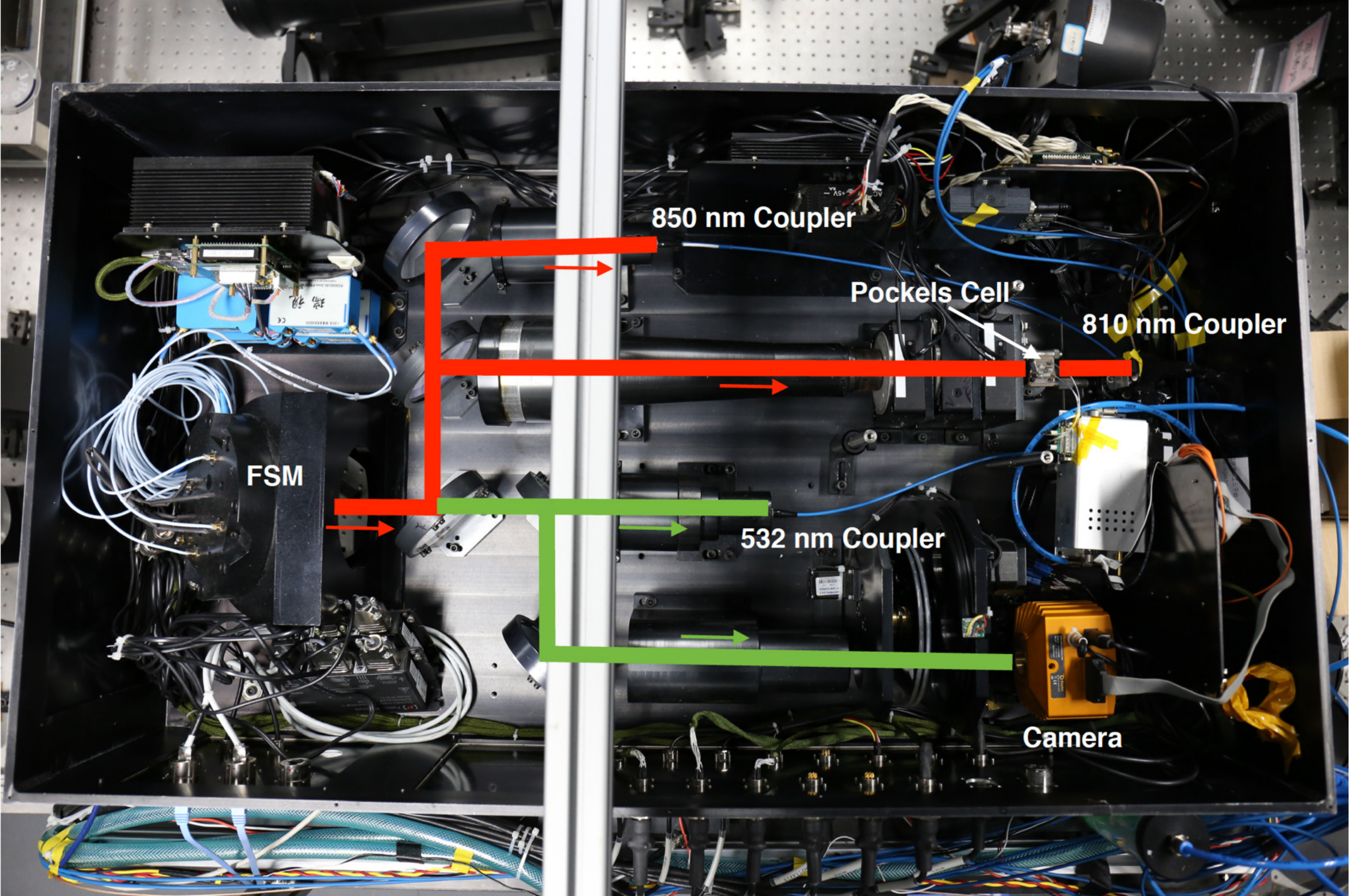}}
\caption{
The photograph of the measurement box at Lijiang station, which represents the typical design for the satellite-based entanglement distribution. \cite{yin2017satellite}}
\label{Fig:optbox-entangle}
\end{figure*}

\begin{figure*}[!t]\centering
\resizebox{15.2cm}{!}{\includegraphics{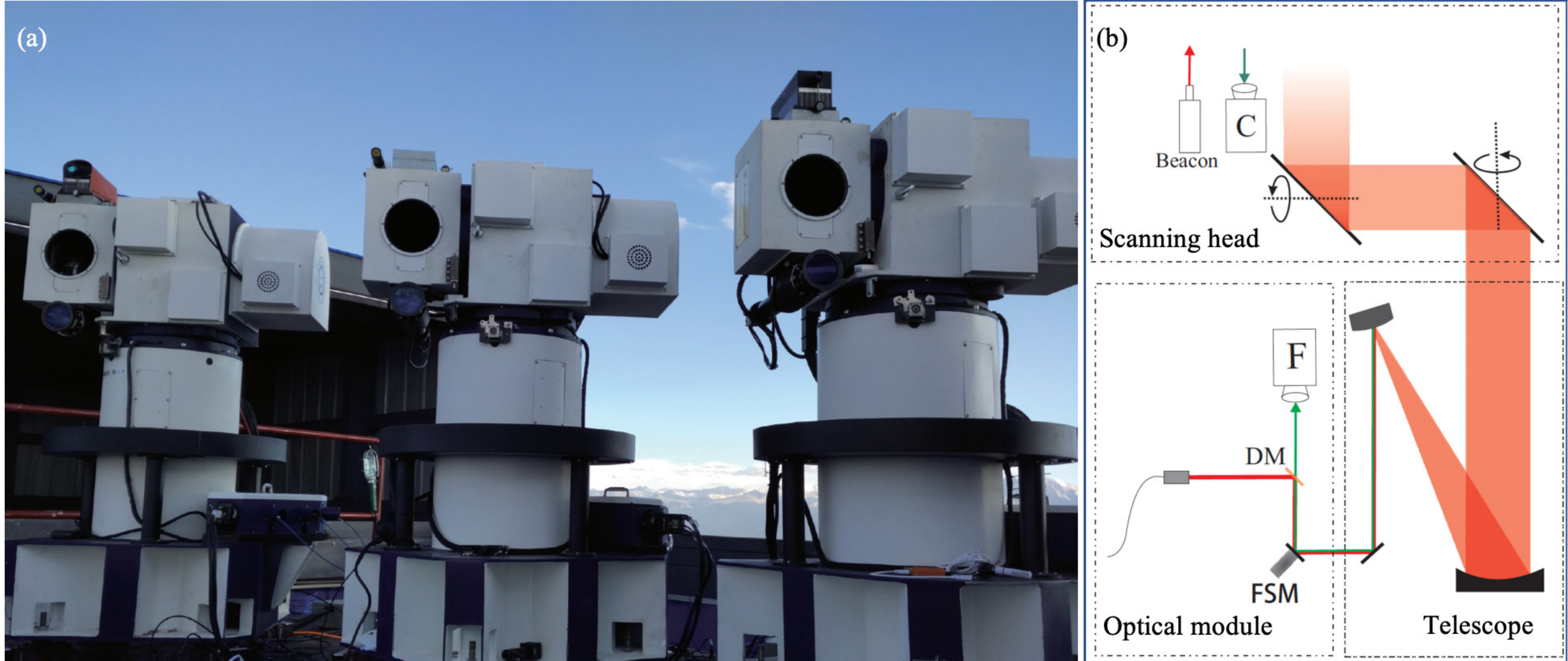}}
\caption{
The transmitting telescope and optical design. \cite{ren2017satellite}
(a) The photograph of transmitting optical antennas, which are placed side by side on a platform and backup each other. Quantum signals generated from the lab at floor 1 are transmitted to three telescopes. The beacon lasers and sync. laser are equipped on the top of the transmitting antennas.
(b) The optical design of the transmitting telescope, which comprises a scanning head, a transmitting telescope, and an optical module.}
\label{Fig:groundstation2}
\end{figure*}

For the mission of the satellite-based entanglement distribution, three ground stations located respectively in Delingha, Urumqi, and Lijiang, are involved.
The distance between Delingha and Lijiang (Nanshan in Urumqi) is approximately 1203 km (1120 km).
Two new telescopes (with a diameter of 1.2 m) with the same design were built in Nanshan and Delingha in 2015, mainly for entanglement distribution experiments, as well as satellite-to-ground QKD experiments.
All the optical elements in two telescopes have a polarization-maintaining property.
The measurement boxes are installed on one of the rotating arms, and they rotate along with the telescopes, as shown in Fig.~\ref{Fig:groundstation1}(a).
The Lijiang ground station was upgraded in 2016 specially for the satellite-based quantum entanglement distribution experiments.
In the Lijiang station, the original telescope with a large diameter of 1.8 m was modified for the quantum satellite-based experiments.
The designs of the measurement boxes in these three stations are similar (see Fig.~\ref{Fig:optbox-entangle}).

Typically, the FSM and the camera are combined to construct a close-loop fine tracking system.
The 850 nm and 532 nm photons are coupled into multimode fibers with $320~\mu m$ core for the synchronization.
Together with a Pockel Cell, an integrated 810 nm module with PBS inside achieved a random polarization analysis of signal photons.
For the Bell test, quantum random number generators (RNG) were employed to afford the random measurement bases choice.

The Ali ground station was completed in 2016, and it has the highest altitude (5100 m) among the five stations.
Three optical telescopes with a diameter of 130 mm were employ as the transmitting antennas for the ground-to-satellite teleportation, as shown in Fig.~\ref{Fig:groundstation2}(a).
To improve the transmitting efficiency, a double off-axis parabolic structure is employed in these telescopes (see Fig.~\ref{Fig:groundstation2}(b)).
All the optical components in the transmitting antenna have polarization-maintaining capabilities.
The wave-plates combinations are employed to correct the unknown transformations applied by the SMFs and automatically compensate for the deviation of the polarized base vector, caused by the movement of the satellite.
The polarization fidelity of the whole system exceeded $99.5\%$.
The FSM and a high-speed CCD constitute the ground fine tracking system, which realizes the pointing and tracking for the satellite with high accuracy.
The tracking accuracy of the whole system is less than 3 $\mu$rad (1 $\sigma$).
Two 671 nm beacon lasers (power: 2 W, divergence angle: 1.2 mrad) are installed on the top of the two transmitting telescopes, respectively, for satellite tracking on the ground.

For time synchronization, the 532 nm beacon laser in the satellite-based transmitter is designed as a pulse laser to perform synchronization; it is a passive Q-switching type laser with a repetition frequency of $\sim$ 10 kHz and an optical pulse width of 0.88 ns.
A part of the laser is guided into a fast photodiode to convert it into an electrical pulse signal.
Both the pulse signal and the GPS PPS signal from the satellite are fed into the time-to-digital convertor (TDC) module of the transmitter.
The acquired data are stored in the memory for further processing.
In the ground station, a part of the 532 nm laser beam is sent to a single-photon detector.
The output signal of the single-photon detector, together with the four single-photon detectors' electrical output pulses and the GPS-PPS signal, is fed into a TDC.
The time synchronization process between the satellite and the ground can be divided into two steps.
First, according to the predicted light-flight-time and the GPS-PPS signal, the received synchronization laser pulse sequence on the ground can be matched with the satellite.
Second, based on the results of step 1, the time between the satellite and the ground will be synchronized.
Finally, a typical temporal distribution of QKD photons with a standard deviation is obtained around 529 ps, with a signal time window of 2 ns \cite{liao2017satellite}.

\section{Satellite-based quantum experiments with Micius}

After the full verification of the feasibility of satellite-based quantum communication, we review in this section the development of a sophisticated satellite named after an ancient  Chinese scientist, Micius.
It was successfully launched in August 16, 2016, in Jiuquan, China, orbiting at an altitude of $\sim$ 500 km.
Coordinating ground observatory stations have been built worldwide to conduct the designed experiments for QKD, entanglement distribution, quantum teleportation, and foundational tests in quantum physics.


\subsection{Satellite-to-ground quantum key distribution}

\begin{figure*}[t]\center
\resizebox{16cm}{!}{\includegraphics{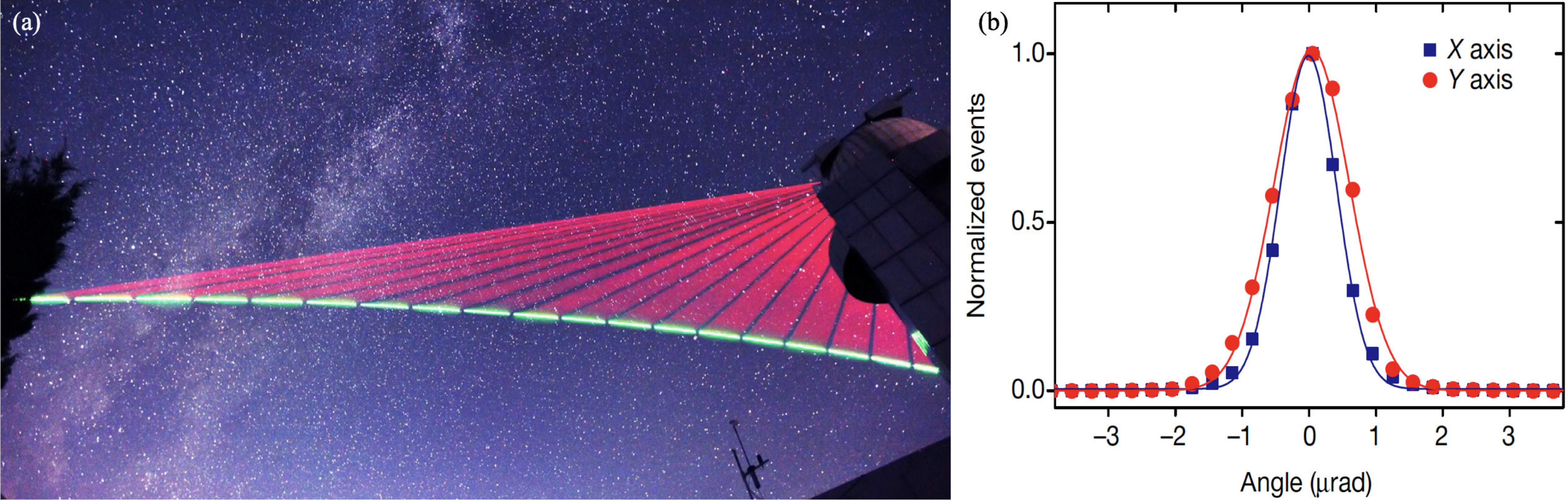}}
\caption{
Establishing a reliable space-to-ground link for the quantum
state transfer. \cite{liao2017satellite}
(a) Overlaid and time-lapse photographs of the beacon lasers for tracking when the satellite flies over the Xinglong ground station. The red and green lasers are sent from the ground and the satellite, respectively, with a divergence angle of 0.9-1.25 mrad.
(b) Distribution of long-time tracking error (shown as the number of detected events normalized by the maximum count in each bin) of the X and Y axes extracted from the real-time images read out from the fast camera.
}
\label{Fig:handshake}
\end{figure*}

After launching satellite Micius, the first goal was to establish a space-ground quantum link and perform QKD from satellite to ground \cite{liao2017satellite}.
The Chinese team observed the first ``handshake'' between the satellite and Xinglong ground station near Beijing ten days after the launch (see Fig.$\,$\ref{Fig:handshake}(a) for an overlaid and time-lapse photographs of tracking laser as the satellite flies over a ground staton).
As discussed in Sec. V.~A, the downlink has reduced beam spreading compared to the uplink because the beam wandering occurs at the end of the transmission path, which is typically smaller than the effect from the beam diffraction.
The 300-mm-aperture telescope equipped in the satellite produced a near-diffraction-limited far-field divergence of about 10 $\mu$rad.
Such a narrow divergence beam from the fast-travelling satellite (with a speed of about 7.6 km/s) requires a fast and precise APT.
Tracking accuracy of approximately 1.2 $\mu$rad (see Fig. \ref{Fig:handshake}b) was achieved, which is much smaller than the beam divergence.
It should be noted that due to the quiet environment in out space, the tracking accuracy is better than the previous ground tests, which deliberately set a more stringent condition.
A diffraction loss of approximately 22 dB was obtained at 1200 km, whereas the loss due to pointing error was below 3 dB.
Additionally, the loss due to atmospheric absorption was 3-8 dB.

\begin{figure*}[t]\center
\resizebox{14cm}{!}{\includegraphics{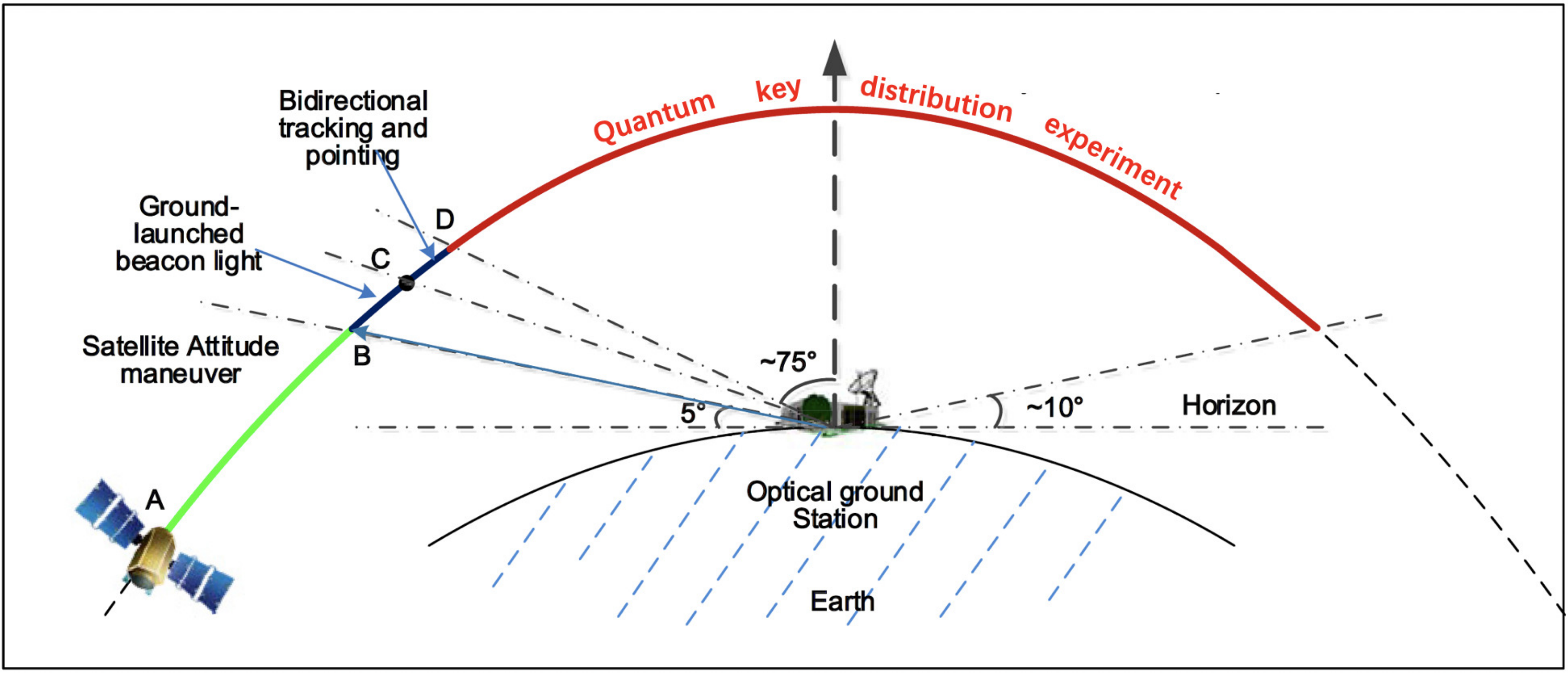}}
\caption{
Tracking and QKD processes during an orbit. \cite{liao2017satellite}}
\label{Fig:oneorbit}
\end{figure*}

The satellite passes each ground station with a sun-synchronous orbit every midnight local time for a duration of about 5 minutes.
Figure $\,$\ref{Fig:oneorbit} shows an experimental procedure for the satellite-to-ground QKD.
First, the Scientific Experiment Plan Center arranged the experiment, if the calculated maximum elevation angle of the satellite to the ground station is greater than 30$^\circ$ (based on predicted satellite orbits) and the weather is forecasted to be clear in the night.
If so, instruction sequence files for the satellite are made and sent to the ground support center	
Then, the instruction file, which is translated to a coding file is executed.
The attitude of the satellite is adjusted to point at the ground station 10 min before the satellite enters the shadow zone.
When the satellite's angle of elevation exceeds 5$^\circ$, its attitude control system ensures that the transmitter is pointing to the ground station with a coarse orientation accuracy of better than 0.5$^\circ$.
Then, the closed-loop APT systems start bidirectional tracking and pointing, ensuring that the transmitter and receiver are robustly locked with a tracking accuracy of $\sim$1.2 $\mu$rad through the whole orbit (see Fig.$\,$\ref{Fig:handshake}(b)).
From a 15$^\circ$ elevation angle, the decoy-state QKD transmitter sends signal and decoy photons together, which are received and decoded by the ground station, until the satellite reaches an angle of elevation of 10$^\circ$ in the other end when a single-orbit experiment ends.

\begin{figure*}[!t]\center
\resizebox{17.6cm}{!}{\includegraphics{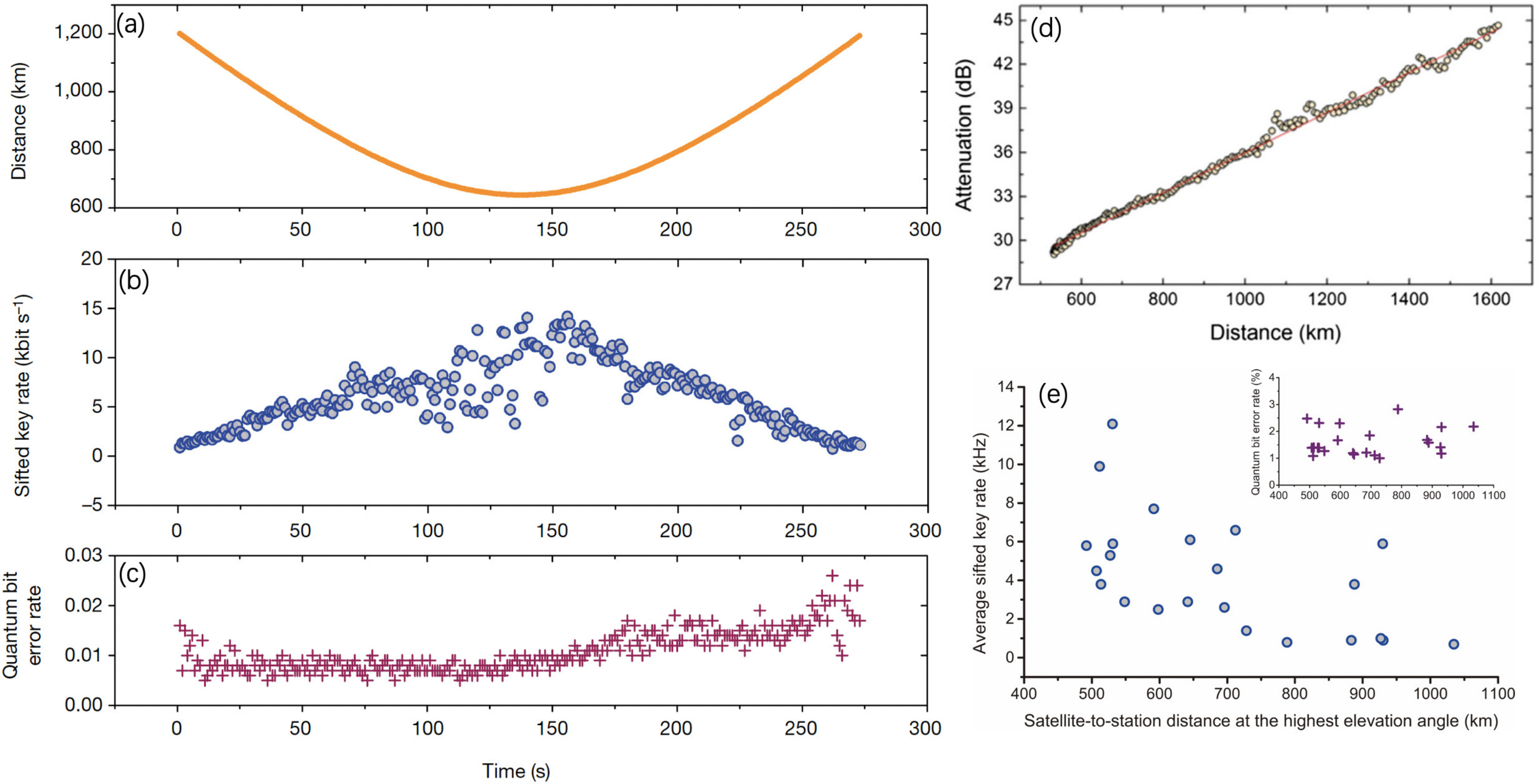}}
\caption{
Performance of satellite-to-ground QKD during one orbit. \cite{liao2017satellite}
(a) The trajectory of the Micius satellite measured from Xinglong ground station.
(b) The sifted key rate as a function of time and physical distance from the satellite to the station.
(c) Observed quantum bit error rate.
(d) The attenuation of the down-link channel with different distances between the satellite and the ground.
(e) Summary of the QKD data obtained for 23 different days. The \textit{x} axis is the shortest satellite-to-station distance, which occurs at the highest elevation angle and varies for different days. The \textit{y} axis is the average sifted key rate.
}
\label{Fig:QKDperformance}
\end{figure*}

\begin{figure}[!t]\center
\resizebox{8.2cm}{!}{\includegraphics{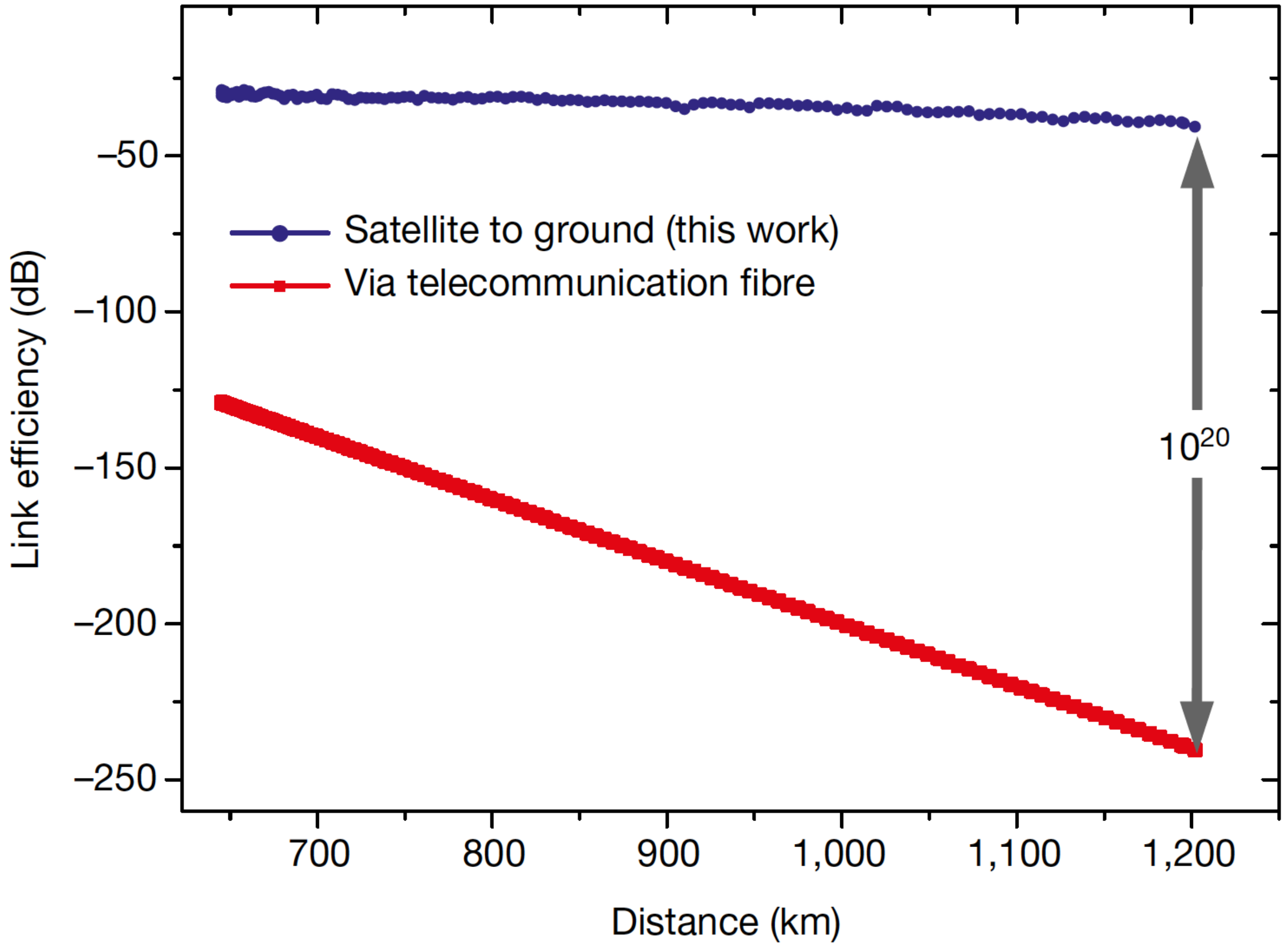}}
\caption{Link efficiencies are shown for direct transmission through telecommunication optical fibers (red) and the satellite-to-ground approach (blue). The link efficiencies for the latter were calculated by dividing the photon intensity that arrived in front of the detectors at the ground station by that at the output of the satellite's transmitter. \cite{liao2017satellite}
}
\label{Fig:QKDlinkeff}
\end{figure}

Apart from being used in the APT, the beacon laser also serves for obtaining the arrival time of the single photons in order to compensate for the space-ground clock drift.
The obtained time synchronization jitter is 0.5 ns, which is useful for filtering the background noise. 
Additionally, a spectral bandpass filter is used in the receiver to reduce the background scattering.
Finally, a motorized half-wave plate is used to dynamically compensate for the time-dependent photon polarization rotation during the satellite passage.

In the experiment, we employed the decoy-state BB84 protocol, a form of QKD that uses weak coherent pulses at high channel loss and is immune to photon-number-splitting eavesdropping.
Since September 2016,  QKD has been performed routinely under good atmospheric conditions.
An example of the relevant QKD data obtained on 19th December 2016 is shown in Fig.~\ref{Fig:QKDperformance}.
The satellite-observatory separation ranged from 645-1200 km.
The experiment collected 3,551,136 detection events in the ground station after 273 s and 1,671,072 bits of sifted keys. From Fig.~\ref{Fig:QKDperformance}(b), the sifted key rate was about 12 kbits/s at 645 km and 1 kbits/s at 1200 km.
This was mainly due to the increase in both the physically separated distance and the effective thickness of the atmosphere near the earth at smaller elevation angles. 
The observed bit error rate had an average of 1.1$\%$.
By performing error correction and privacy amplification, the secure final key was 300,939 bits when the statistical failure probability was set to be 10$^{-9}$, corresponding to a key rate of  1.1 kbits/s.

Meanwhile, similar QKD experiments were routinely performed with other ground stations, such as Delingha.
The typical LEO satellite-to-ground channel attenuation was calibrated during one orbit, which varies from 29 dB at 530 km to 44 dB at 1600 km, as plotted in Fig.~\ref{Fig:QKDperformance}(d).
Figure~\ref{Fig:QKDperformance}(e) shows a summary of the QKD experiments performed for 23 days, where the physical distance between the satellite and the ground station varied each day.

The performance of the satellite-based QKD can then be compared with the conventional method of direct transmission through telecommunication fibers.
Figure~\ref{Fig:QKDlinkeff} shows the extracted link efficiency at the distance from 645 km to 1200 km from the observed count rate, with the theoretically calculated link efficiency using fibers with 0.2 dB/km loss.
At 1200 km, the satellite-based QKD within the 273 s coverage time demonstrated a channel efficiency of about 20 orders of magnitudes higher than using the optical fiber.
Compared with the data in Fig.~\ref{Fig:QKDlinkeff}, using a 1200 km fiber, even with a perfect 10-GHz single-photon source, and the ideal single-photon detectors with no dark count, we can obtain only 1-bit sifted key over six million years.

\begin{figure*}[!t]\center
\resizebox{14 cm}{!}{\includegraphics{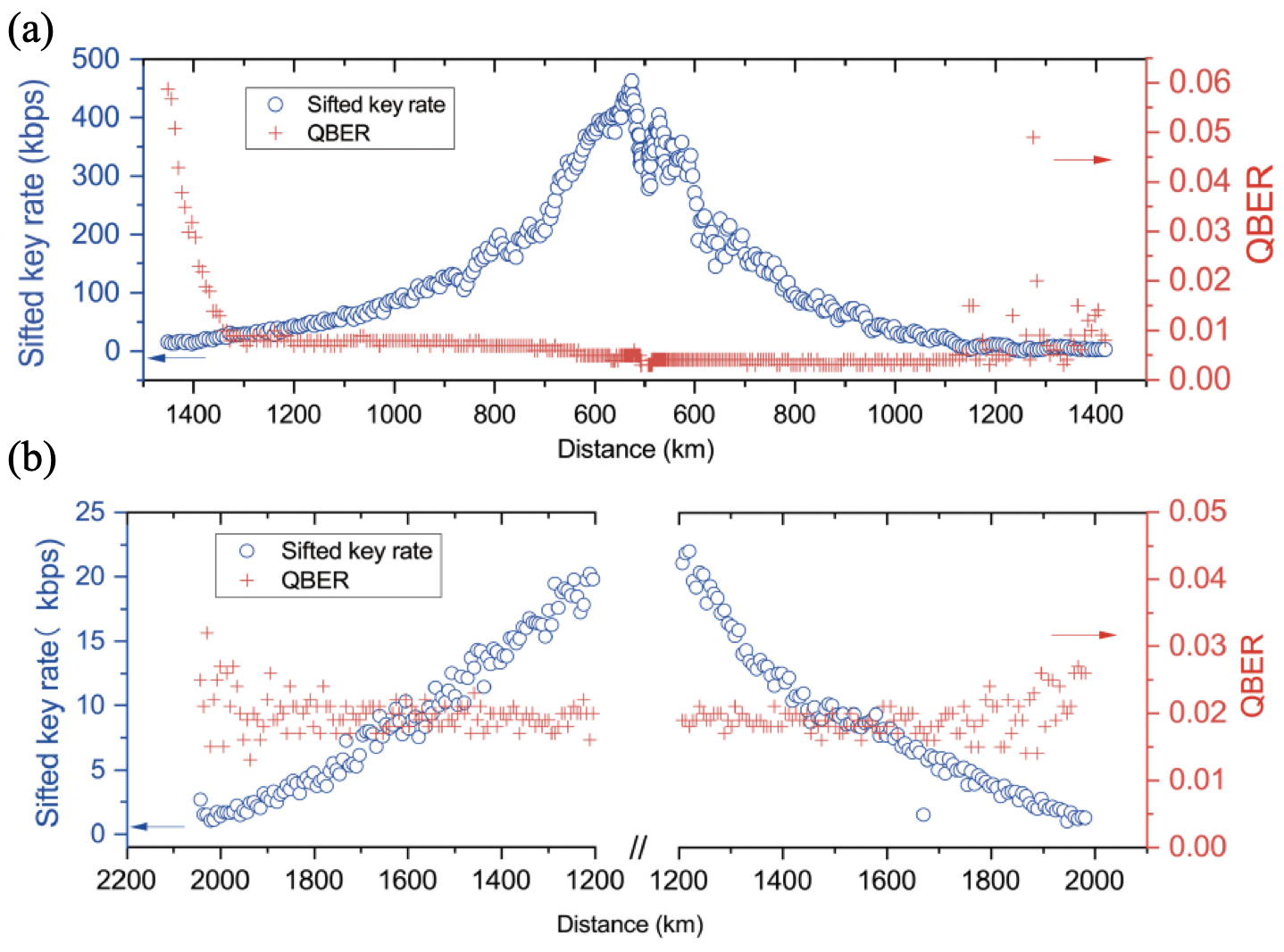}}
\caption{
Performance of high-speed satellite-to-ground QKD. \cite{IntegratedQKDNature2021} (a) The sifted key rate and observed QBER as a function of physical distance from the satellite to the Nanshan station.(b) The test for long-distance satellite-to-ground QKD. Sifted key rate and observed QBER at a distance beyond 1200 km.
}
\label{highspeedsatqkd}
\end{figure*}

Improving the final key rate is always one of main goals for the practical QKD.
Recently, about three years after the first satellite-to-ground QKD was implemented, it achieved an average secret key rate of 47.8 kbps for a typical satellite pass, which is more than 40 times higher than previous results \cite{liao2017satellite}, as shown in Fig~\ref{highspeedsatqkd}.
Such great improvement of the final key rate is due to the following: 
1) the signal state ratio, increased from 0.5 to 0.72, the Z-base ratio increased from 0.5-0.889 at the satellite and 0.5-0.9 on the ground station, thereby enhancing the key rate by 2.34 times; 
2) the repetition frequency increased from 100-200 MHz; 
3) the ground telescope increased from 1-1.2 m, corresponding to an increment about 1.5 times; 
4) the QBER is reduced and the raw key size increased to about 2 times; 
5) the ground coupling efficiency increased from 14\%-40\%, corresponding to an increment of about 3 times \cite{IntegratedQKDNature2021}.

\subsection{Satellite-based entanglement distribution}

The second planned mission of the Micius satellite was a bidirectional distribution of its space-borne entangled photons to two distant locations on the earth \cite{yin2017satellite}.
Long-distance entanglement distribution is essential for both foundational tests of quantum physics and scalable quantum networks.
Owing to channel loss, however, the previously achieved distance was limited to $\sim$ 300 km \cite{Inagaki:13}.
This is mainly due to the photon loss in the channel (optical fibers or terrestrial free space), which normally scales exponentially with the channel length.
For example, using a bidirectional distribution of an entangled source of photon pairs with 10-MHz count rate directly through two 600-km telecommunication fibers with a loss of 0.2 dB/km, we can only obtain 10$^{-17}$/s two-photon coincidence events.

\begin{figure}[!t]\center
\resizebox{8.5cm}{!}{\includegraphics{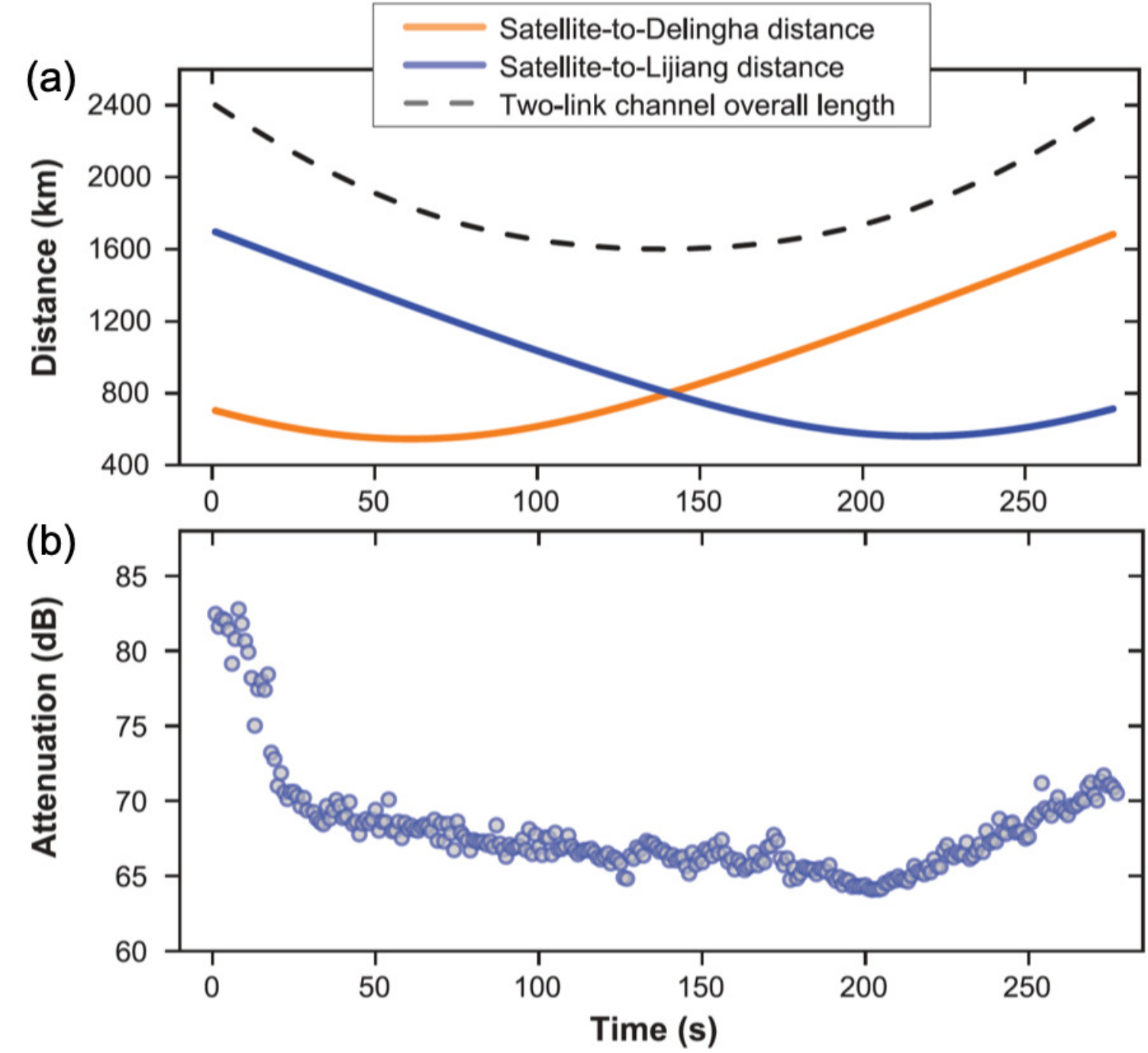}}
\caption{
(a) A typical two-downlink transmission from the satellite to Delingha and Lijiang that lasted for about 275 s in one orbit. The distance from the satellite to Delingha varies from 545 to 1680 km. The distance from the satellite to Lijiang varies from 560 to 1700 km. The overall length of the two-downlink channel varies from 1600 to 2400 km.
(b) The measured two-downlink channel attenuation in one orbit, using the high-intensity reference laser co-aligned with the entangled photons. The highest loss is ~82 dB at the total distance of 2400 km, when the satellite has just reached a 10$^{\circ}$ elevation angle observed from Lijiang station. Since its telescope has a diameter of 1.8 m (the largest) and thus has a higher receiving efficiency than other stations, when the satellite flies over Lijiang at an elevation angle of more than 15$^{\circ}$, the channel loss remains relatively stable, from 64 to 68.5 dB. \cite{yin2017satellite}
}
\label{Fig:entanglementresult1}
\end{figure}

For the mission of entanglement distribution, three ground stations are cooperating with the satellite, located at Delingha in Qinghai, Nanshan in Urumqi, Xinjiang, and Gaomeigu Observatory in Lijiang, Yunnan.
The physical distance between Delingha and Lijiang (Nanshan) is 1203 km (1120 km).
The separation between the orbiting satellite and these ground stations varies from 500 km to 2000 km.
At Delingha, Lijiang and Nanshan station, the receiving telescopies has diameters of 1200 mm, 1800 mm, and 1200 mm, respectively.
The entanglement distribution was achieved both between Delingha and Lijiang, and between Delingha and Nanshan.
The experiment involving Delingha and Lijiang is described below.
Figure~\ref{Fig:entanglementresult1}(a) plots the physical distances from the satellite to Delingha and Lijiang during one orbit, and the sum channel length of the two downlinks.

The satellite (Fig.$\,$\ref{Fig:entanglesource}) emits 5.9 million entangled photon pairs per second, which are then sent out using two telescopes.
Cascaded multi-stage close-loop APT systems are designed in both the transmitters and receivers, establishing two independent satellite-to-ground quantum links simultaneously.
Using a reference laser on the satellite, the overall two-downlink channel attenuation can be measured in real time, which varies from 64 dB to 82 dB (Fig.~\ref{Fig:entanglementresult1}(b)).
A slight asymmetry is observed in the attenuation curve--when the satellite moves closer to Lijiang.
Furthermore, the link efficiency is higher because Lijiang station has a larger-size aperture telescope.

The experiment observed an average two-photon count rate of 1.1 Hz, with a signal-to-noise ratio of 8:1.
Compared to the previous entanglement distribution method by direct transmission of the same two-photon source using the common commercial telecommunication fibers with a loss of 0.2 dB/km (the best performance fiber with a loss of 0.16 dB/km), the effective link efficiency of the satellite-based approach within the 275 s coverage time is 17 (12) orders of magnitude higher than the one with the same transmission distance in fibers.

The received photons were analyzed by a half-wave plate, a polarizing beam splitter, and a Pockel cell controlled by 4 Mbits/s random numbers.
To verify whether both photons are still entangled after passing an overall distance ranging from 1600 km to 2400 km, we obtained 134 coincidence counts during an effective time of 250 s in satellite-orbit shadow time.
From the \textit{H/V} and diagonal basis measurements, the state fidelity of the two photons distributed over 1203 km was estimated to be 0.869~$\pm$~0.085.

\begin{figure}[!t]\center
\resizebox{8cm}{!}{\includegraphics{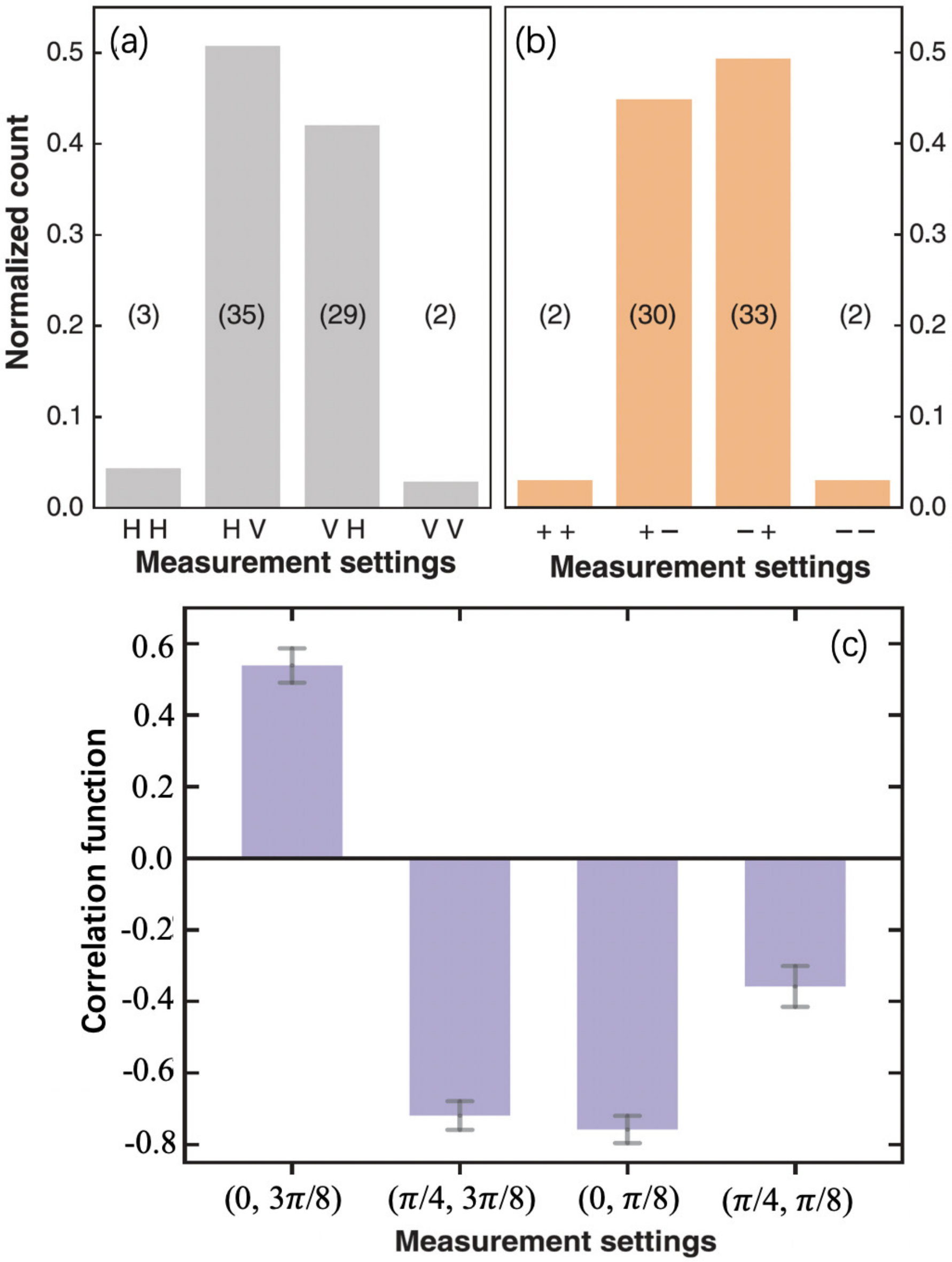}}
\caption{
(a) Normalized two-photon coincidence counts in the measurement setting of the H/V basis.
(b) Normalized counts in the diagonal basis.
(c) Correlation functions of a CHSH-type Bell inequality for entanglement distribution. Error bars are one standard deviation, calculated from propagated Poissonian counting statistics of the raw photon. \cite{yin2017satellite}
}
\label{Fig:entanglementresult2}
\end{figure}

The distributed entangled two photons were then employed for a Bell test.
The experimental configuration and Pockels cells used were fast enough to close the locality and freedom-of-choice loopholes.
The Bell test ran 1167 trials during an effective time of 1059 s, with the observed data summarized in Fig.$\,$\ref{Fig:entanglementresult2}.
A violation of the Bell inequality of 2.374~$\pm$~0.093 by 4 standard deviations was obtained.
The result again confirms the nonlocal feature of entanglement and excludes the models of reality, which sits on the notions of locality and realism, in a new space scale with thousands of kilometers.

\subsection{Entanglement-based quantum key distribution}

In the experiment of entanglement distribution between Lijiang and Delingha over 1200 km, the observed two-photon count rate of 1.1 Hz and signal to noise ratio of 8:1 was sufficient for violating Bell's inequality.
However, there the key rate and the quantum bit error rate (8.1$\%$) were insufficient for performing the entanglement-based quantum cryptography \cite{E91}.
Entanglement-based QKD is particularly attractive because of its inherent source-independent security \cite{KoashiPreskill_03, MXF:EntanglementPDC:2007} where the security can be established without any assumption on trusted relay. 

Inside laboratories, the record length of QKD was about 500 km in coiled fiber~\cite{MDI-404km, Boaron421km, TF-QKD500km}.
The satellite-to-ground decoy-state QKD reviewed in Section IX.A achieved a point-to-point distance of 1200 km \cite{liao2017satellite}, which, however, was not for two ground users.
Without using trusted replays, the QKD distance for two ground users was about 100 km over terrestrial free space \cite{ursin2007entanglement, yin2012-100km}.

After the first satellite-based entanglement distribution \cite{yin2017satellite, Yin2017entQKD}, a later experiment \cite{entanglementQKDNature} performed between the ground station of Delingha and Nanshan with a physical separation of 1120 km.
The receiving efficiencies were considerably improved using a higher efficiency telescope and follow-up optics.
Both ground stations used newly built telescopes with diameters of 1.2 m.
In the telescopes, the main lens was re-coated and the beam expander was re-designed.
In the follow-up optics, the collection efficiency was enhanced by optical pattern matching, particularly, through shortening the optical path by reducing spectral splitting to avoid the beam spread.
With these technical improvements, the authors observed an average two-photon count rate of 2 Hz (corresponding to an increase of the two-photon link efficiency by a factor of 4), which significantly increased the obtained key rate and decreased the quantum bit error rate from 8.1$\%$ to 4.5$\%$.

\begin{figure*}[!t]\center
\resizebox{16cm}{!}{\includegraphics{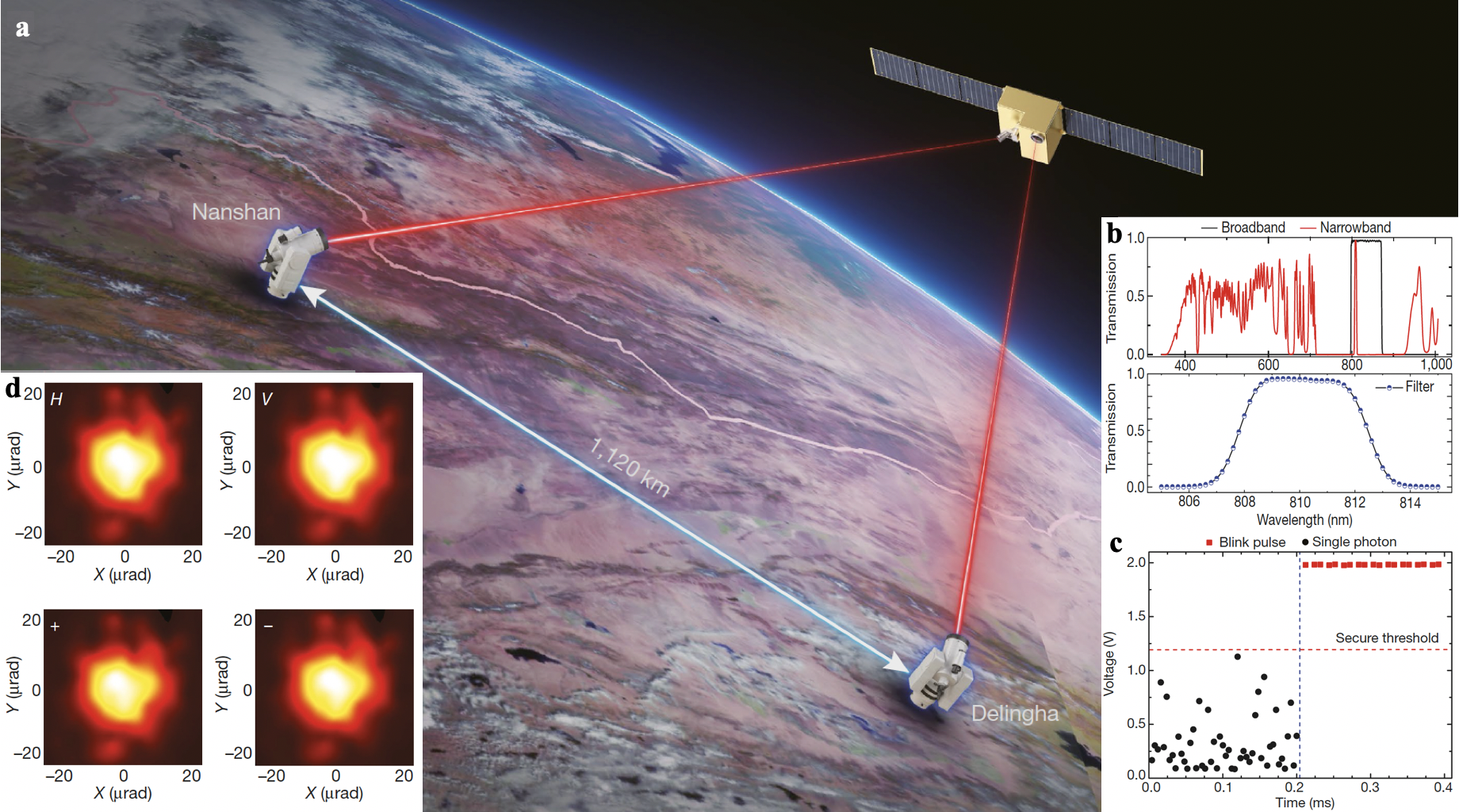}}
\caption{
Overview of the experimental set-up of quantum key distribution based on the entanglement distribution. \cite{entanglementQKDNature} (a) An illustration of the Micius satellite and two ground stations. Image credit: Fengyun-3C/Visible and Infrared Radiometer, with permission (2020). (b), (c), and (d) are monitoring and filtering against side channels. (b) The transmission of broad-bandwidth and narrow-bandwidth filters. (c) The output of monitoring circuit with/without blinding attack. Without blinding attack, the outputs are random avalanching single-photon-detection signals (black dots). With blinding attack (starting from 0.2 ms), the output signals are at around 2 V, which is clearly above the security threshold, thus triggering the security alarm. (d) The system detection efficiency of the four polarizations in the spatial domain. With the spatial filter, the four efficiencies are identical.
}
\label{Fig:engQKD}
\end{figure*}

A special effort of \cite{entanglementQKDNature} was made to ensure its implementation is practically secure against all known side channels.
Due to the source-independent nature of the entanglement-based QKD, the system was immune to any loophole in the source, and all left is to ensure the security on the detection sides in both ground stations.
In general, the side channels, known and to be known, on the detection primarily violate the assumption of fair-sampling.
Note that the concept of ``fair-sampling" here refers to the consistency of different receiving detectors on spatial and spectral freedom of the arrival beams, which differs from the experimental requirement of addressing the ``fair-sampling loophole" of the Bell test.
Experimentally, Yin \textit{et al}. ensured the validity of the fair-sampling by filtering in different degrees of freedom, including frequency, spatial and temporal modes. 
Also, countermeasures were taken for the correct operation of the single-photon detectors, as shown in Fig.~\ref{Fig:engQKD}.
They considered all known detection attacks, including the detector-related attacks \cite{Zhao:TimeshiftExp:2008, Lydersen:Hacking:2010, Weier:DeadtimeAttack:2011}, wavelength-dependent attack \cite{li2011attack}, spatial-mode attack \cite{Sajeed2015attack}, and other possible side-channels.
For example, for the side channels targeting at the operation of detectors, such as blinding attacks \cite{Lydersen:Hacking:2010}, additional monitoring circuits were used to monitor the anode of the load resistance in the detection circuit to counter the blinding attack.
For the time-shift attack \cite{Zhao:TimeshiftExp:2008} and the dead-time attacks \cite{Weier:DeadtimeAttack:2011}, the countermeasure was to operate the detector in free-running mode, in which the detector records all the detection events and post-selects the detection windows such that the detection efficiency is guaranteed to be at a nominal level.
Consequently, the secret key, generated by this QKD system, is secured under realistic devices.

By running 1021 trials of the Bell test during an effective collection time of 226 s, Yin \textit{et al}. observed that the parameter \textit{S} was $2.56\pm0.07$ with a violation of local realism by 8 standard deviations.
Having violated the Bell's inequality, they demonstrated the entanglement-based QKD using the protocol presented by Bennett, Brassard, and Mermin in 1992 (BBM92), where both Alice and Bob took measurements randomly along the \textit{H/V} and +/- basis \cite{Bennett:BBM92:1992}.
Due to the efforts to ensure the fair-sampling assumption, the practical security of the BBM92 protocol is compatible with the E91.

Within 3100 sec data collection time, 6208 initial coincidences were obtained, which gave 3100 bits of sifted keys with 140 erroneous bits.
The quantum bit error rate was 4.5$\%$~$\pm$~0.4$\%$.
After error correction and privacy amplification, the secure key rate of 0.43 bits/s in the asymptotic limit of infinite long key and a finite secret-key rate of 0.12 bits/s were obtained.
More details on the final key rate are discussed in Ref.~\cite{PhysRevLett.126.100501}.
The secure key rate was 11 orders of magnitude higher than that would be obtained by direct transmission of entangled photons over 1120 km through the best commercial fibers.
The results increase the secure distance of practical QKD for ground users by 10 times to the order of a thousand kilometers, representing a key step toward the Holy Grail of cryptography.
Note that with the newly developed entangled photon source with a 1 GHz generation rate \cite{Caofreewill} increased the secure key rate by about 2 orders of magnitude directly.

\subsection{Ground-to-satellite quantum teleportation}

\begin{figure}[!t]\center
\resizebox{8.64cm}{!}{\includegraphics{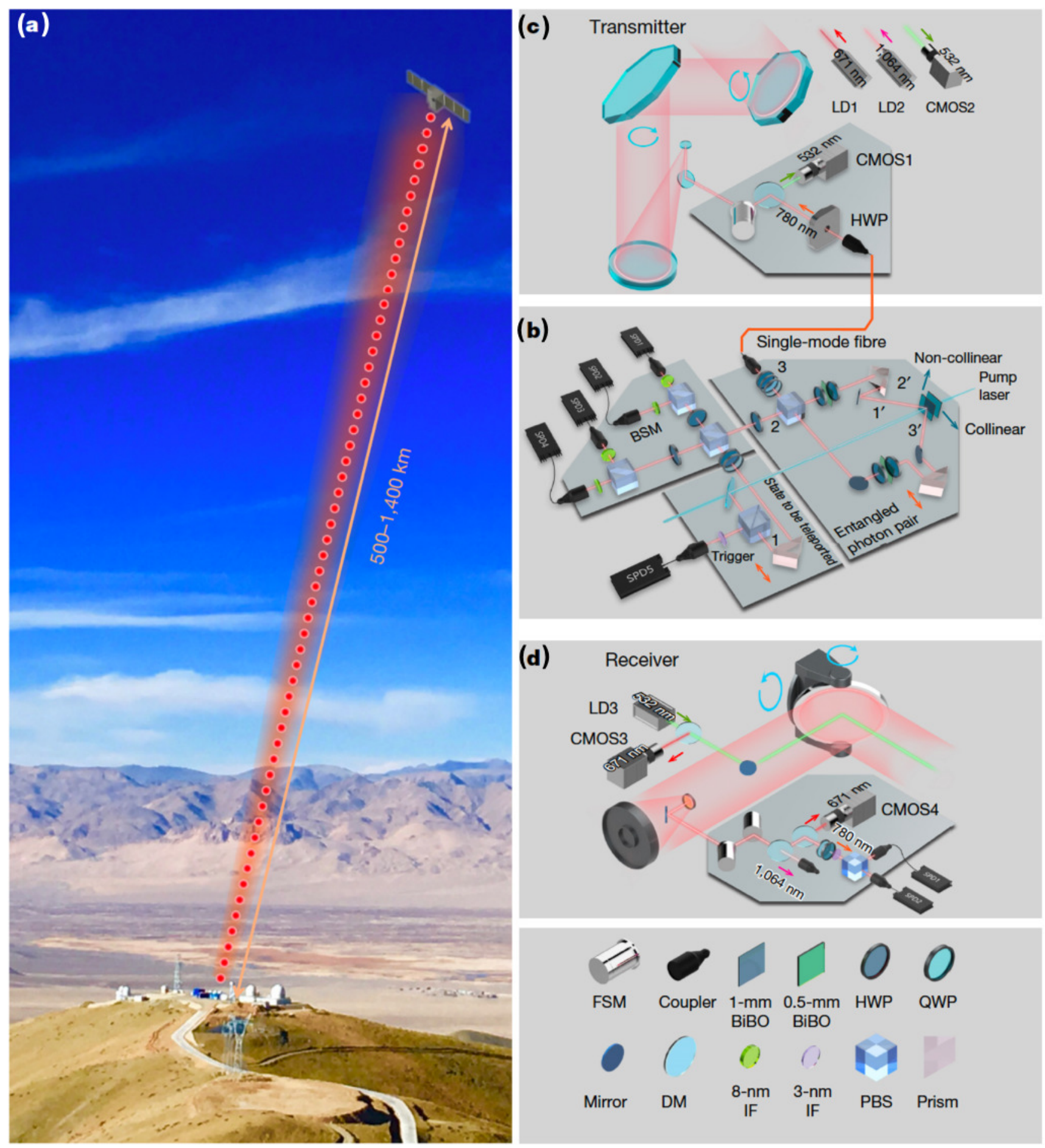}}
\caption{
Overview of the set-up for ground-to-satellite quantum teleportation of a single photon over distances of up to 1,400 km. \cite{ren2017satellite}
(a) A schematic of the satellite is overlaid on a photograph of the Ngari ground station in Tibet. The separation between the satellite and the ground station varies from about 500 km to 1,400 km during quantum teleportation.
(b) The compact multi-photon set-up for teleportation at the ground station.
(c) The transmitter at the ground station.
(d) The receiver on the satellite.
}
\label{Fig:tele1}
\end{figure}

The third mission of the Micius satellite was to perform quantum teleportation of a single photon from an observatory ground station in Ngari to the satellite, an uplink (for an overview, see Fig.$\,$\ref{Fig:tele1}(a)) \cite{ren2017satellite}.
The uplink teleportation experiment has two additional challenges compared with the previous downlink work.
First, the teleportation of an independent single photon requires a multi-photon interferometry with a coincidence count rate several orders of magnitude lower than typical single- or two-photon experiments.
Second, the atmospheric turbulence in the uplink channel occurs at the beginning of the transmission path, which causes beam wandering and broadening that increases the amount of spreading of the traveling beams.

\begin{figure}[!t]\center
\resizebox{8.64cm}{!}{\includegraphics{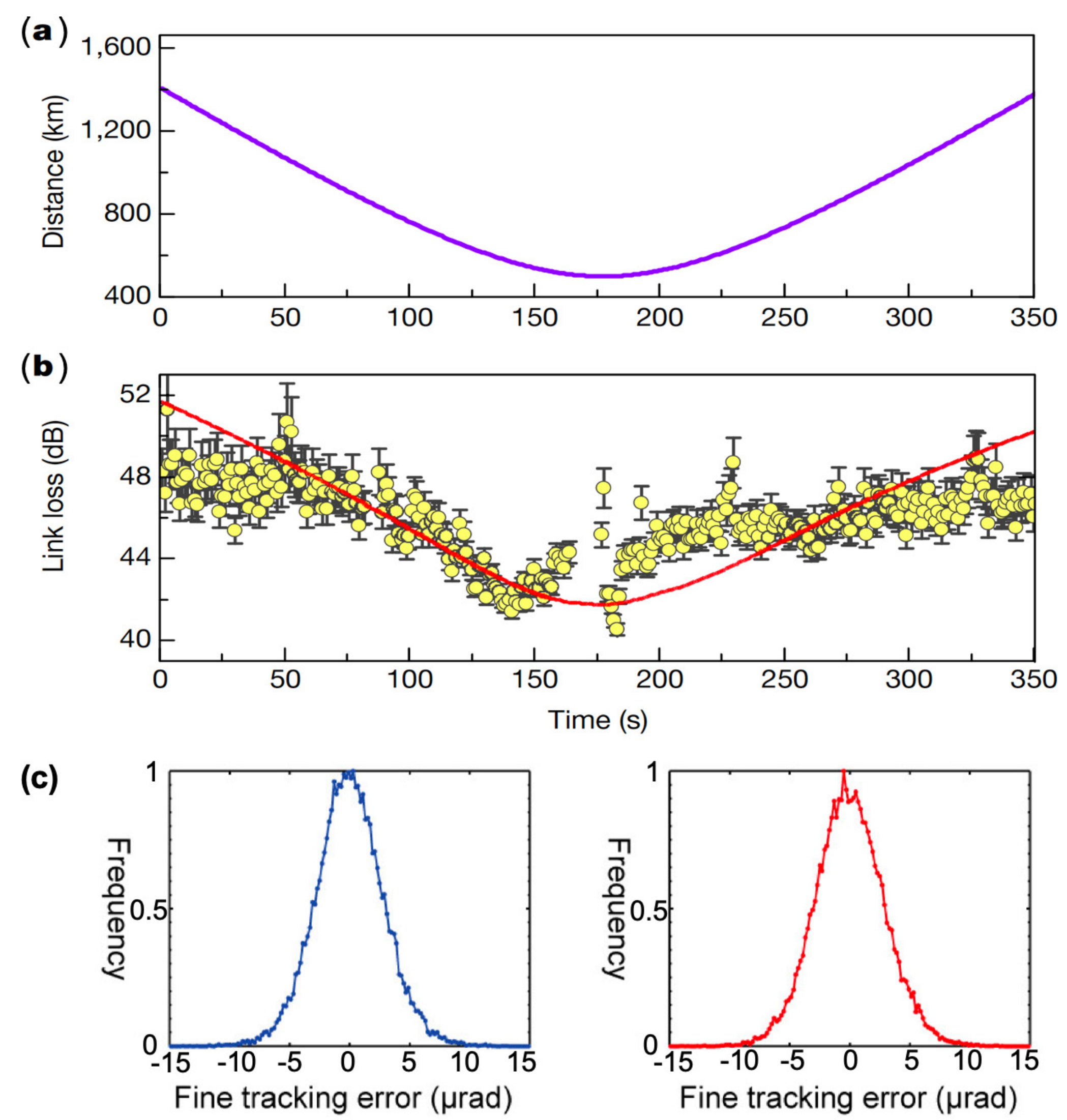}}
\caption{
Distance from the ground station to the orbiting satellite and the measured attenuation during one orbit. \cite{ren2017satellite}
(a) The trajectory of the Micius satellite measured from the Ngari ground station over one orbit with a duration of 350 s.
(b) The measured ground-to-satellite channel loss using a strong reference laser as a function of time. The highest loss is about 52 dB at a distance of 1,400 km (when the satellite is at an angle of $14.5^{\circ}$). The lowest loss is about 41 dB at a distance of around 500 km (when the satellite is at an angle of $76.0^{\circ}$). The red curve is a model that considers the effect of distance variation. Error bars show one standard deviation, calculated from Poissonian counting statistics. Owing to the structure of the altazimuth telescope at the ground station, the rotation speed of the optical transmitter has to be increased as a function of the increasing altitude angle of the satellite. When the satellite reaches the top altitude angle, the speed that is required can be very large and beyond the ability of the APT system. The tracking accuracy is therefore reduced with increasing rotational speed, leading to larger measured channel attenuation when the satellite is closer to the ground station. As a result, the trend in the data appears more compressed than the model. This model does not fully capture all of the features of the measured data of channel loss.
(c) Performance of the APT system on the \textit{X} (horizontal; blue) and \textit{Y} (vertical; red).
}
\label{Fig:tele2}
\end{figure}

A very compact design of ultra-bright four-photon sources, which used both collinear and noncollinear SPDC (see Fig.$\,$\ref{Fig:tele1}(b)), was employed to meet the extreme condition of the field experiment in Ngari.
The four-photon interferometry system was integrated into a compact platform with a dimension of 460 mm$\times$510 mm$\times$100 mm and weighing less than 20 kg.
The pump laser was used for two identical multi-photon modules built sequentially, where the multiplexed four-photon count rate was 8200/s.
Note that with the newly developed SPDC source \cite{zhong201812} improved the four-photon count rate by a factor of $\sim$10.

The teleported single photons from a single-mode fiber were transmitted through a 130-mm-diameter off-axis reflecting telescope (Fig.$\,$\ref{Fig:tele1}(c)), and received by a 300-mm-diameter telescope in the satellite (Fig.$\,$\ref{Fig:tele1}(d)).
Both the transmitter and receiver were equipped with APT systems to optimize the uplink efficiency.
Figure~\ref{Fig:tele2} shows the time trace of channel attenuation measured during one orbit of the satellite passing through the Ngari station.
The physical distance between the ground station and the satellite varies from a maximum of 1400 km (at an altitude angle of 14.5$^\circ$, the starting point of our measurement) to a minimum of 500 km (at the highest altitude angle of 76.0$^\circ$, when the satellite passes through the ground station above the top).
Here, the channel loss of the uplink falls from 52 dB to 41 dB, measured using a high-intensity reference laser.

An important technical note is that when exposed to radiation in the space environment, the dark count of the single-photon detectors increases significantly.
To mitigate this problem, the detectors are carefully shielded and cooled down to $-50^\circ C$.
This reduces the dark counts to less than 150 Hz over three months.
The teleportation data, with an overall 911 four-photon counts, was collected in 32 orbits.
For the set of the six input states on a mutually unbiased basis, the teleportation state fidelities gave an average of 0.80~$\pm$~0.01.

\subsection{Satellite-relayed intercontinental quantum key distribution}

\begin{figure*}[!t]\center
\resizebox{16cm}{!}{\includegraphics{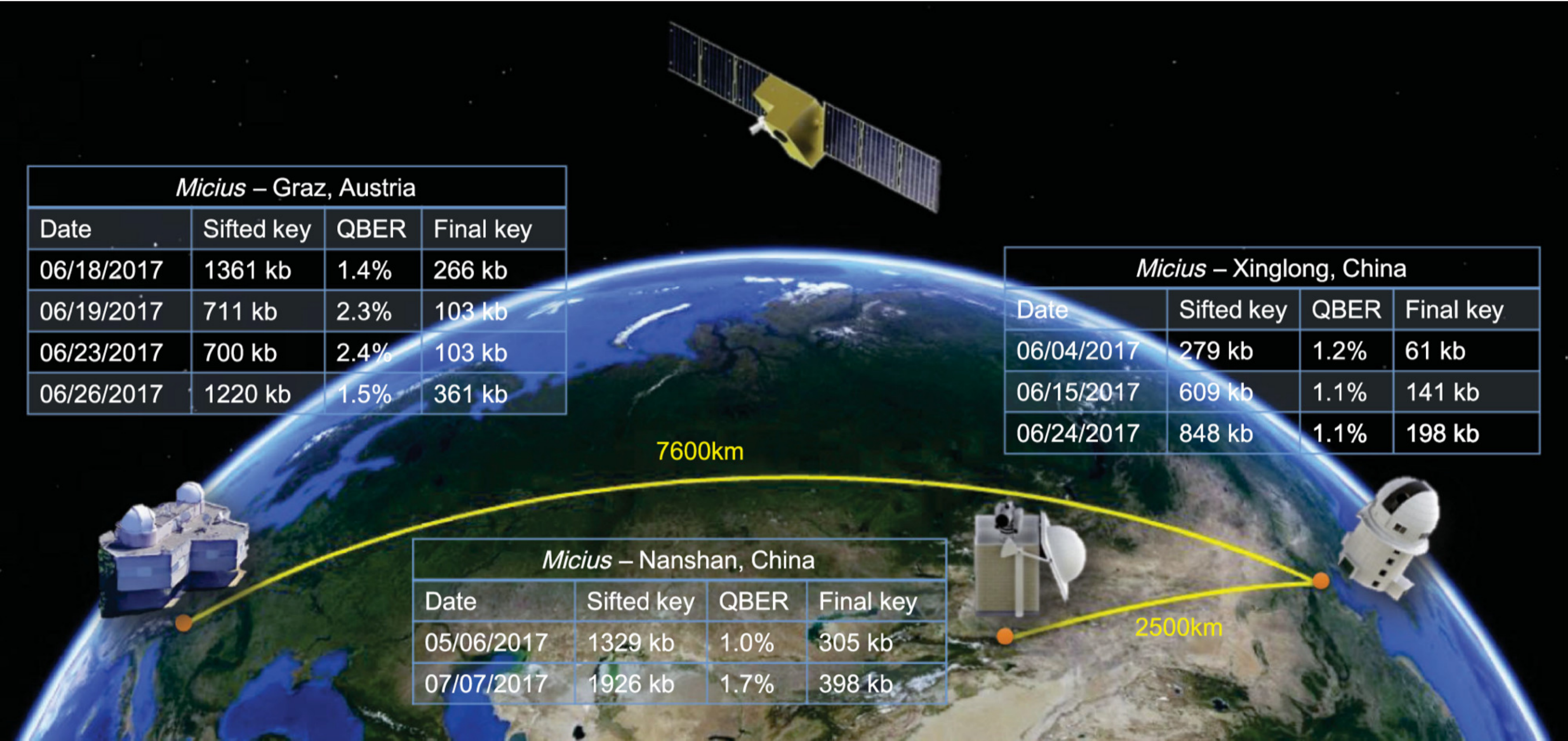}}
\caption{
Illustration of the three cooperating ground stations (Graz, Nanshan, and Xinglong). Listed are all paths used for key generation and the corresponding final key length. \cite{Liao2018relay}
}
\label{Fig:intercontinental1}
\end{figure*}

The Micius satellite can be further exploited as a trustful relay to conveniently connect any multiple points on Earth to form a network for high-security key exchange.
To further demonstrate the Micius satellite as a universal and robust platform for quantum experiments with different ground stations on Earth, the satellite down-link to Nanshan ground station near Urumqi and Graz ground station near Vienna were performed successfully.
Typical satellite-to-ground QKD performances between May and July 2017 are summarized in Fig.~\ref{Fig:intercontinental1}, with the final key length ranging from 400 to 833 kbits \cite{Liao2018relay}.

\begin{figure}[!t]\center
\resizebox{8.4cm}{!}{\includegraphics{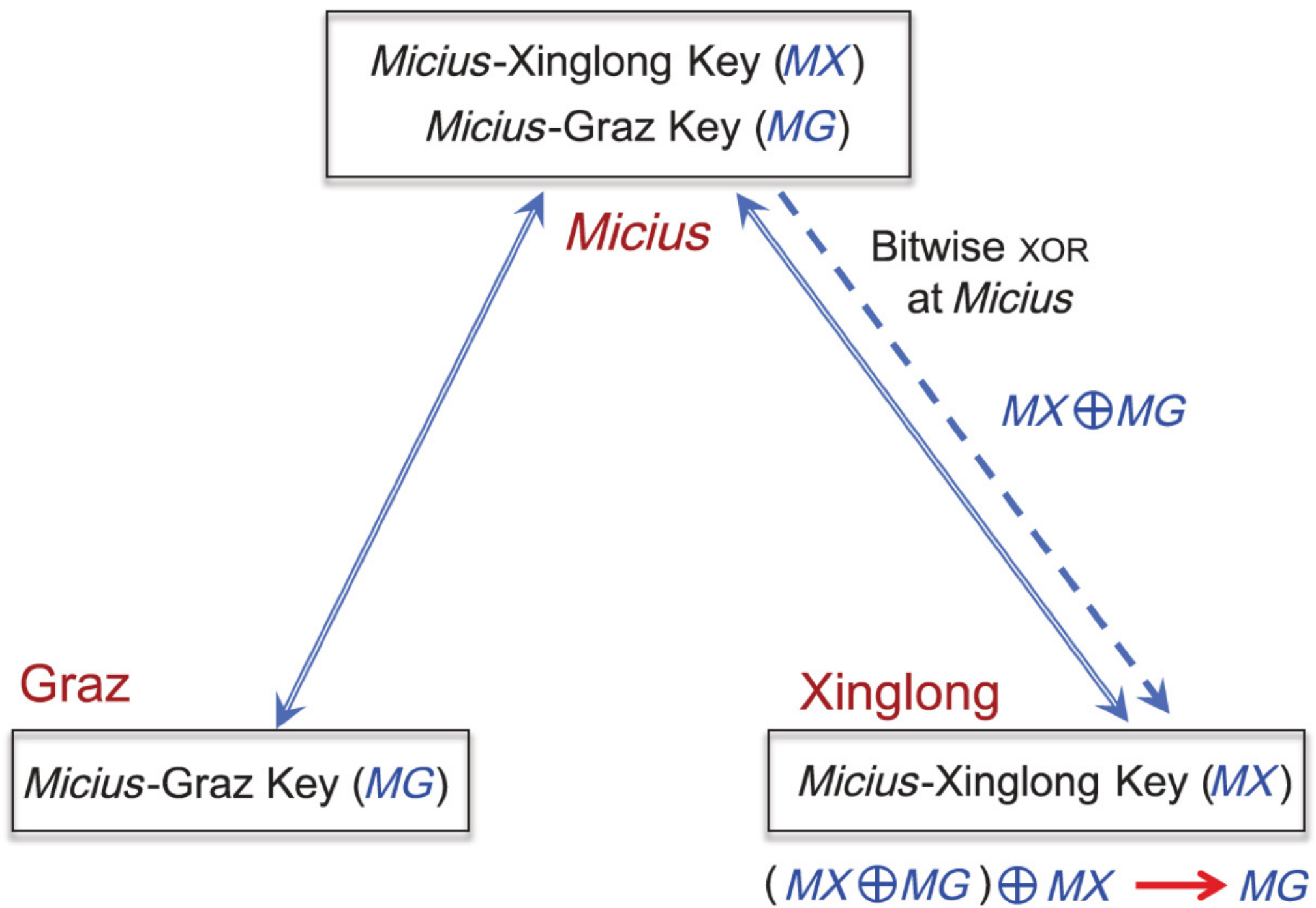}}
\caption{
Schematic of key exchange procedure between Graz and Xinglong with the satellite as a trusted relay \cite{Liao2018relay}. After Micius distributed a key with Graz (MG) and Xinglong (MX), it performs a bitwise exclusive OR between those keys (MX $\oplus$ MG) and sends this combined key via a classical channel towards Xinglong station. Combining the XORed key at Xinglong with MX leads to the same key (MG) on both sides.
}
\label{Fig:intercontinental2}
\end{figure}

Then, upon request from the ground command, the satellite acts as a trusted relay to establish secure keys among any two ground stations.
Figure~\ref{Fig:intercontinental2} shows an example of exchanging keys between the Xinglong and Graz stations.
Let us denote the random keys shared between Micius and Xinglong as MX, and between Micius and Graz as MG. Micius can simply perform a bitwise exclusive OR operation ($\oplus$) between MX and MG of the same string length, which then yields a new string: MX$\oplus$MG.
Then, the new string can be sent through a classical communications channel to Xinglong or Graz, which decodes other original keys using another exclusive OR (i.e. MG=(MX$\oplus$MG)$\oplus$MX).
This process can be easily understood since Micius uses MX to encrypt MG and Xinglong decrypts the cipher text to recover MG, shared with Graz.
Such a key is known only to both communicating parties and the satellite, but no fourth party.

For a demonstration, a 100 kB secure key was established between Xinglong and Graz.
Approximately 10 kB of the key was used to transmit a picture of Micius (with a size of 5.34 kB) from Beijing to Vienna, and a picture of Schr\"odinger (with a size of 4.9 kB) from Vienna to Beijing, using one-time-pad encoding.
The other 70 kB of the secure key was combined with the advanced encryption standard (AES)-128 protocol and used in a video conference between Beijing and Vienna for 75 minutes with total data transmission of about 2 GB.

\subsection{Probing gravity-induced decoherence}

The Micius satellite also provides the feasibility for testing the entanglement decoherence induced by the gravitation of the earth \cite{Joshi2018}.
Quantum mechanics and relativity form the bedrock of modern physics.
The general theory of relativity predicts a kind of exotic spacetime structure called the closed time curve (CTC) \cite{Friedman1990}.
CTC is interesting because it violates causality and in principle can be formed from the quantum fluctuations of spacetime itself \cite{Morris1988,Politzer1992,Deutsch1991,Hartle1994,Hawking1995}.
To theoretically describe the quantum fields in both exotic spacetime containing CTCs and ordinary spacetime, Ralph \textit{et al}. reported the event formalism of quantum fields \cite{Ralph2009}.
This theory predicts that the different evolutions of quantum fields may probabilistically induce time decorrelation of two entangled photons passing through different regions of curved spacetimes, which are able to keep the entanglement in standard quantum theory.

Considering the curved spacetime brought by the earth's gravitation, the decoherence effect can be tested via distributing entanglement between ground station and satellite.
The probability of entanglement losing is characterised by the decorrelation factor \emph{D}, which is given by:

\begin{equation}
D\approx{\rm exp}(-0.5 \Delta_t^2/d_t^2),
\label{d0}
\end{equation}
where $d_t$ denotes the photon coherence time and $\Delta_t$ is derived from the effect of the curved spacetime \cite{Joshi2018}:

\begin{equation}
\Delta_t \approx   \int^{r_e + h}_{r_e} {{M}\over{r}} (1 + {{2 M}\over{r}} + {{r_e^2 \tan^2 (90 - \theta)}\over{r^2}})^{1/2} dr,
\end{equation}
where $r_e$ is the earth's radius, $h$ is the satellite altitude, $m$ is the mass of the earth expressed in units of length and $\theta$ is the altitude angle.

\begin{figure}[htbp]
\centering
\includegraphics[width=8cm]{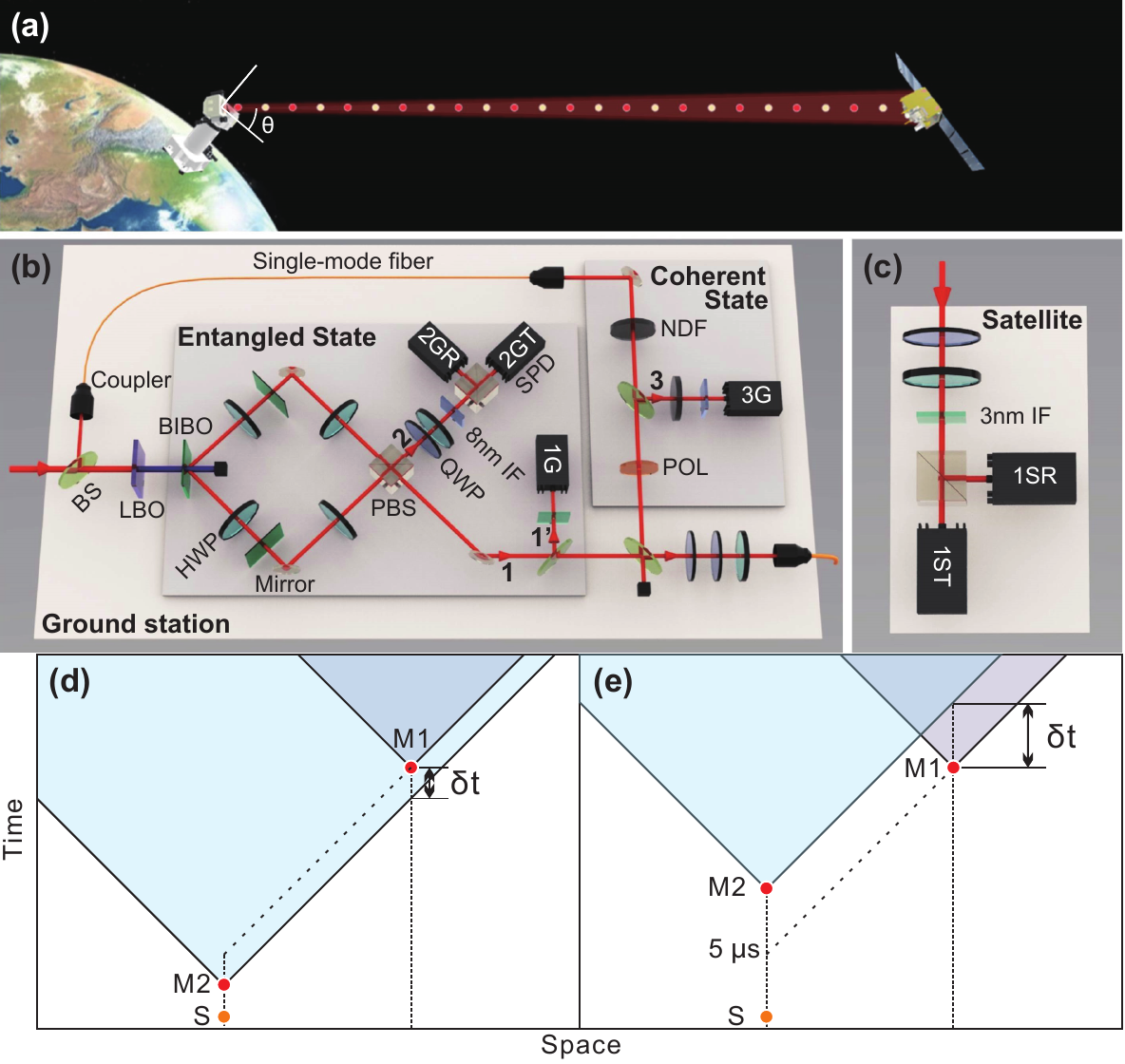}
\caption{
The setup of experimental test of gravity induced entanglement decoherence. \cite{Xu2019}
}
\label{fig:decoherence1}
\end{figure}

In the implementation, polarization entangled photon pair was prepared in the Ngari ground station, as shown in Fig.~\ref{fig:decoherence1}.
The photon in path 2 is detected on the ground after passing through the ordinary spacetime, while its twin is received by the satellite \emph{Micius} after propagating in the curved spacetime. 
Since gravity cannot induce the decoherence of classical correlation, it is possible to use the coherent laser a reference.
The entangled photons are combined with the faint coherent laser pulses in path 1 before transmitting,.
Then, the transmitted coherent photons are classically correlated with the photons in path 3 on the ground.
Two trains of entangled and coherent photons are shifted by half a pulse interval ($\sim$ 6 ns), allowing the satellite to distinguish the photons by their arrival times.

Altitude of the satellite \emph{Micius} is about 500 km.
According to Eq.~\ref{d0}, it can be shown that the decorrelation factor \emph{D} is only a function of the altitude angle $\theta$ for a given photon source.
To estimate the factors $D(\theta)$ in the experiment, the observed coincident counts was compared with the expected counts in the standard quantum theory.
$C_{\rm exp, EPR}(\theta)$ and  $C_{\rm exp, COH}(\theta)$ denote the measured two-photon coincidence events of entangled and coherent states, respectively.
From the standard quantum theory, the coincident counts of entangled photon pairs are $C_{\rm SQT, EPR}(\theta) = \eta_2 S_{\rm EPR}(\theta)$, where $S_{\rm EPR}(\theta)$ is the number of entangled photons detected on the satellite and $\eta_2$ is the efficiency of detecting photons in path 2.
For the faint coherent laser pulses, coincidence counts are estimated as $C_{\rm SQT, COH}(\theta)=S_{\rm COH}(\theta)S_{3}t_p/t_{\Delta\theta}$, where $S_{\rm COH}(\theta)$ and $S_3$ are the numbers of detected coherent photons at the satellite and in path 3, respectively, $t_p$ is the repeat frequency of pulse laser and $t_{\Delta\theta}$ is data collection time.
Thus, the decorrelation factors then can be given as

\begin{equation}
D_{\rm EPR}(\theta)=C_{\rm exp, EPR}(\theta)/C_{\rm SQT, EPR}(\theta)
\end{equation}
for entangled photon pairs and

\begin{equation}
D_{\rm COH}(\theta)=C_{\rm exp, COH}(\theta)/C_{\rm SQT, COH}(\theta)
\end{equation}
for faint coherent pulse lasers.

\begin{figure}[htbp]
\centering
\includegraphics[width=8.8cm]{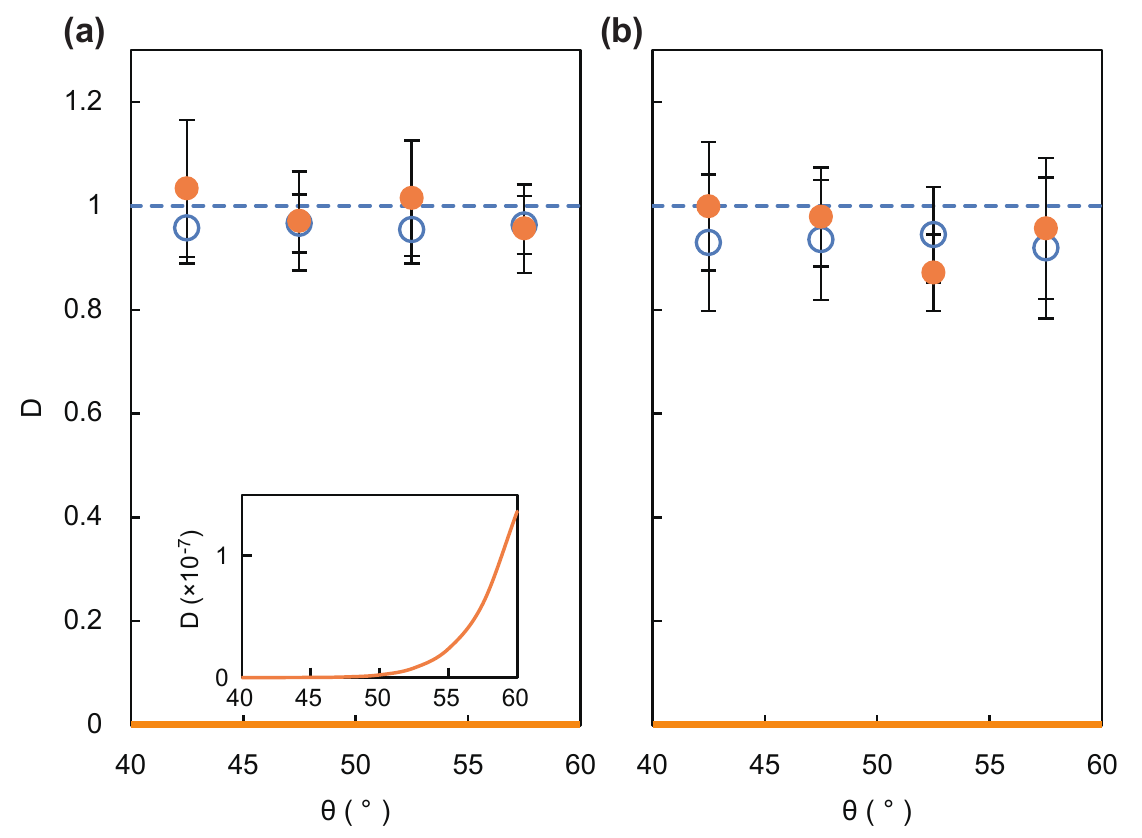}
\caption{
Experimentally estimated decorrelation factors in different altitude angle ($\theta$) of the satellite without (left) and with (right) fulfilling the non-signaling condition. \cite{Xu2019}
}
\label{fig:decoherence2}
\end{figure}

The experiment can be implemented with and without fulfilling the no-signaling condition to account for the quantum collapse models.
By using 1 km fiber to delay the photons in the ground station, the detection events of entangled photons on the ground and satellite are separated spacelike (see Fig. \ref{fig:decoherence2}). 
By collecting data when the altitude angle of satellite varies from $40^\circ$ to $60^\circ$, the estimated decoherence factors for both spacetime settings are shown in Fig.  \ref{fig:decoherence2} \cite{Xu2019}.

The experimental results are consistent with the standard quantum theory and hence do not support the event formalism. 
However, it may be explained by a weaker decoherence effect. 
By using the clock local to the detector as the global reference, the expression of $\Delta_t$ is given by

\begin{equation}
\Delta_t = \int^{r_e + h}_{r_e} ({{M}\over{r}} - {{M}\over{r_e + h}})(1 + {{2 M}\over{r}} + {{r_e^2 \tan^2 (90-\theta)}\over{r^2}})^{1/2} dr.
\label{d}
\end{equation}
The corresponding $D(\theta)$ is between 0.96 and 0.98 for $40^\circ<\theta<60^\circ$ . 
The future testing of such model may be performed using a satellite in higher orbit.

\section{Other quantum satellite projects}

\begin{figure*}[bt]
	\centering
	\includegraphics[width=0.8\textwidth]{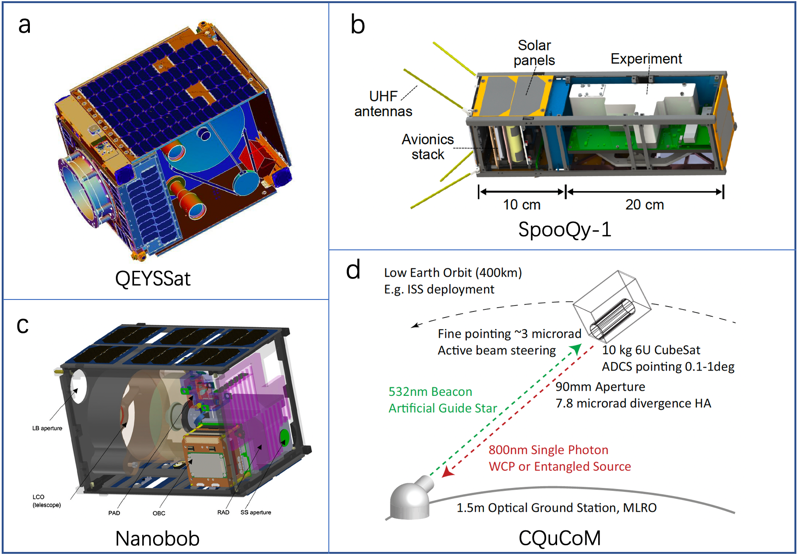}
	\caption{
	Other quantum satellite plans besides Micius. (a) The Quantum Encryption and Science Satellite (QEYSSat) project in Canada \cite{Jennewein2014QEYSSAT, Pugh_2017}. (b) The 3U Cubesat involving an entangled photon source developed by the group in National University of Singapore \cite{VillarEngSource2020}. (c) The CubeSat-based mission concept Nanobob proposed by researchers of France and Austria \cite{2018Nanobob}. (d) The CubeSat Quantum Communications Mission (CQuCoM) jointly undertaking by a joint research team \cite{2017CubeSat}.
	}
	\label{othersatplan}
\end{figure*}

The positive results and exciting prospect kick-started an international race on quantum experiments in space. 
Many satellite projects for quantum communications have been approved and supported, as shown in Fig.~\ref{othersatplan}.
For instance, the Quantum Encryption and Science Satellite (QEYSSat) project in Canada has been studied by the Canadian Space Agency since 2010, and it has received \$1.5 million and \$30 million funding in 2017 and 2019, respectively. 
Its mission concept was developed in partnership with Honeywell Aerospace. 
In contrast to many other missions, it proposes a quantum uplink, placing the receiver on the microsatellite and keeping the quantum source on the ground \cite{Jennewein2014QEYSSAT, Pugh_2017}.
Recently, it was reported in \cite{NASAplan} that NASA plans to build a quantum satellite link, which was called ``Marconi 2.0". The main idea behind Marconi 2.0 is to establish a space-based quantum link between Europe and North America by the mid- to late-2020s.

In addition to the traditional ``big-space'' paradigm of satellite, many other teams worldwide have started a new paradigm based on nanosatellites, even the CubeSat standard \cite{2017CubeSat}.
The CubeSat Quantum Communications Mission (CQuCoM), which is jointly undertaken by the University of Strathclyde, Austrian Academy of Sciences, Clyde Space Ltd, Technical University of Delft, Ludwing-Maximilian University, University of Padua (in collaboration with ASI-Italian Space Agency), and the National University of Singapore, will perform satellite-to-ground entangled photon transmission and QKD using a CubeSat platform deployed from the International Space Station (ISS). 
CubeSat employed in CQuCoM will be a 10-kg and 6-litre mass-volume envelope, which will first be carried up to the ISS on a regular resupply mission (Dragon, Cygnus, HTV, ATV, Progress and Soyuz) and then deployed into orbit using the NanoRacks CubeSat Deployer (NRCSD) mounted upon the Japanese Experimental Module Remote Manipulator System (JEMRMS). 
The CQuCoM calls for two missions, the first to demonstrate the pointing mechanism with a high brightness transmission source that can also be used for a weak coherent pulse (WCP) source-based QKD, and the second mission to distribute entanglement between space and ground. 
The CubeSat exploits advances in nanosatellite attitude determination and control systems (ADCS) as coarse pointing by rotating the satellite body to align the transmitting telescope with the ground station. 
The CQuCoM would be a pathfinder for advanced nanosatellite payloads and operations, and would establish the basis for a constellation of low-Earth orbit (LEO) trusted relays for QKD service provision \cite{2017CubeSat}.

Another CubeSat-based mission concept, Nanobob, was proposed by researchers of France and Austria in 2018. 
They studied the feasibility of implementing ground-to-space optical quantum communication by placing the quantum source on the ground and the 12U CubeSat with the ``Bob'' detection system only. 
In addition to its main scientific aim of demonstrating space-based QKD using a CubeSat and uplink configuration, the NanoBob mission has extra technological aims, such as accurate clock synchronization and the fast classical optical communication with approximately 1 Gbits/s \cite{2018Nanobob}.

A group at the National University of Singapore has been committed to designing and developing quantum sources based on nanosatellites and CubeSat platforms. 
They developed a correlated photon pair source as the pathfinder for their future plan of space applications. 
Their first attempt was unsuccessful when the launched vehicle (CRS Orb-3) failed shortly after takeoff, although the payload was successfully recovered intact and found to be fully operational \cite{Tang2016source}. 
Their second attempt was successful. 
The source was launched onboard the Galassia CubeSat (PSLV C29) to an orbit of approximately 550 km at the end of 2015, which laid the foundation of their future space-based quantum experiment mission \cite{Tang2016, Grieve2018SpooQySats}. 
Recently, they developed an entangled photon-pair source onboard a 3U CubeSat, SpooQy-1, which was launched successfully to the ISS in April 2019. The CubeSat was then deployed into orbit from ISS on June 17, 2019 \cite{VillarEngSource2020}.

Furthermore, some feasibility tests for using smaller-sized or high orbit satellites have been reported \cite{Takenaka2017Satellite, Gunthner:17, Vallone2015, Dequal7000km, Vedovato2017Extending}. 
Bedington \textit{et al}. provided a table of notable satellite QKD proposals \cite{2017satelliteProgress}.

In addition to the university consortia and national agencies, the international space race also involves private companies, e.g., QKDSat (ArQit) \cite{ArQit} and QUARTZ \cite{QUARTZ}. 
Also, a more ambitious quantum communication infrastructure project is taking shape in Europe. 
The European Quantum Communication Infrastructure (EuroQCI) initiative aims to build a secure quantum communication infrastructure that will span the entire European Union (EU), including its overseas territories. 
Since June 2019, all 27 EU Member States have signed the EuroQCI declaration, indicating their commitment to the EuroQCI initiative. 
The EuroQCI will include a terrestrial segment relying on fiber communications networks linking strategic sites at national and cross-border level, and a space segment based on satellites. 
It will link national quantum communication networks across the EU and provide global coverage \cite{EuroQCI}.

\section{Outlook}

Although Section IX has shown that the Micius satellite greatly enhances the scale and capability of quantum experiments in space, Micius only marks the beginning.
For the Chinese quantum satellite plans, there are two goals in the next 5 to 10 years. 
The first one is to develop 3 to 5 small LEO satellites dedicated to QKD missions, which will provide more practical and efficient QKD services.
The second goal is to develop a medium-Earth-orbit (MEO)-to-geosynchronous-orbit (GEO) quantum science satellite that involves several ambitious scientific objectives. 
Compared to the LEO satellites, high-orbit satellites can provide much longer service time and wider coverage. 
The combination of a high-orbit satellite and multiple LEO satellites can form a quantum constellation for global services. 
Furthermore, with such a new generation space platform, researchers plan to realize the high-precision satellite-ground time-frequency transfer and GEO satellite-based optical clocks to verify the technology of wide-area optical frequency standard; further research includes fundamental test of quantum physics and its interface of general relativity to deepen the understanding of the basic laws of nature.

\subsection{Daytime quantum communications}

\begin{figure*}[!t]\center
\resizebox{15cm}{!}{\includegraphics{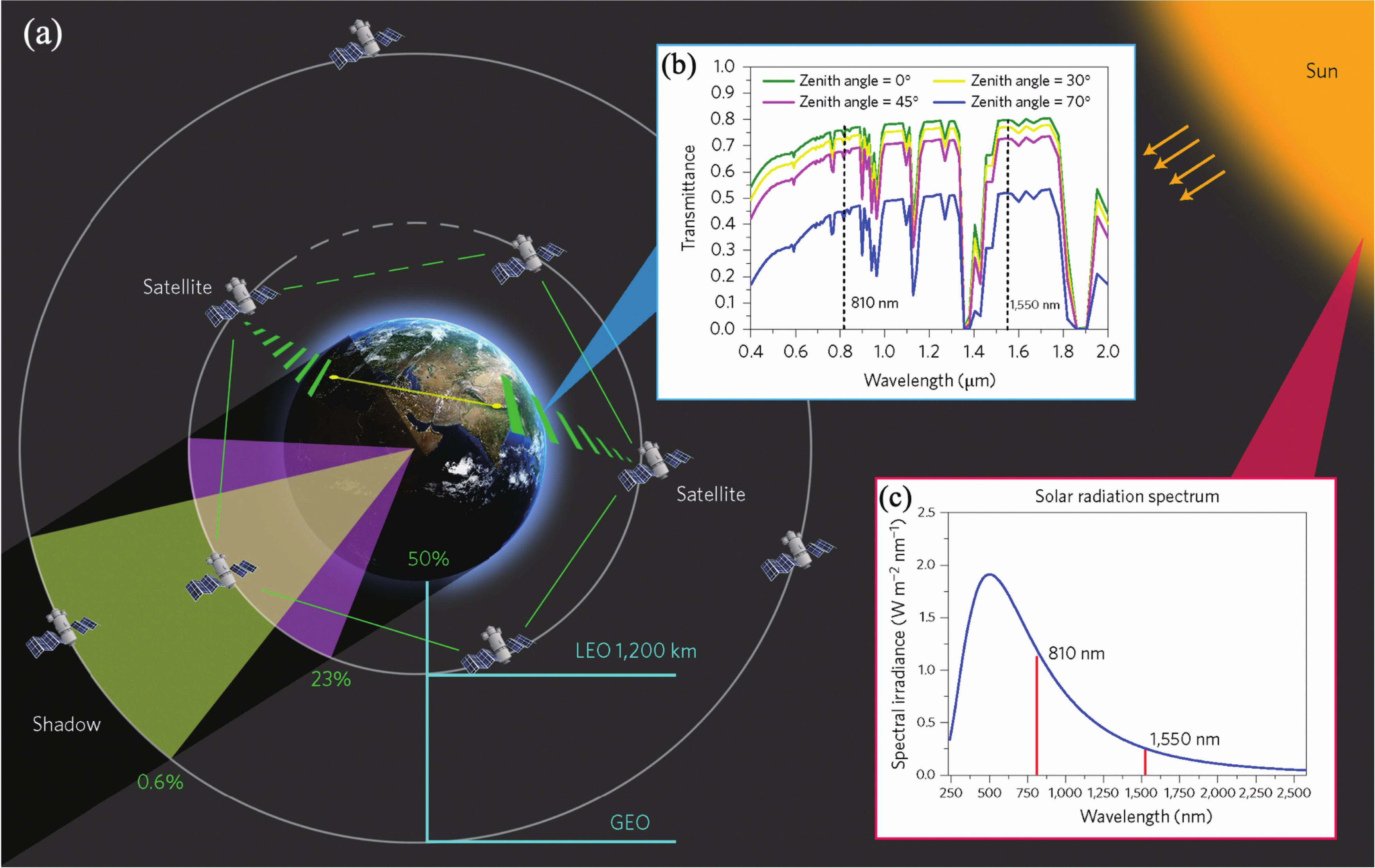}}
\caption{Satellite-constellation-based global quantum network \cite{Liao2017daylight}. (a) A global quantum network needs many LEO satellites or several geosynchronous orbit satellites to compose a satellite constellation. The time of a satellite in the Earth shadow area is inversely proportional to the orbit height of the satellite.
(b) Transmittance spectra from visible to near infrared light in atmosphere at selected zenith angles.
(c) Solar radiation spectrum from visible to near infrared light.}
\label{Fig:Daylight:SC}
\end{figure*}

There is much space for improvement.
One of the main drawbacks of the current satellite-based quantum communication missions is that they can only work at night, which greatly limits their practical application.
In satellite-based classic communication, the Iridium system \cite{Pratt1999An} functions to provide worldwide connectivity.
Similarly, quantum satellite constellation, which is composed of a few dozen satellites, can provide global real-time quantum communication.
Such a satellite constellation is expected to operate with both LEO and high-Earth-orbit (HEO) satellites, such as GEO satellites.
The probability of a satellite being in the Earth's shadow zone reduces rapidly with increasing orbit height (Fig.~\ref{Fig:Daylight:SC}).
A LEO satellite system has a $\sim 70\%$ probability of being in the sunlight area, and for GEO satellites, this probability increases to $\sim 99\%$ \cite{Gilmore2002}.
Therefore, a step toward the quantum satellite constellation is to demonstrate daylight free-space quantum communication.

The main challenge is the strong background noise from the scattered sunlight, which is typically five orders of magnitude greater than the background noise during night time.
To this end, as reviewed in Section II, early indoor and outdoor tests \cite{Jacobs1996freespace, Buttler1998freespace, Buttler2000freespace, Hughes2002freespace, Kurtsiefer2002} over distances from 75 m to 23.4 km suggested improving the signal-to-noise ratio by a combination of detection timing, narrowband filters, and spatial filtering.
Later, many research teams further employed ultranarrow bandwidth spectral filtering using multipassing etalon and Rb vapor filter and improved the time resolution \cite{shan2006Free, Rogers2006Free, Hockel:09, Peloso_2009, Restelli2010Improved}.

A preliminary verification of free-space QKD in daylight under conditions of high channel loss ($\sim$48 dB) over 53 km was reported by \cite{Liao2017daylight}.
To increase the signal-to-noise ratio, first, Liao \textit{et al}. chose a working wavelength of 1,550 nm.
Compared to 800 nm, the telecom-band wavelength has the transmission slightly higher, and Rayleigh scattering $\sim$14 times smaller.
Further, the sunlight intensity at 1,550 nm is $\sim$5 times weaker than that at 800 nm. 
Second, free-space single-mode fiber-coupling was developed with an efficiency of  30\% for the indoor test and 5\% in the outdoor.
The field of view for the receiving system is reduced below 10 $\mu$rad to reduce the background noise.
Finally, ultralow-noise up-conversion single-photon detectors were used with a built-in spectral filtering employing volume Bragg grating with a bandwidth of 0.16 nm.
Such narrowband filtering reduces noise by a factor of $\sim$ 100 compared to the 3-10 nm filters used in previous experiments at night.
A combination of the three key toolbox enabled a decoy-state QKD with a final key rate of 20-400 bits per second, where the variation was mainly due to the atmospheric environment.

For higher-orbit satellites, especially those working in day time, due to the longer distances and the associated diffraction loss, new techniques need to be developed to increase the link efficiency in the future, including large-size telescopes, better APT systems, and wavefront correction through adaptive optics \cite{10.1117/1.OE.55.2.026104, Gruneisen:15, 10.1117/1.OE.56.12.126111, PhysRevApplied.16.014067, Yang:20, Gong:18}.

\subsection{Satellite-constellation-based quantum networks}

A low-earth-orbit (LEO) satellite alone is not enough to support the construction of the global-scale quantum communication network.
In general, attention should be paid to two aspects: increasing the number of satellites and raising the orbital altitude.
It is necessary to build a quantum constellation combining LEO satellites and HEO satellites, as shown in Fig.~\ref{Fig:Daylight:SC}(a).

\begin{figure}
\centering
\includegraphics[width=8.5cm]{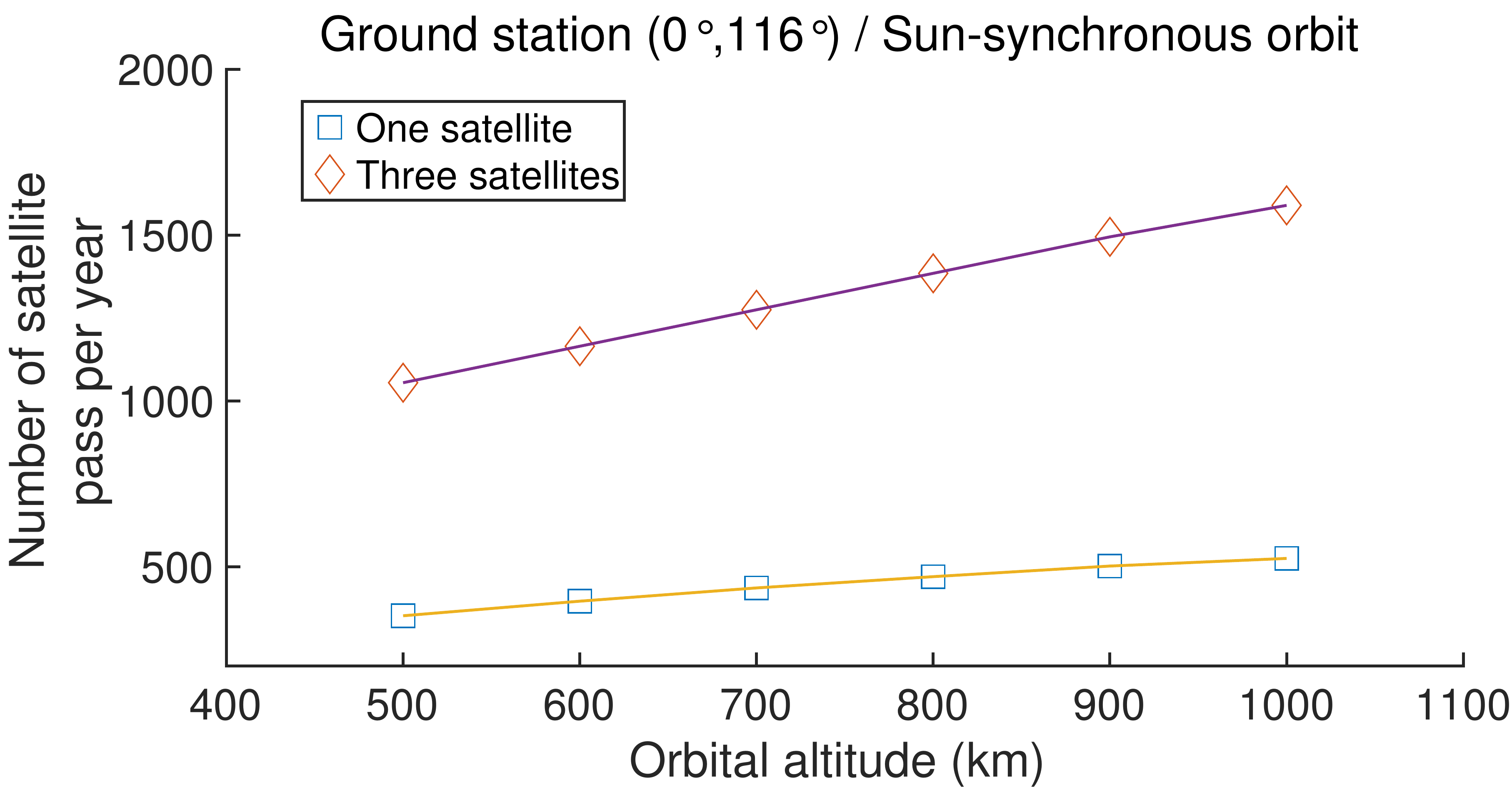}
\caption{
Simulation on the number of times of the satellite passes through the low latitude region.
}
\label{fig:satellitepass}
\end{figure}

There is also a need to develop many cheaper satellites in LEO to cover the earth and a few HEO satellites to provide 24-hr service for some important regions, as shown in Fig.~\ref{Fig:Daylight:SC}.
On the number of LEO satellites, as shown in Fig.~\ref{fig:satellitepass}, it suggests that about three low-orbit satellites are needed to ensure all the major regions of the world can be covered with enough passages per year.
Considering the altitude of LEO satellites, the higher the orbit, the more sufficient QKD times can be guaranteed for the ground stations at a specific time.
For instance, if there is a three-satellite constellation with an orbit altitude of 800-1000 km, the average number of satellite passes in major regions of the world is about 3.7 times per day, and the average effective QKD time for each ground station at an elevation above $25^{\circ}$ is $\sim$ 5 min, which can cover more than 100 ground stations in major areas of the world and guarantee the QKD of each ground station once a week.

Therefore, for the future scenarios, one of the simplest quantum constellations may include at least three low-orbit satellites and one HEO satellite.
In this configuration, the LEO satellites are responsible for the daily needs of numerous ordinary users, for instance, $\sim$ 100 users, and the HEO satellites can provide long-term uninterrupted services for a few important areas and users.
According to the Micius data, since LEO satellites have a large margin in channel efficiency, miniaturization and low-cost designs can be considered for both satellite payloads and ground receiving stations. 
When we consider HEO-satellite-based QKD, due to the high channel loss, we need prioritize ensuring performance in the design of payloads and ground stations.

\begin{figure}
\centering
\includegraphics[width=8.5cm]{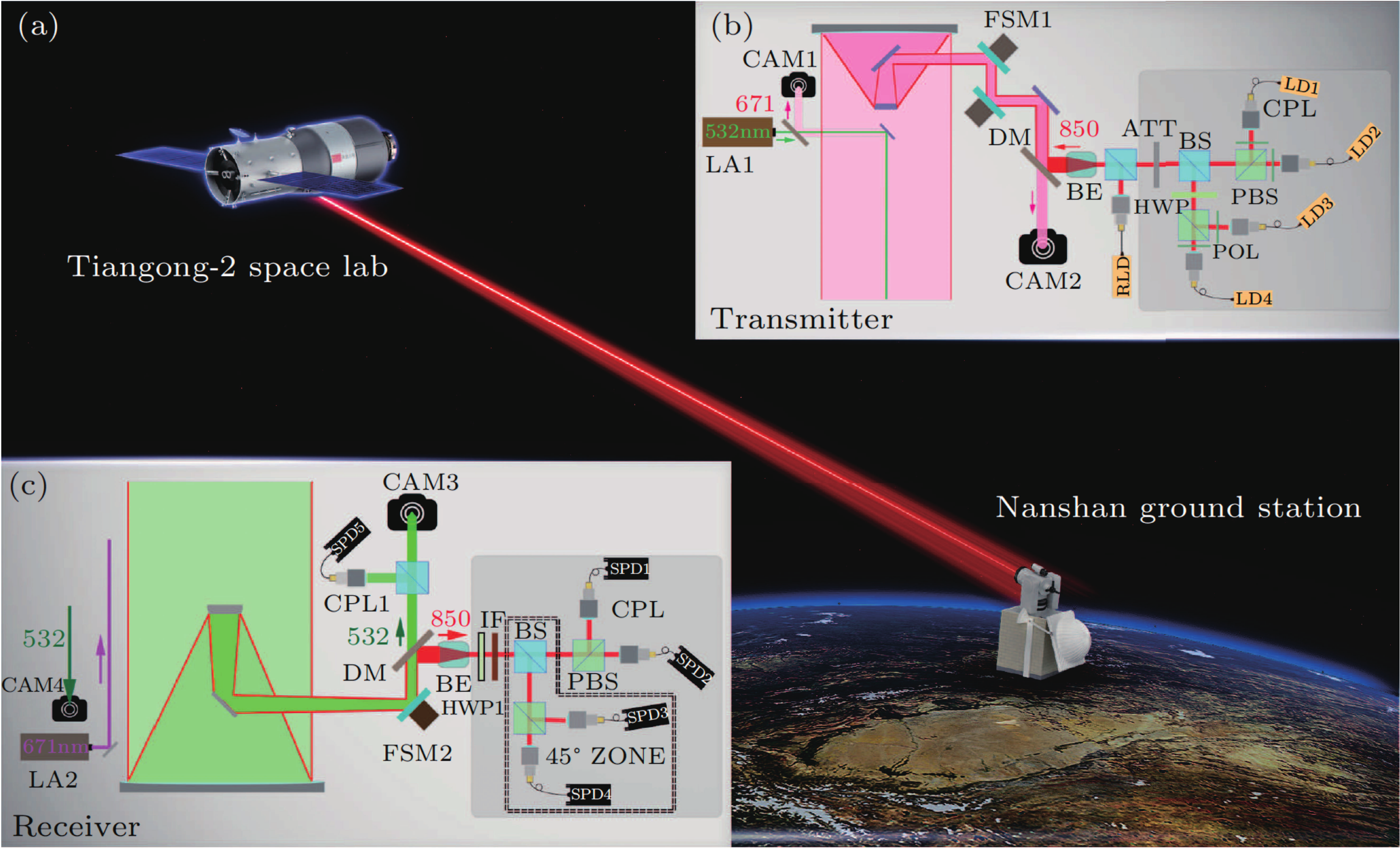}
\caption{
Schematic diagram of QKD from Tiangong-2 space lab to ground \cite{liaoCPL2017}. (a) Overview of the space-to-ground QKD. (b) Schematic of the decoy-state QKD transmitter. (c) Schematic of the decoy-state QKD decoder in the Nanshan ground station equipped with a 1200-mm-aperture telescope. LA1: green laser (532 nm), CAM1: coarse camera, CAM2: fine camera, LD: laser diode, RLD: reference laser diode, FSM1: fast steering mirror, HWP: half-wave plate, POL: polarizer, PBS: polarization beam splitter, BS: beam splitter, ATT: attenuation, LA2: red laser (671 nm), CAM3: fine camera, CAM4, coarse camera, CPL: coupler, DM: dichroic mirror, IF: interference filter, FSM2: fast steering mirror, BE: beam expander, SPD: single photon detector.
}
\label{fig:tiangong-2}
\end{figure}

For LEO satellites, it is economical to consider small-sized and low cost QKD payloads, which can be assembled on satellites with different sizes, such as microsatellites and space stations.
In this regard, Liao \textit{et al}. have also made preliminary attempts to develop a small-sized payload for space-to-ground QKD, from Tiangong-2 space lab to Nanshan ground station \cite{liaoCPL2017}.
The 57.9-kg payload integrates a tracking system, a QKD transmitter along with modules for synchronization, and a laser communication transmitter.
In the space laboratory, a 50-MHz vacuum + weak decoy-state optical source was sent through a reflective telescope with an aperture of 200 mm, as shown in Fig.~\ref{fig:tiangong-2}.
In the experiment, the communication distance was within the range between 388 km and 719 km, the quantum bit error rate (QBER) was $1.8\%$ and the final key rate was $\sim$ 91 bps when the quantum channel was established.

\begin{figure}
\centering
\includegraphics[width=8 cm]{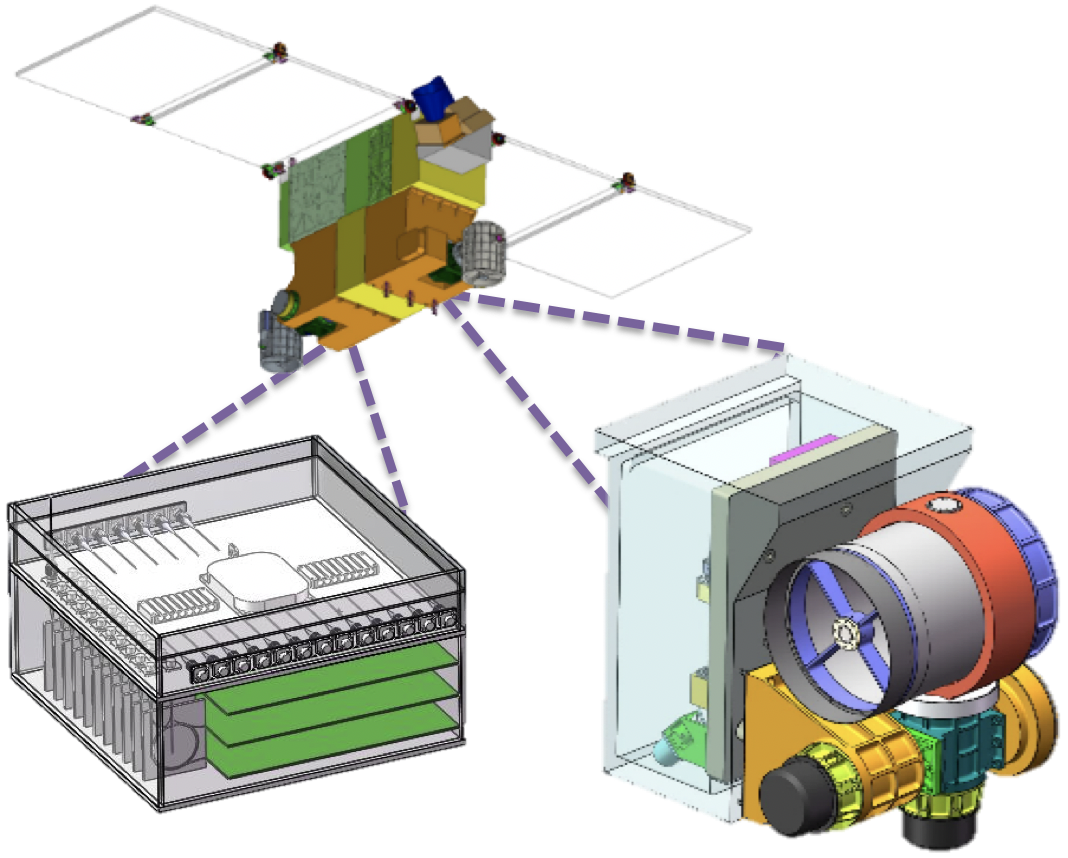}
\caption{
Preliminary design of small satellites in China. It will be focus on the QKD using downlink channels. The total weight will be less than 100 kg. The orbit height will be 800 km. There will be 3 or 5 small quantum satellites for the first step.
}
\label{fig:smallsatellite}
\end{figure}

Such compact and low-cost payloads used in Tiangong-2 can be assembled on satellites of various sizes to construct a satellite-constellation-based quantum network, as shown in Fig.~\ref{fig:smallsatellite}.
The performance of QKD and the size of the payload can still be improved. For instance, the size of the telescope can be reduced to 100 mm, the divergence angle of the source narrowed to the diffraction limit, and the decoy-state source rate increased to 1 GHz.
With such improvements, the weight of the payload is reduced below 20 kg, and the final key rate is increased to $\sim$ 10 kbps.

\begin{figure*}
\centering
\includegraphics[width=16 cm]{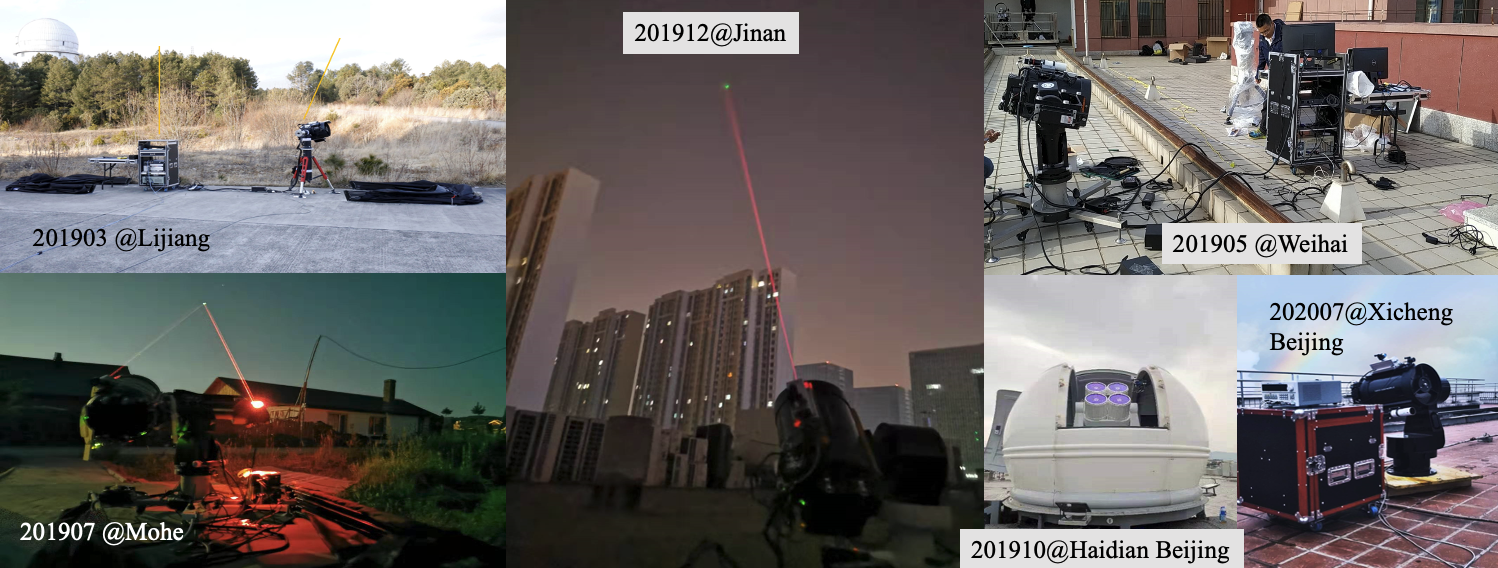}
\caption{
Experimentally demonstrating the feasibility of satellite-to-ground QKD with compact ground stations. Up to 2020, QKD with Micius satellite and the compact ground station, in many cities of China, such as Beijing, Jinan, Weihai, Lijiang, Mohe, and so on. The typical sifted key rate is $\sim$ 2 kbps.
}
\label{fig:compactstationexp}
\end{figure*}

In addition, for the practical space-ground integrated quantum communication network, the number of users is far greater than the QKD payloads in the sky.
The typical ground station for Micius satellite is too big and heavy for large-scale applications for more users.
The ground station should be redesigned to be smaller, lighter, and cheaper for the requirements of the practical quantum constellation.
Recently, the feasibility of performing the satellite-to-ground QKD using the compact ground station (less than 100 kg, 280 mm diameter) has been verified by Pan's group in many cities of China \cite{compactgroundstation}, as shown in Fig.~\ref{fig:compactstationexp}.
A typical sifted key rate can be achieved to 2 kbps.

Besides QKD, quantum teleportation also plays a key in future quantum networks, and it will require that the long-distance entanglement distribution should be implemented before the Bell-state measurement.
For the Micius satellite-based teleportation~\cite{ren2017satellite}, the entangled-photon source and the Bell-state measurement are performed at the same location on the ground.
A next step towards a real network is to develop an entangled-photon source with a long coherent time $T_{c}$ and reduce the arrival-time jitter $T_{j}$ between independent photons to achieve $T_{c} > T_{j}$.
In this case, semiconductor chip based sources of deterministic single- \cite{ding2016demand, wang2019towards} and entangled-photon source \cite{wang2019demand}, possessing high purity, indistinguishability and efficiency simultaneously, have coherence time in the order of a few hundred picoseconds and can offer a more efficient and viable solution.
Furthermore, teleportation can be used to transfer the quantum state of a flying single photon to a long-lived matter qubit to realize quantum memory at a distance~\cite{Sherson:2006dt, Chen2008Memory, Bussi2014Quantum}.
Teleportation can also be employed in entanglement swapping and distributed quantum computing schemes.
Using the above long-lived quantum memories and efficient light-matter interfaces, more sophisticated space-ground-scale teleportation will soon be realized, and will play an important role in a future distributed quantum network.

Through these further efforts, we could envision a global quantum communication infrastructure with quantum constellation and ground-based fiber networks, as shown in Fig. 46. Fiber-based network on the ground provides secure communication services for distance cities. Meanwhile, a quantum constellation with LEO and the high-orbit satellites connecting key nodes on the fiber networks and movable nodes, even ships in the ocean. Based on the above analysis, we suggest that the simplest quantum constellation should include at least three low-orbit satellites and one high-orbit satellite. In this configuration, assuming that not less than 100 ground stations need to be covered, each ground station need more than 50 times of QKD links with satellite per year, and can obtain about 2 Mbits for each satellite passage. Then, each station can obtain 100 Mbits per year, and the quantum constellation can output about 10 Gbits of keys per year totally, which can support the basic function of voice communication. Also, HEO satellites can provide 24-hr QKD services at a key rate of 1 kbps for some important areas, which can provide the basic needs of text communication.

\begin{figure*}
\centering
\includegraphics[width=15cm]{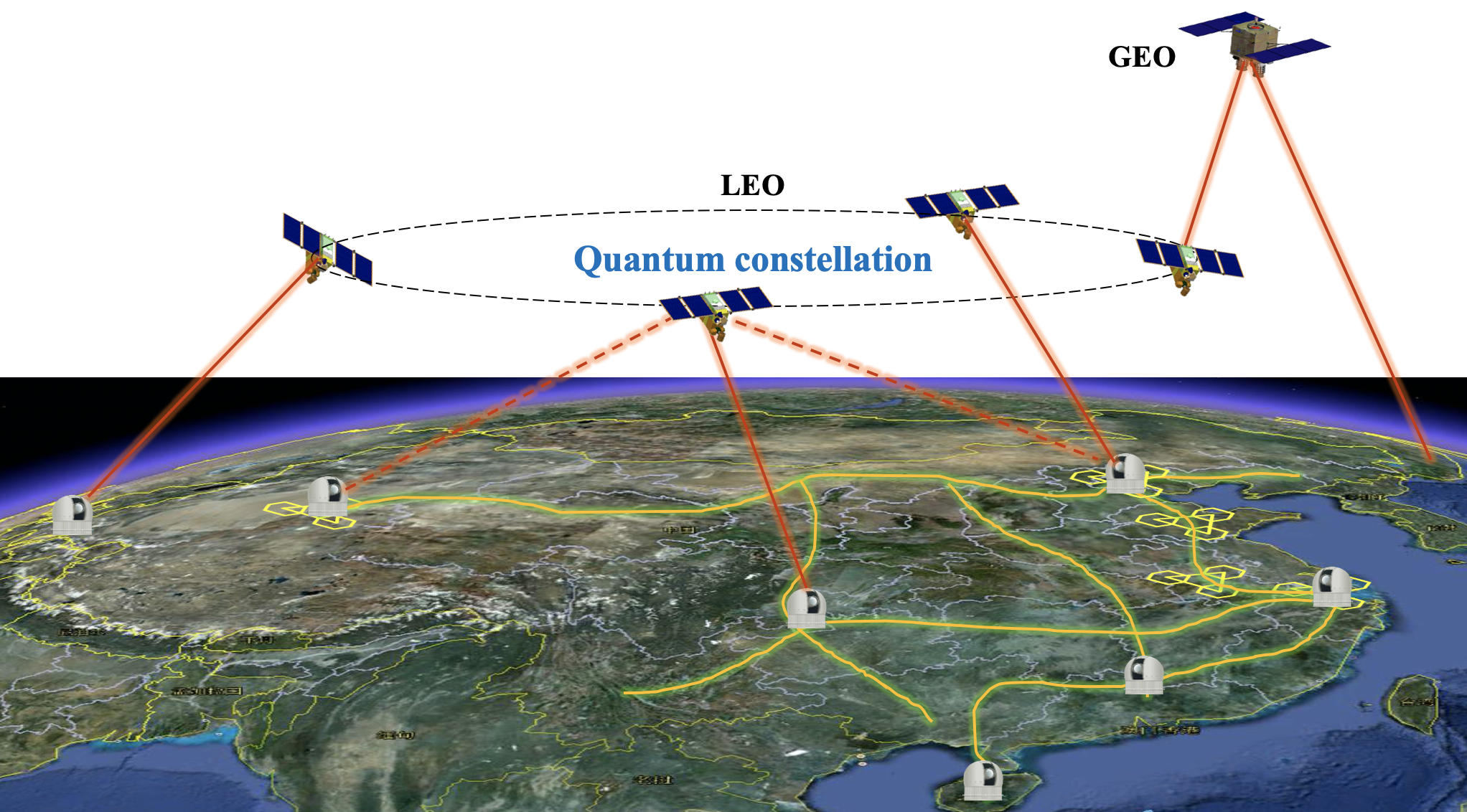}
\caption{
The roadmaps towards the global quantum communication network. Intra-city metropolitan networks will be created using fibers. Quantum repeaters can connect the metropolitan networks. Long-distance and intercontinental quantum communication will be realized via satellite-based quantum channels.
}
\label{fig:network}
\end{figure*}

\subsection{Fundamental test of quantum physics at space scale}
The success of the satellite Micius shows not only the feasibility of realizing the global-scale quantum secure communication network, but also a new way for performing fundamental tests in quantum physics at the space scale.
With the Micius satellite, the first step toward this space-scale fundamental research has been achieved; for example, Bell test over 1200 km \cite{yin2017satellite}, and satellite testing of a gravitationally induced quantum decoherence model \cite{Xu2019}.
Furthermore, many more profound scientific explorations of quantum physics will need higher orbital altitude satellite.
In this subsection, we briefly introduce several space-scale fundamental experiments on quantum physics based on the plan of the HEO satellite of China.

\textbf{
Further experiments based on entanglement distribution at larger space scale}. 
According to Einstein's local realism, the maximum value of Bell inequality is 2 \cite{Bell1964, CHSH1969}. 
However, according to quantum mechanics, the maximum value of this quantity can reach $2\sqrt{2}$. 
For decades, physicists have performed several experiments, and all of them have confirmed the correctness of quantum mechanics, though some loopholes still exist and should be addressed
One loophole is the freedom-of-choice loophole \cite{Bellnonlocality2014RMP}, that is, the random number generators that determine the choice of measurement bases for the Bell test can be prior correlated, and thus, the choice of measurement bases are not truly independent and random. 
Another loophole is the so-called collapse locality loophole, or Schrodinger's cat loophole \cite{Kent2017collapse}.
In this case, according to the Schrodinger-cat \textit{gedanken experiment}, it is arguable that the measurement outcome, for example, the cat state in a closed black box is not defined until it is registered by a human consciousness. 
This implies that the realized ``events'' have never been space-like separated.

\begin{figure*}
\centering
\includegraphics[width=15cm]{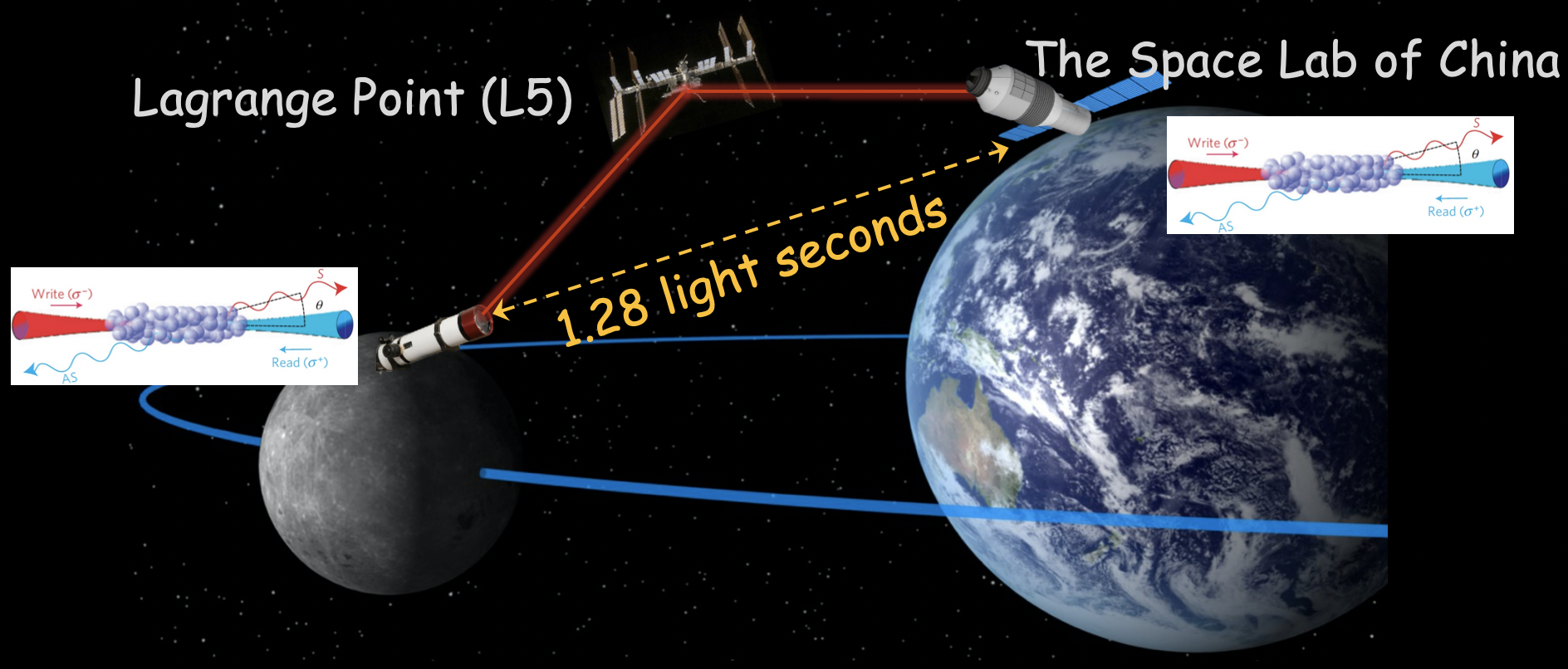}
\caption{
Scheme for conducting Bell test between Earth and the Moon. There are five Lagrangian points in the Earth-Moon system, denoted by L1, L2, L3, L4, and L5. Since only L4 and L5 are stable, and have the most appropriate space arrangement of the five points, they were chosen for the position of the entanglement source satellite. This satellite contains two telescopes, one aims at the Moon and the other at Earth. Two large telescopes must also be built, one on or near the Moon and one on Earth, to create entanglement distribution channels. The human observers and quantum memory, combining the event-ready scheme, should be employed to realize the loophole-free Bell test.
}
\label{fig:belltestearthmoon}
\end{figure*}

A possible solution to address these two loopholes is to perform Bell-test experiments with human observers \cite{Bellspeakable}.
In this way, the measurement basis would be chosen by human's free will, and the measurement outcomes can be defined by human consciousness. 
Since such experiments require the quantum signal transit time to exceed that of human reactions, which is typically 100 ms, one must ensure entanglement distribution at a distance in the order of one light-second.
To address the freedom-of-choice and collapse-locality loopholes, performing the entanglement distribution between Earth and Moon may be a possible solution, as suggested by \cite{Caofreewill}. 
Then, the entangled photon pairs will be sent to Earth and Moon from one of the Earth-Moon Lagrangian points, as shown in Fig.~\ref{fig:belltestearthmoon}.
The distance between the two detectors would be greater than one light-second, and both loopholes would be closed. 
Furthermore, using an event-ready scheme \cite{entswapping1993,eventready2003} and a quantum memory technique \cite{Yang-memory} could increase the heralding efficiency significantly, making it possible to introduce human recorders to address the collapse loophole. 

\textbf{Test the interface of quantum mechanics and general relativity.}
The emergence of quantum mechanics and general relativity has radically changed our understanding of nature.
However, any theory that integrates quantum mechanics with general relativity encounters great challenges.
Among the four basic interactions currently known, the electromagnetic, weak and strong interactions have been quantized and unified.
Only the question of how to quantize gravitational action is pending.
Testing the interplay of quantum mechanics and general relativity helps establish a grand unified theory of four basic interactions.

The results from a recent work \cite{Xu2019} show consistency with the descriptions of the standard quantum theory and do not support the predictions of event formalism.
However, this does not necessarily rule out other approaches since it may be explained by a weaker decoherence effect.
Therefore, this type of experiment can be expanded naturally to the higher orbit satellite to test other gravity-related models.

Meanwhile, there is another scheme to test the interface of two theories, quantum mechanics and general relativity.
It's to probe quantum interference within the frame of general relativity, that is, the optical version of the Colella, Overhauser, and Werner (COW) experiment \cite{rideout2012fundamental, Zych2011Quantum, Zych2012General}.
The first experiment on measuring the effects of gravity on the quantum wave-function of a single particle was achieved by Colella, Overhauser, and Werner using a neutron beam interferometer \cite{COW1975}.
Then, COW-like experiments were repeated with an increase in precision over several years \cite{Peters1999Measurement}.
However, the phase-shifts observed in these interferometric experiments are fully compatible with nonrelativistic quantum mechanics in the presence of the Newtonian gravitational potential.
On the other hand, all previous tests of general relativity can be described within the framework of classical physics \cite{Zych2012General}.
Nevertheless, the quantum interference of photons provides a promising way to probe quantum mechanics in curved space-time.
Single-photon interference is a phenomenon that can prove the wave-particle duality and complementarity in quantum mechanics.
In a COW-like experiment with single photons, the Newtonian limit of gravity is incomplete to explain the interference result without the theory of equivalent of mass and energy, which is one conceptual pillar of general relativity.
Furthermore, if the concept of the time dilation is introduced to the single-photon COW experiment, and the difference in the time dilation of each arm is comparable with the photon's coherence time, the visibility of the quantum interference will drop.
This predicted effect of gravitation-induced decoherence would provide the most possible test of the genuine general relativistic notion of proper time in quantum mechanics.

\begin{figure}[htbp]
\centering
\includegraphics[width=8.5cm]{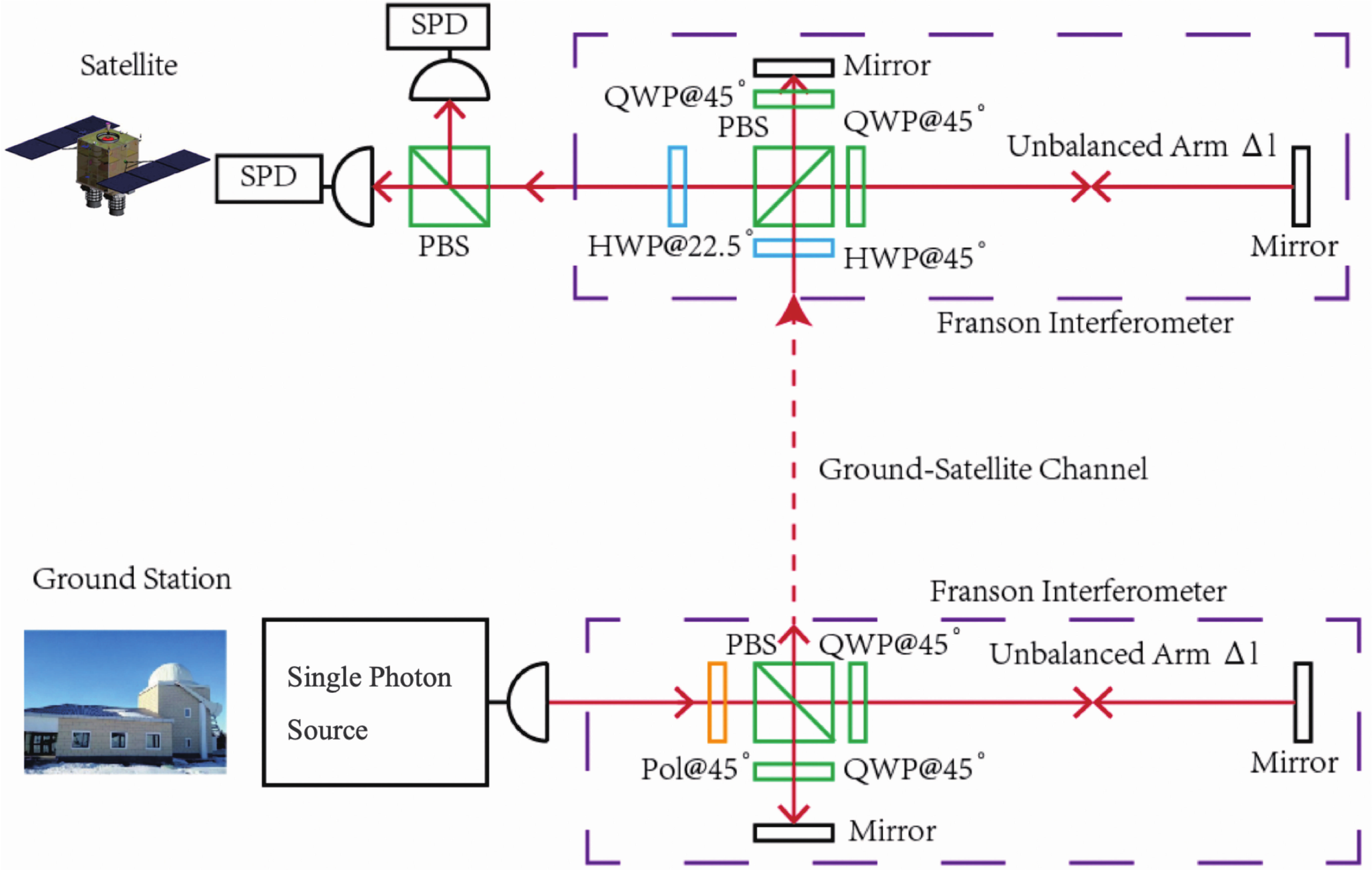}
\caption{The scenario of the single-photon COW experiment based on the high earth orbit satellite. Before entering Franson interferometer on the ground, the genuine single photon is prepared as $\vert + \rangle$ state. After passing through the interferometer on the ground and the ground-satellite free-space channel, a half wave plate changes the polarization of the photon, which guarantees that if photons pass through the long arm of interferometer on the ground, they must pass through the short arm in the satellite, and vice versa. HWP: half-wave plate; QWP: quarter-wave plate; PBS: polarization beam splitter; M: mirror.
}
\label{fig:opticalcow}
\end{figure}

A typical and possible scheme of the single-photon COW experiment can be implemented by combining two identical unbalanced Michelson interferometers, and equipping them in the ground station and the high-earth-orbit satellite respectively, as shown in Fig.~\ref{fig:opticalcow}.
In such an experimental configuration, both the effects of gravitational induced redshift and time dilation can be tested, depending on the orbit altitude, the length of unbalance arm $\sim \Delta l$, and the measurement accuracy of the interferometer.

\textbf{Wide-area quantum-secure and high-precision optical time-frequency transfer.}
Combining quantum communication with time-frequency transmission will develop new meaningful research interests.

\begin{figure}[htbp]
\centering
\includegraphics[width=7cm]{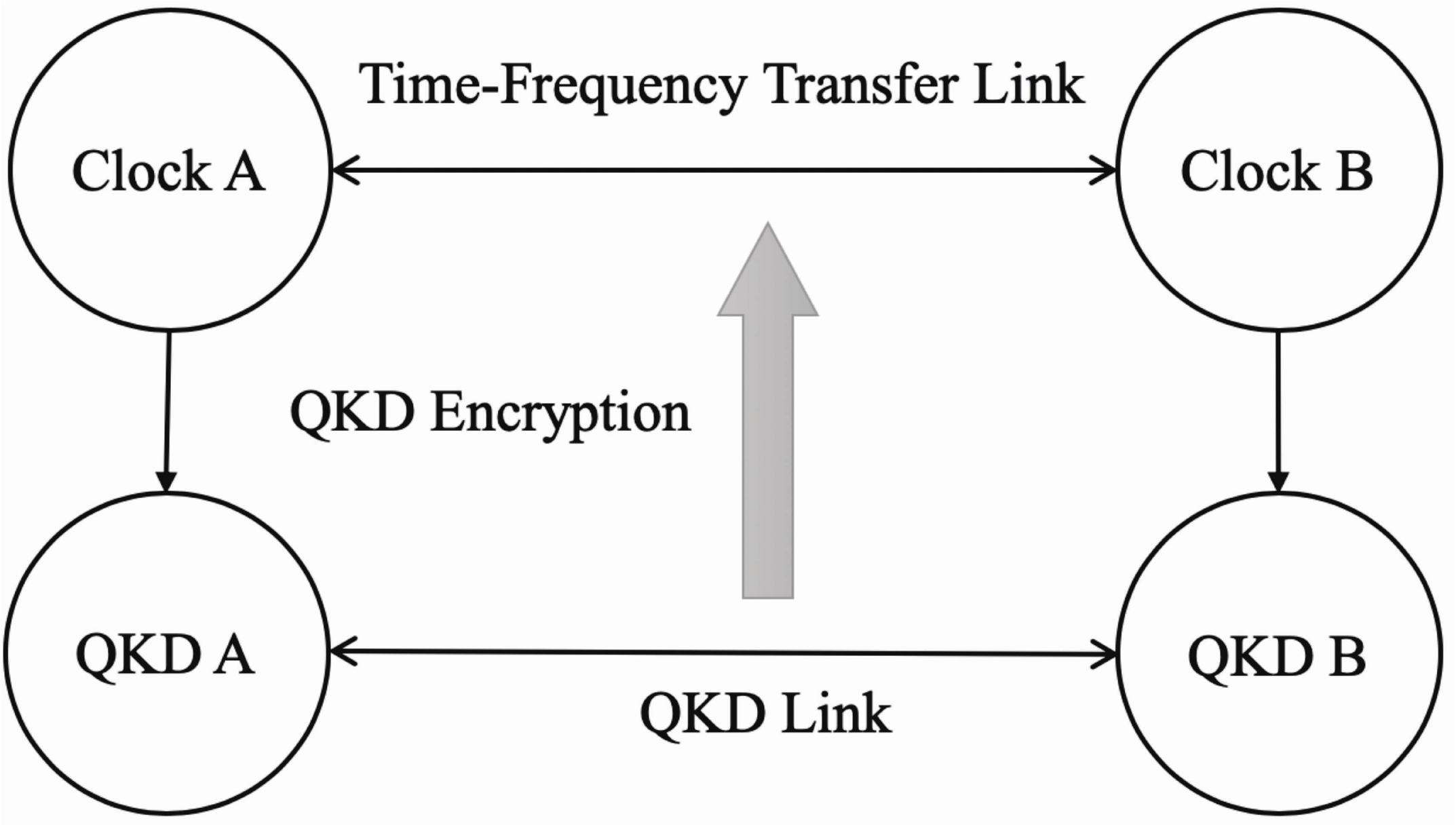}
\caption{
The schematic diagram of the quantum-secure time-frequency transfer.
}
\label{fig:timetransfer}
\end{figure}

From the perspective of time-frequency transmission, wide-area high-precision time-frequency transfer is an essential component for constructing a physical network of optical time scales, which plays an important role in fundamental science \cite{Riehle2017Optical, Kolkowitz2016prd, Marra2018Ultrastable} and real-life applications \cite{Mills1991}.
For instance, it can improve the precision dramatically of navigation and timing \cite{Ludlow2015rmp, Gill2011, Komar2014}, clock-based geodesy \cite{Ludlow2015rmp}, testing the effects of special and general relativity \cite{Ludlow2015rmp, Cacciapuoti2009Space, M2008Geodesy, Schiller2009Einstein}, and even search for dark matter \cite{Derevianko2013Hunting}.
However, security issues, such as man-in-the-middle attacks \cite{Treytl2007Traps}, remains unaddressed. 
It leads to secure time-frequency transfer evolving as a crucial problem in time-frequency applications.
Inspired by the information-theoretic security of QKD, there is a need to extend the application of QKD to time-frequency transmission to improve its security, as shown in Fig.~\ref{fig:timetransfer}.
Recently, a satellite-based quantum-secure time transfer scheme based on the two-way quantum key distribution in a free-space link was experimentally demonstrated, which can be regarded as the first step towards an enhanced infrastructure for a time-transfer network \cite{timetransferQKD2020}.

More importantly, the technologies developed from satellite-based quantum communication, such as high-stable and -efficiency satellite-ground optical links, promote the development of the satellite-based optical time-frequency transfer.
Traditional satellite-based links exhibit an optimum frequency instability of approximately $1\times 10^{-15}$ for a day, which is mainly limited by the resolution of the microwave carrier.
Optical-based links naturally become an important means to further improve the accuracy of time-frequency dissemination.
Many important works on high precision and stability optical time-frequency transfer with free-space and fiber link have been reported \cite{Smith2006Two, 2008TIME, Giorgetta2013Optical, Predehl2012A, Sinclair2018}.
Recent experiments have shown that the stability of $10^{-18}$ level can be achieved at 3000 s with an average loss of 72 dB, corresponding to the loss of a satellite-ground link \cite{Shen:21}.

A combination of QKD and time-frequency transfer can also further improve the clock synchronization accuracy of QKD, promoting the realization of wide-area quantum repeaters, measurement-device-independent QKD and twin-field QKD.
Thus, for the plan of high orbit satellite, it will be a very meaningful direction to develop quantum-secure and high-precision optical time-frequency transfer between the satellite and the ground, which will create new possibilities toward a global-scale quantum time-frequency transfer networks.

\textbf{Space-based ultra-high precision optical frequency standard.}
Optical atomic frequency references, or optical atomic clocks, have better stability and total uncertainty than the microwave atom frequency standards, and are promising for providing next-generation frequency standards.
On the $26^{th}$ General Conference on Weights and Measures (CGPM), scientists have reached an important resolution to define all units in the International System of Units (SI) by nature constants in quantum physics, which opened a new era for quantum metrology.
Time is one of the fundamental quantities in SI. 
It is the most precisely measurable quantity by humans and the most important one. 
Since the distance can be measured by the propagation time of light, ultimately the precision of distance measurements is limited by the precision of time measurements.

The unit of time in SI is the second, which is defined based on an atomic hyperfine transition in neutral cesium (Cs) atoms. 
Nowadays, the primary time standard is defined by cold Cs atomic fountains in laboratories, which leads to the inaccuracy of several parts in $10^{16}$. 
It provides the time-frequency signals and the international atomic time (TAI), serving as a worldwide time reference. 
Atomic clocks are now an essential tool in modern society, especially in navigation systems such as GPS, GLONASS, Galileo, and Beidou systems.

Although the state-of-the-art technology in atomic clocks already by far has the lowest inaccuracy of any physical unit, the optical clocks, which use optical rather than microwave transitions, are being developed in laboratories worldwide and are at the forefront of frequency metrology. 
They outperform microwave clocks' instability and inaccuracy by two orders of magnitude, raising discussion on the redefinition of the SI second (27th CGPM, 2026).
The outstanding performance of optical atomic clocks paves the way for their applications in different fields apart from metrology.
They cover the fields of time and frequency metrology (comparison of distant clocks \cite{Takano2016Geopotential}), quantum communications (synchronization of quantum phase for wide-area \cite{Lucamarini2018Overcoming}), geophysics (mapping of the gravitational potential of the Earth \cite{Mehlst2018Atomic}), and potential applications in fundamental physics (tests of General Relativity and its foundations \cite{Chou2010Optical, Kolkowitz2016prd}).

With the rapidly improving performance of optical clocks, it is possible to only take full advantage of it by operating the clocks in space \cite{Riehle2017Optical, Origlia2018}, since the frequency of the clocks on Earth is influenced by the Earth's gravitational potential at the location of the clock, which may cause a drift in time e.g., from tidal effects.
Therefore, the ultrahigh-precision optical frequency standard system for space platforms is the key to achieving more precise time reference and ultrahigh-precision time synchronization, which provides important support for state-of-art technologies and basic scientific research.
At present, a research trend in optical frequency standards is to develop neutral-atom optical clocks for space platforms, and several optical clock plans for low-orbit space platforms have been established.
However, optical clocks in the low-orbit space platform are affected by the uneven distribution of the Earth's surface and geological activities, which can impede the space optical frequency standard from reaching ultra-high precision and accuracy (below $10^{-19}$).
To develop an ultrahigh-precision optical frequency standard with a stability of $10^{-21}$ or better, an optical clock plan for high-orbit platforms is desired.

Utilizing the future high-orbit satellite, it's possible develop an optical clock to promote the stability to the $10^{-21}$ level or even better to achieve ultra-high-precision optical frequency standards, and to establish a global new time reference.
Such ultrahigh-precision optical clocks meet the ultra-high-precision time synchronization requirements for wide-area quantum communication.
Furthermore, multiple space-based ultrahigh-precision optical clocks are promising for detecting gravitational waves and dark matter, which particularly complement existing gravitational wave detection techniques by providing efficient detection in the 1 Hz-10 Hz frequency band \cite{Kolkowitz2016prd}.

The work covered in this review is only the dawn of an emerging field of quantum experiments in space scale.
We expect to see both practical applications, such as global-scale quantum communications, and exciting fundamental research that was otherwise impossible to conduct on Earth.

\section{Acknowledgement}

We thank Xiao-Hui Bao, Han-Ning Dai, Yingqiu Mao, Qi Shen, Ping Xu, and Juan Yin for helpful discussions and numerous assistance.
This work is supported by the National Key R\&D Program of China (Grants No. 2017YFA0303900), the National Natural Science Foundation of China (Grants No. U1738201, No. U1738202, No. U1738203, No. U1738204), the Anhui Initiative in Quantum Information Technologies, and the Chinese Academy of Science.


%

\end{document}